\documentclass[prd,twocolumn,aps,showpacs,nofootinbib,amsmath,amssymb,floatfix,superscriptaddress,showkeys]{revtex4-1}
\usepackage{multirow}
\usepackage{graphicx}
\usepackage{epstopdf}
\usepackage{dcolumn}
\usepackage{bm}
\usepackage{threeparttable}
\usepackage{subfigure}
\usepackage{color,txfonts}
\usepackage{ulem}
\usepackage{diagbox}
\usepackage{booktabs}
\usepackage{makecell}
\usepackage{array}
\usepackage{longtable}
\usepackage{float}

\definecolor{blue0}{rgb}{0,0,0.6}
\usepackage[colorlinks,linkcolor=blue0,anchorcolor=blue0,citecolor=blue0,urlcolor=blue0]{hyperref}

\newcommand{\mnras}{Mon. Not. R. Astron. Soc.}

\newcommand{\aap}{Astron. Astrophys}
\newcommand{\aj}{Astron. J.}

\newcommand{\apjl}{Astrophys. J.}

\UseRawInputEncoding
\begin{document}
\title{Comparing hierarchical black hole mergers in star clusters\\and active galactic nuclei}

\author{\href{https://orcid.org/0000-0003-3306-5217}{Guo-Peng Li}}
\email[]{lgp@st.gxu.edu.cn}
\affiliation{Guangxi Key Laboratory for Relativistic Astrophysics, School of Physical Science and Technology, Guangxi University, Nanning 530004, China}

\author{\href{https://orcid.org/0000-0003-1474-293X}{Da-Bin Lin}}
\email[]{lindabin@gxu.edu.cn}
\affiliation{Guangxi Key Laboratory for Relativistic Astrophysics, School of Physical Science and Technology, Guangxi University, Nanning 530004, China}

\author{Yong Yuan}
\affiliation{School of Physics Science and Technology, Wuhan University, No.299 Bayi Road0, Wuhan, Hubei, China}

\date{\today}

\begin{abstract}
Star clusters (SCs) and active galactic nuclei (AGNs) are promising sites for the occurrence of hierarchical black hole (BH) mergers.
We use simple models to compare hierarchical BH mergers in two of the dynamical formation channels.
We find that the primary mass distribution of hierarchical mergers in AGNs is higher than that in SCs, with the peaks of $\sim$$50\,M_{\odot}$ and $\sim$$13\,M_{\odot}$, respectively.
The effective spin ($\chi_{\rm eff}$) distribution of hierarchical mergers in SCs is symmetrical around zero as expected and $\sim$$50\%$ of the mergers have $|\chi_{\rm eff}|>0.2$.
The distribution of $\chi_{\rm eff}$ in AGNs is narrow and prefers positive values with the peak of $\chi_{\rm eff}\ge0.3$ due to the assistance of AGN disks.
BH hierarchical growth efficiency in AGNs, with at least $\sim$$30\%$ of mergers being hierarchies, is much higher than the efficiency in SCs.
Furthermore, there are obvious differences in the mass ratios and effective precession parameters of hierarchical mergers in SCs and AGNs.
We argue that the majority of the hierarchical merger candidates detected by LIGO-Virgo may originate from the AGN channel as long as AGNs get half of the hierarchical merger rate.
\end{abstract}

\maketitle

\section{Introduction}\label{Sec.A}

At least one binary black hole (BBH) merger event in the gravitational-wave transient catalog (GWTC,~\cite{GWTC-1-2019PhRvX...9c1040A,GWTC-2LVK-2021PhRvX..11b1053A,GWTC-2.1-2021arXiv210801045T,GWTC-32021arXiv211103606T}) reported by the LIGO-Virgo-KAGRA (LVK) Collaboration is likely a hierarchical merger~\cite{O'Leary-2016-Meiron-ApJ...824L..12O,Fishbach-2017-Holz-ApJ...840L..24F,
Gerosa-2017-Berti-PhRvD..95l4046G,Kimball-2021-Talbot-ApJ...915L..35K,Mould-2022-Gerosa-PhysRevD.106.103013}.
Hierarchical mergers are expected to occur in dense stellar environments such as star clusters (SCs, e.g., nuclear star clusters, NSCs and globular clusters, GCs) and active galactic nuclei (AGNs)~\cite{Gerosa-2021-Fishbach-NatAs...5..749G}.

A second-generation (2G) black hole (BH) formed by merging a 1G BBH (1G+1G) formed from the collapse of stars can be retained by the host if the escape speed of the host stands larger than its kick recoil velocity imparted by the loss of linear momentum.
Then, the 2G BH will pair with another BH to form a 2G BBH (2G+1G or 2G+2G), merge within a Hubble time, and therefore produce a 3G BHs.
Repeatedly, there might be the occurrence of higher-generation mergers.
A $N$-G BBH (or merger) is referred to that one is a $N$-G BH and the other is a $M$-G BH ($N\ge M$), which will merge to produce a $(N+1)$-G BH (see Fig.~1 of Ref.~\cite{Mahapatra-2022-Gupta-arXiv220905766M}).
For example, a 3G BBH refers to a 3G+1G, 3G+2G, or 3G+3G, whose outcome is a 4G BH.
Hierarchical mergers have been extensively discussed in SCs (e.g.,~Refs.~\cite{Rodriguez-2019-Zevin-PhRvD.100d3027R,Kimball-2020-Talbot-ApJ...900..177K,
Baibhav-2021-Berti-PhRvD.104h4002B,
Mapelli-2021-Santoliquido-Symm...13.1678M,Mapelli-2021-Dall'Amico-MNRAS.505..339M,Li-2022A&A...666A.194L})
and AGNs (e.g.,~Refs.~\cite{Yang-2019-Bartos-PhRvL.123r1101Y,Gayathri-2020-Bartos-ApJ...890L..20G,
Tagawa-2021-Kocsis-ApJ...908..194T,Li-2022PhRvD.105f3006L}),
which can efficiently pollute the pair-instability (PI) mass gap (between $\sim$$50{-}120\,M_{\odot}$) predicted by PI supernovae~\cite{Heger-2003-Fryer-ApJ...591..288H} and Pulsational PI supernovae~\cite{Woosley-2007-Blinnikov-Natur.450..390W}.
It is also an alternate pathway to explain the growth of intermediate-mass black holes (IMBHs) in dense stellar environments~\cite{Quinlan-1987-Shapiro-ApJ...321..199Q,Fragione-2020-Loeb-ApJ...902L..26F,Fragione-2022-Kocsis-ApJ...927..231F,
Prieto-2022-Kremer-arXiv220807881G}.

Reference~\cite{Zevin-2022-Holz-ApJ...935L..20Z} studied the retention efficiency of BBH merger remnants in dense stellar clusters by considering three hierarchical merger branches: NG+1G, NG+NG, and NG+$\le$NG (NG refers to the BH generation).
By seeding, growing, and pruning the three hierarchical branches, they found that if escape velocities reach $\sim$$300\,{\rm km~s^{-1}}$, then the fraction of detectable hierarchical mergers with a source-frame total mass of $\geq$$100\,M_{\odot}$ will exceed the observed upper limit of the LVK analysis~\cite{LVK-GWTC-3-2021arXiv211103634T}.
Therefore, they stressed that some unknown mechanisms are needed to avoid a `cluster catastrophe' of overproducing BBH mergers if such environments dominate
the BBH merger rate.

NG+1G mergers are expected to preferentially occur in AGNs because of migration traps in high-density gas disks within about 300 Schwarzschild radii from the central supermassive BH~\cite{McKernan-2012-Ford-MNRAS.425..460M,Bellovary-2016-Mac-ApJ...819L..17B,Secunda-2019-Bellovary-ApJ...878...85S}.
Because merger remnants could continue to reside in migration traps and merge again with another 1G BH that aligned with the AGN disk and migrated to traps within the disk~\cite{McKernan-2018-Ford-ApJ...866...66M,Yang-2019-Bartos-PhRvL.123r1101Y,Li-2022PhRvD.105f3006L}.
While the occurrence of NG+NG mergers is preferentially in SCs because of mass segregation (e.g.,~Refs.~\cite{Scaria-1981-Bappu-JApA....2..215S,Nony-2021-Robitaille-A&A...645A..94N,Pavl-2022-Vesperini-MNRAS.515.1830P,
Vitral-2022-Kremer-MNRAS.514..806V}).
Because more massive NG BHs would concentrate on the dense core of SCs, where they will preferentially form NG+NG binaries in dynamical interactions~\cite{Rodriguez-2019-Zevin-PhRvD.100d3027R}.
NG+$\le$NG mergers include but are not limited to the mergers of NG+1G and NG+NG, which is representative of a steady-state limit~\cite{Zevin-2022-Holz-ApJ...935L..20Z}.

Previous studies focused on hierarchical BH mergers in a single formation channel or multiple channels without AGNs (e.g.,~Refs.~\cite{Gerosa-2019-Berti-PhRvD.100d1301G,Rodriguez-2019-Zevin-PhRvD.100d3027R,Yang-2019-Bartos-PhRvL.123r1101Y,Fragione-2020-Silk-MNRAS.498.4591F,
Liu-2021-Lai-MNRAS.502.2049L,Mapelli-2021-Santoliquido-Symm...13.1678M,
Mapelli-2021-Dall'Amico-MNRAS.505..339M,Mahapatra-2021-Gupta-ApJ...918L..31M,Mahapatra-2022-Gupta-arXiv220905766M,
Li-2022A&A...666A.194L,Li-2022PhRvD.105f3006L,Zevin-2022-Holz-ApJ...935L..20Z},
but~\cite{Doctor-2020-Wysocki-ApJ...893...35D,Tagawa-2021-Haiman-MNRAS.507.3362T}).
In this paper, we compare hierarchical BH mergers in SCs and AGNs using simple models that are similar in construction to previous work~\cite{Gerosa-2021-Giacobbo-ApJ...915...56G,Tagawa-2021-Haiman-MNRAS.507.3362T,Zevin-2022-Holz-ApJ...935L..20Z,
Mahapatra-2022-Gupta-arXiv220905766M}.
Because hybrid Monte Carlo and/or N-body simulations of dense stellar environments are extremely difficult to investigate the relevant parameter space of hierarchical mergers due to the computational cost.
The rest of this paper is organized as follows.
In Sec.~\ref{SecB} we describe our model framework.
In Sec.~\ref{SecC} we show our results in both SCs and AGNs.
Finally, in Sec.~\ref{SecD} we discuss our assumptions, and escape velocities and delay times, and we conclude with implications in Sec.~\ref{SecE}.

\section{Models}\label{SecB}

Following Ref.~\cite{Zevin-2022-Holz-ApJ...935L..20Z}, we consider three hierarchical BBH merger branches: NG+1G, NG+NG, and NG+$\le$NG.
We use numerical relativity fits to calculate each merger remnant's
total mass~\cite{Barausse-2012-Morozova-ApJ...758...63B},
final spin~\cite{Hofmann-2016-Barausse-ApJ...825L..19H},
and kick velocity~\cite{Campanelli-2007-Lousto-ApJ...659L...5C} (see also summaries of~Refs.~\cite{Gerosa-2016-Kesden-PhRvD..93l4066G,Mahapatra-2021-Gupta-ApJ...918L..31M}).
Table~{\ref{MyTabA}} lists the summary of our models we will cover below.

\subsection{First-generation BHs}\label{SecB1}

We adopt a 1G BH mass distribution in dense stellar environments as $p(m)\propto m^{-\alpha}$.
The range of BH masses $m \in [5\,M_{\rm \odot},50\,M_{\rm \odot}]$ is adopted, which is determined by the lower and PI mass gap.
We adopt $\alpha = 2.3$ in SCs corresponding to the Kroupa initial mass function~\cite{Kroupa-2001MNRAS.322..231K}; $\alpha = 1$ within AGN disks because the disks harden the initial BH mass function~\cite{Yang-2019-Bartos-ApJ...876..122Y}.

We assume a uniform spin magnitude distribution: U(0,$\chi_{\rm max}$) with $\chi_{\rm max}=0.2$ in SCs~\cite{LVK-GWTC-3-2021arXiv211103634T}.
Spin tilt angles for all BH generations are isotropically drawn over a sphere.
However, the spin of BHs in AGN disks may be significantly altered under accretion.
The misalignment angle $\theta$ between the spin and the orbital angular momenta changed with ${\rm cos} \theta \rightarrow 1$ or $-1$~\cite{Yi-2019-Cheng-ApJ...884L..12Y}.
Whereas the vast majority should have $\theta \le \pi/2$ because gas accretion from AGN disks will tend to torque the BH spin into alignment with the gas~\cite{Bogdanovi-2007-Reynolds-ApJ...661L.147B,Mould-2022-Gerosa-PhysRevD.106.103013,McKernan-2022-Ford-MNRAS.514.3886M},
which also causes that spin magnitudes are going to be higher overall under accretion~\cite{Yi-2019-Cheng-ApJ...884L..12Y,McKernan-2020-Ford-MNRAS.494.1203M}.
For simplicity, we neglect the case of ${\rm cos} \theta <0$, which should be a very few part and not make a difference to our results.
Therefore, we adopt $\chi_{\rm max}=0.4$, and ${\rm cos} \theta$ between 0 and 1 according to a distribution uniform in $p({\rm cos} \theta) \propto {\rm cos} \theta$ in AGN disks.
Here, a higher-spin distribution in AGN disks made is because the black holes there should have relatively high spins due to gas accretion.
We also adopt $\chi_{\rm max}=0.01$ and 0.4 and $\chi_{\rm max}=0.2$ and 1 in SCs and AGN disks, respectively, for comparison, which involves the same spin distribution in both SCs and AGN disks.

We draw the primary component BH mass ($m_1$) of a 1G binary (i.e., 1G+1G) according to the above distributions.
Then, we pair it with another component BH according to $m_2 = m_1q$ ($m_1 \ge m_2 \ge 5\,M_{\odot}$, $q$ is the mass ratio of a BBH) with $p(q) \propto q^{\beta}$, and adopt $q$ between the bounds [0.3,~1].
We consider two values for $\beta$: 1.08 for SCs and 0 for AGNs.
$\beta \sim 1.08$ is inferred from the GWTC-3 by the LVK Collaboration~\cite{LVK-GWTC-3-2021arXiv211103634T}.
$\beta = 0$ represents random pairing, which is expected in AGN disks because of runaway mergers in migration traps~\cite{Yang-2019-Bartos-PhRvL.123r1101Y,Li-2022PhRvD.105f3006L}.
We also consider $\beta = 5$ for SCs indicating `strong' mass segregation, and we adopt $\beta = 1.08$ for AGN disks for comparison if migration traps are inefficient.

We note that a predictively initial BH mass distribution from the Power Law + Peak model of Ref.~\cite{LVK-GWTC-3-2021arXiv211103634T} is also considered by Ref.~\cite{Zevin-2022-Holz-ApJ...935L..20Z}.
The difference between these two distributions is the latter allows BH masses to be in the PI mass gap because it probably includes merger remnants, which means it is not representative of a true distribution of 1G black hole masses.
Therefore, we do not consider it in our models.
Reference~\cite{Mahapatra-2022-Gupta-arXiv220905766M} considered $\beta = -1$ that prefers asymmetric binaries, although it is in disfavor of the observed results.
However, they have shown that if the pairing prefers equal-mass binaries, then the 2G and 3G mergers are consistent with two of the subdominant peaks of the predictive BH mass spectrum from the Flexible Mixture model~\cite{Tiwari-2021CQGra..38o5007T,Tiwari-2022ApJ...928..155T}.

\begin{table}
\centering{
\caption{Summary of the models.}\label{MyTabA}
\begin{threeparttable}
\begin{tabular}{lcccccc}
\hline
Model & $\alpha$ & $\chi_{\rm max}$ & Spin direction &$\beta$ & $V_{\rm esc}~[{\rm km~s^{-1}}]$ & $t_{\rm min}~[{\rm Myr}]$\\
\hline
SC\_1 & 2.3 & 0.2 & Isotropic & 1.08 & 100 & 10  \\
SC\_2 & 2.3 & 0.2 & Isotropic & 1.08  & 50 & 10 \\
SC\_3 & 2.3 & 0.2 & Isotropic & 1.08  & 200 & 10  \\
SC\_4 & 2.3 & 0.2 & Isotropic & 1.08  & 300 & 10  \\
SC\_5 & 2.3 & 0.2 & Isotropic & 1.08 & 500 & 10  \\
SC\_6 & 2.3 & 0.2 & Isotropic & 1.08  & 100 & 0.1  \\
SC\_7 & 2.3 & 0.2 & Isotropic & 1.08  & 100 & 100 \\
SC\_8 & 2.3 & 0.2 & Isotropic &  5 & 100 & 10  \\
SC\_9 & 2.3 & 0.01 & Isotropic & 1.08  & 100 & 10  \\
SC\_10 & 2.3 & 0.4 & Isotropic & 1.08  & 100 & 10  \\
AGN\_1 & 1 & 0.4 & Anisotropic & 0 & $\infty$ &  0.1  \\
AGN\_2 & 1 & 0.4 & Anisotropic & 0  & $10^3$ &  0.1  \\
AGN\_3 & 1 & 0.4 & Anisotropic & 0  & $\infty$ & 0.01  \\
AGN\_4 & 1 & 0.4 & Anisotropic & 0  & $\infty$ & 1  \\
AGN\_5 & 1 & 0.4 & Anisotropic & 0  & $\infty$ & 2  \\
AGN\_6 & 1 & 0.4 & Anisotropic & 1.08  & $\infty$ & 0.1  \\
AGN\_7 & 1 & 0.2 & Anisotropic & 0  & $\infty$ & 0.1  \\
AGN\_8 & 1 & 1 & Anisotropic & 0  & $\infty$ & 0.1  \\
\hline
\end{tabular}
\footnotesize{Column 1: Name of the model. `SC\_$i$ ' represents the SC-like environment; `AGN\_$i$ ' represents the AGN-like environment.
Column 2: The mass index $\alpha$.
Column 3: The maximum initial spin $\chi_{\rm max}$.
Column 4: The spin direction for all BH generations. `Isotropic' represents spin tilt angles are isotropically drawn over a sphere; `Anisotropic' represents the
misalignment angle $\theta$ obeying a  distribution uniform in $p({\rm cos} \theta) \propto {\rm cos} \theta$ spanning from 0 and 1.
Column 5: The mass-ratio index $\beta$.
Column 6: The escape velocity $V_{\rm esc}$. $V_{\rm esc} = \infty$ represents that the kicks of merger remnants are neglected.
Column 7: The delay times $\Delta t$ between the subsequent mergers.}
\end{threeparttable}}
\end{table}

\begin{figure*}
\centering
\includegraphics[width=0.8\textwidth]{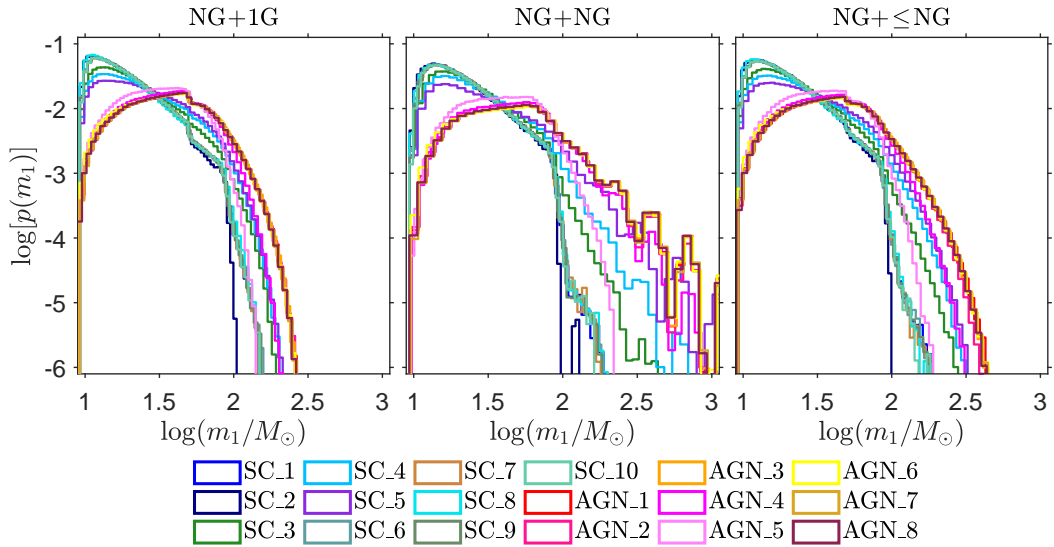}
\caption{The probability density distribution of the primary masses ($m_1$) of hierarchical BH mergers.
The columns show the three hierarchical branches (i.e., NG+1G, NG+NG, and NG+$\le$NG).
The different lines in each pixel show the eighteen models listed in Table~\ref{MyTabA}.
Each line plotted contains the contributions of all hierarchical merger generations and the fraction of each merger generation is obtained from Table~\ref{MyTabB} in Appendix~\ref{appA}.
}
\label{MyFigA}
\end{figure*}
\subsection{Constraining hierarchical growth efficiency}\label{SecB2}

We constrain the growth efficiency of hierarchical mergers by escape velocities and delay times.

\begin{itemize}
    \item{
    We drop all subsequent mergers if $V_{\rm kick} \ge V_{\rm esc}$, where $V_{\rm kick}$ is the kick velocity of the merger remnant and $V_{\rm esc}$ is the escape velocity of the host.
    The kick velocities inferred from the GWTC events can lie in a wide range: $\sim$$50{-}2000\,{\rm km\,s^{-1}}$~\cite{Mahapatra-2022-Gupta-arXiv220905766M,Varma-2022-Biscoveanu-PhRvL.128s1102V}.
    In comparison, the escape speed  is $\sim$$2{-}100\,{\rm km\,s^{-1}}$ for GCs~\cite{Antonini-2016-Rasio-ApJ...831..187A},
    $\sim$$10{-}600\,{\rm km\,s^{-1}}$ for NSCs~\cite{Antonini-2016-Rasio-ApJ...831..187A},
    and up to $\sim$$1000\,{\rm km\,s^{-1}}$ in AGN disks within an inner radii.
    The kicks of merger remnants in AGNs are generally neglected by the previous works~\cite{Tagawa-2020-Haiman-ApJ...898...25T,Yang-2020-Bartos-ApJ...901L..34Y,Li-2022PhRvD.105f3006L}, because of the large orbital velocities $\sim$$2\times 10^4\,{\rm km\,s^{-1}}$ and the small kick magnitude due to BH spins are largely aligned or antialigned with the disk~\cite{McKernan-2020-Ford-MNRAS.494.1203M}.
    }

    \item{
    BBH mergers can occur before the present day. We draw the delay times between the subsequent mergers according to a distribution uniform in $p(\Delta t) \propto {\Delta t }^{-1}$ with $\Delta t = t_{N_{\rm merg}+1}-t_{N_{\rm merg}}$~\cite{Dominik-2012-Belczynski-ApJ...759...52D,Mapelli-2021-Santoliquido-Symm...13.1678M}.
    For SCs, we span $\Delta t $ from $t_{\rm min}= 10\,{\rm Myr}$ to $t_{\rm max}= 1.4\times 10^4\,{\rm Myr}$~\cite{Carlo-2020-Mapelli-MNRAS.497.1043D,Zevin-2022-Holz-ApJ...935L..20Z}.
    The time efficiency of BBH formation and merger is significantly high under the assistance of AGN disks.
    In AGN disks, the characteristic time of the migration for BHs is $10^5\,{\rm yr}$~\cite{McKernan-2012-Ford-MNRAS.425..460M,Bartos-2017-Kocsis-ApJ...835..165B},
    which is also much larger than the merger time of $\lesssim$$10^4\,{\rm yr}$~\cite{Bartos-2017-Kocsis-ApJ...835..165B,Secunda-2019-Bellovary-ApJ...878...85S}.
    We assume that delay times in AGN disks spanning from $t_{\rm min}= 0.1\,{\rm Myr}$ to $t_{\rm max}= 10\,{\rm Myr}$.
    For comparison, we also adopt that $t_{\rm min}= 0.1\,{\rm Myr}$ and $100\,{\rm Myr}$ for SCs~\cite{Mapelli-2021-Santoliquido-Symm...13.1678M}
    and $t_{\rm min}= 0.01\,{\rm Myr}$, $1\,{\rm Myr}$, and $2\,{\rm Myr}$ for AGNs. We discard the mergers that occurred $10\,{\rm Myr}$ later in AGNs.
    }

\end{itemize}

\begin{figure*}
\centering
\includegraphics[width=0.8\textwidth]{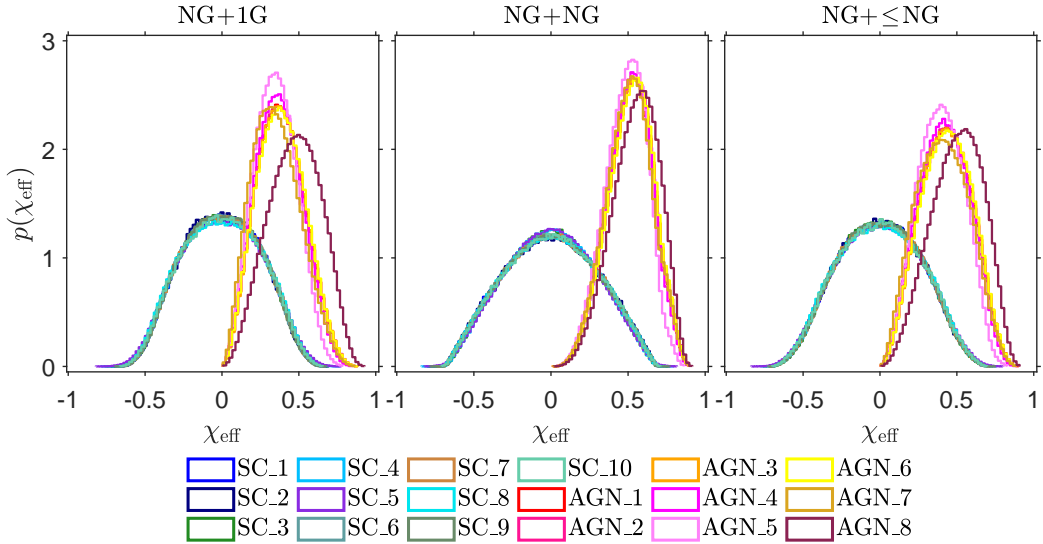}
\caption{Same as Fig.~\ref{MyFigA}, but the probability density distribution of the effective spins ($\chi_{\rm eff}$) of hierarchical BH mergers.}
\label{MyFigB}
\end{figure*}

\subsection{Synthesizing hierarchical mergers}\label{SecB3}

We first produce $N_{\rm 1G}=10^6$ 1G BHs according to the previously described mass distributions and spin distributions and pair them according to the previously described mass-ratio distribution.
We calculate their kick velocities and merger times to select the remnants (i.e., 2G BHs) with the number of $N_{\rm 2G}=N_{\rm 1G}-N_{\rm 1G}^{\prime}$  that were retained by the host and occurred before the present day.
We randomly pair these 2G BHs with 1G BH population and 2G BH population for NG+1G and NG+NG mergers, respectively.
For NG+$\le$NG mergers, we pair each 2G BH with a BH with the generation $M$ ($M\le N$).
The probability of the generation $M$ obeys $p(M) \propto 2^{-(M-1)}$~\cite{Zevin-2022-Holz-ApJ...935L..20Z}.
For example, a NG BH is twice as likely to merge with a 1G BH than a 2G BH, and four times as likely to merge with a 1G BH than a 3G BH.
We constrain the fraction of mergers with generation $N$ to $f(N)\le 2^{-N}$.
For example, there has at most  100 2G mergers and 50 3G mergers if only 200 1G mergers occur.
The merger generation with $N$ contains $N$ merger types: NG+1G, NG+2G, \dots, and NG+NG.
We repeat the above method to obtain the higher-generation merger population and stop our iteration until all BHs but one have been either ejected or accreted.

\section{Results}\label{SecC}

\subsection{Mass distribution}\label{SecC1}

We show the primary BH mass distribution of hierarchical mergers (i.e., excluding 1G mergers) in Fig.~\ref{MyFigA}.
There is a distinct difference between the masses of hierarchical mergers in SCs and AGNs, in which the distributions with wide ranges in AGNs are higher than that in SCs due to the hard initial mass spectrum and efficient hierarchical mergers (see Table~\ref{MyTabB} in Appendix~\ref{appA}).
The peaks of the distributions in SCs are $\sim$$11{-}15\,M_{\odot}$ as similar with Ref.~\cite{Mahapatra-2022-Gupta-arXiv220905766M},
while that in AGNs can reach up to $\sim$$50\,M_{\odot}$ being consistent with Ref.~\cite{Yang-2019-Bartos-PhRvL.123r1101Y}.
The NG+1G mergers have relatively low masses because one of each of them came from a 1G BH that has a mass of $\le$$50\,M_{\odot}$.
Whereas the NG+NG mergers have relatively high masses because the binaries are in favor of symmetric masses.

We find that the hierarchical mergers for all the different cases can efficiently pollute the PI mass gap and IMBHs, especially in AGN disks.
We see that the escape velocities play an important role for hierarchical merges in SCs.
The small escape velocity represents the inefficiency of hierarchical merges, which causes low merger masses; the larger the escape velocity, the higher the masses.
The high-mass end of the distributions for the cases with different escape velocities has significant differences; in particular, the masses of the NG+NG mergers can reach up to $\gtrsim$$1000\,M_{\odot}$.
The pairing probability of $\beta=5$ (SC\_8) could upraise the mass distribution at the high-mass end.

For the hierarchical merges in AGNs, the mass distributions for all the different cases (excluding AGN\_5) are no significant differences.
Because all the mergers could be retained in migration traps, and the delay times are relatively short with the assistance of AGN disks, resulting in almost the same fraction in the same merger generation for the different model (see Table~\ref{MyTabB} in Appendix~\ref{appA}).
This also results in $\sim$$30{-}50\%$ of the merging BBHs being hierarchical mergers.

\subsection{Spin distribution}\label{SecC2}

\begin{figure*}
\centering
\begin{minipage}{0.49\linewidth}
\includegraphics[height=0.65\textwidth, width=0.95\textwidth]{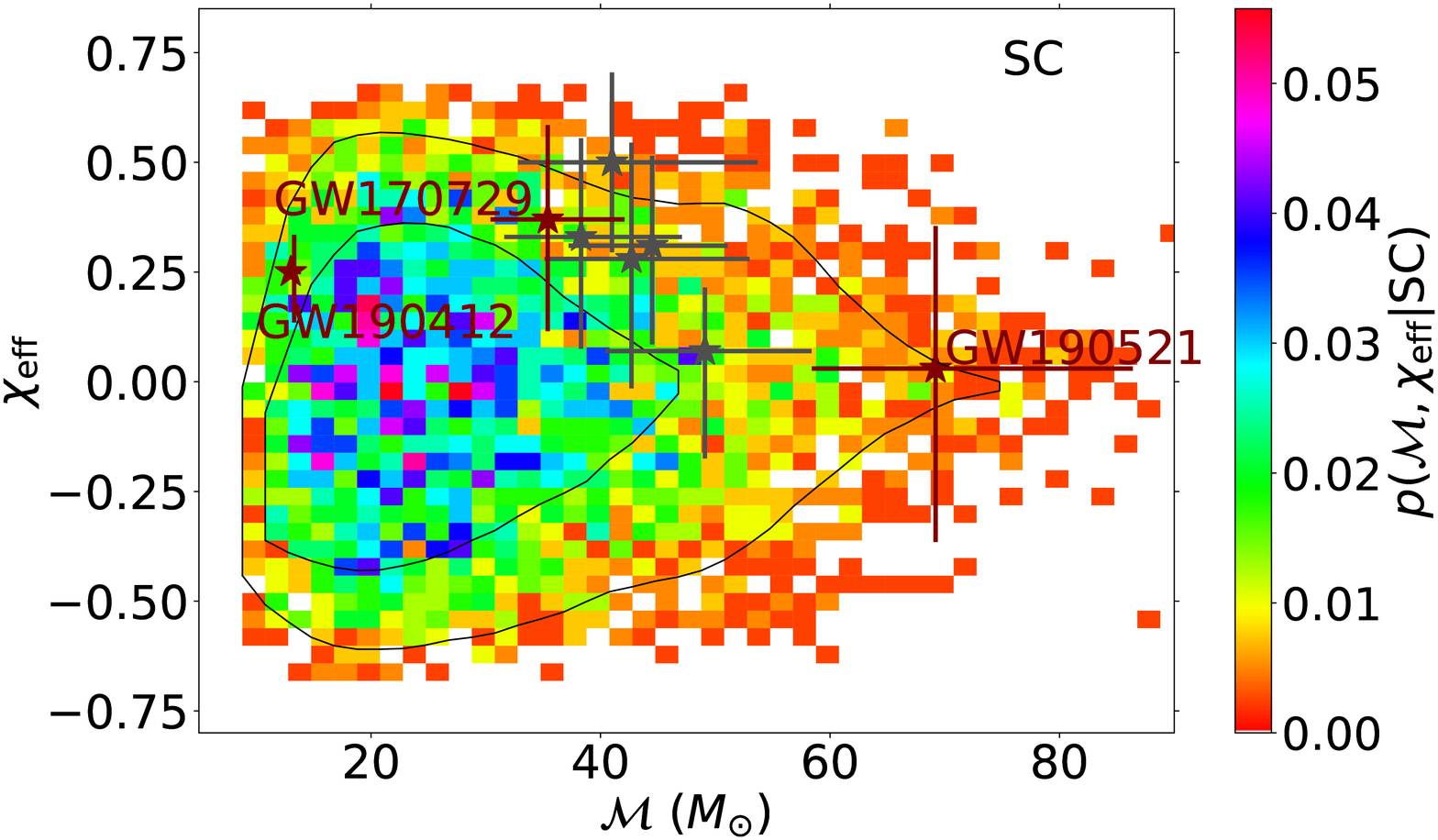}
\end{minipage}
\begin{minipage}{0.49\linewidth}
\includegraphics[height=0.65\textwidth, width=0.95\textwidth]{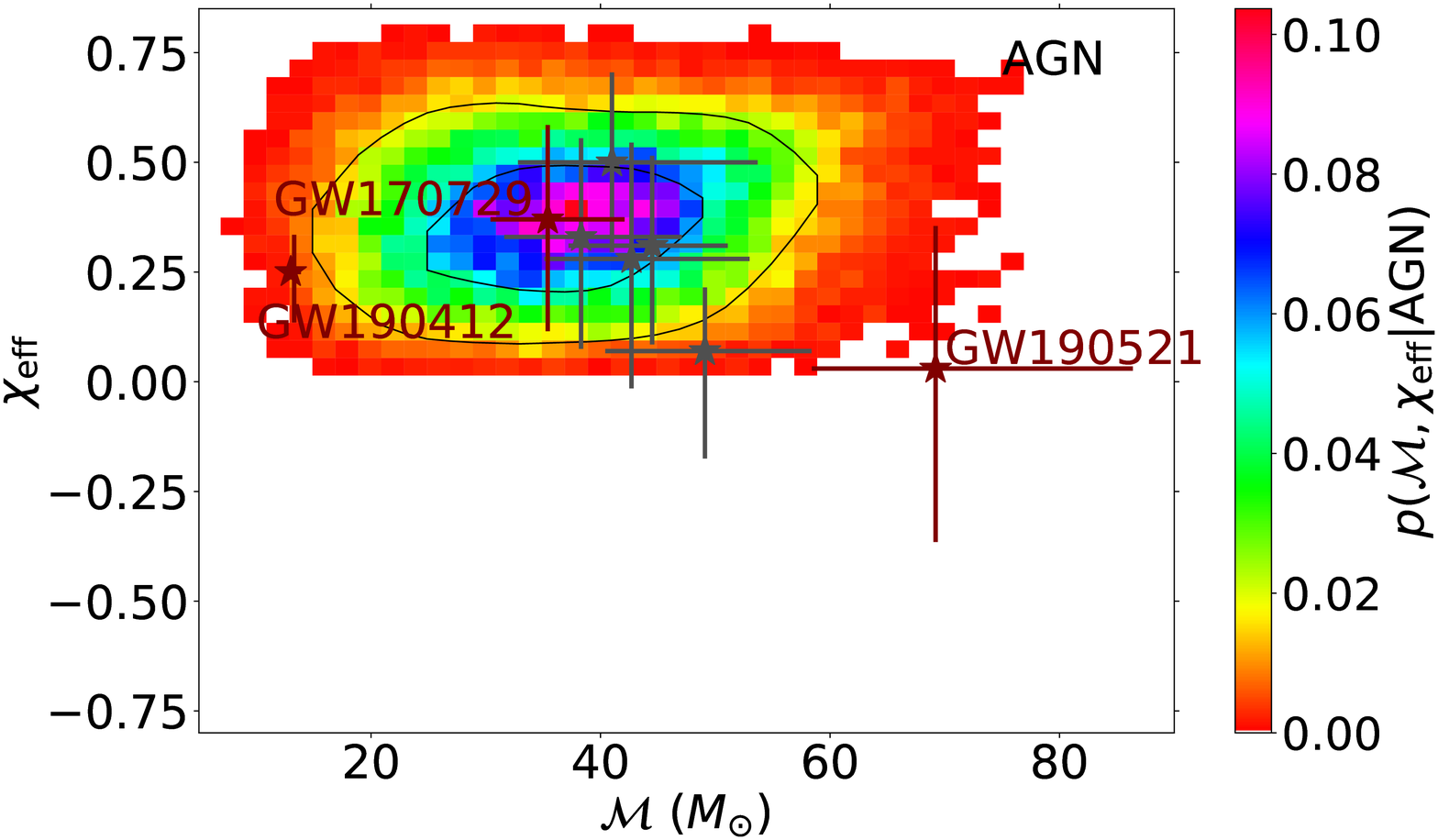}
\end{minipage}
\caption{2D probability densities of the chirp mass ($\mathcal{M}$) and effective spin ($\chi_{\rm eff}$) of hierarchical BH mergers in SCs (left) and AGNs (right).
The two fiducial models (i.e., SC\_1 and AGN\_1) are adopted for SCs and AGNs, respectively.
The black solid lines show $50\%$ and $90\%$ confidence regions.
The hierarchical branches with NG+NG and NG+1G are assumed in SCs and AGNs, respectively.
Because we expect that the mergers of NG+NG and NG+1G dominate the hierarchical merger rates in SCs and AGNs, respectively.
We also show eight promising GW candidate events (star symbols) for hierarchical mergers, in which GW170729~\cite{GWTC-1-2019PhRvX...9c1040A}, GW190412~\cite{GW190412-2020PhRvD.102d3015A}, and GW190521~\cite{GW190521-2020PhRvL.125j1102A} are highlighted.}
\label{MyFigC}
\end{figure*}

In Fig.~\ref{MyFigB}, we plot the probability density distribution of the effective spins ($\chi_{\rm eff}$) of hierarchical BH mergers.
$\chi_{\rm eff} = (m_1 \chi_1 {\rm cos}\theta_1+m_2 \chi_2 {\rm cos}\theta_2)/(m_1+m_2)$, where $m_i$, $\chi_i$, and $\theta_i$  are the mass, the dimensionless spin, and the misalignment angle, respectively, of each BH in a merged BBH.

We see that the distributions in SCs are symmetrical around zero as expected due to random spin directions.
However, they have a wide range from $\sim$$-0.75$ to $\sim$$0.75$ with $\sim$$50\%$ of the mergers have $|\chi_{\rm eff}|\ge0.2$ because the final spins of 1G mergers concentrate on 0.69, which the similar results were obtained by Refs.~\cite{Rodriguez-2019-Zevin-PhRvD.100d3027R,Mapelli-2021-Dall'Amico-MNRAS.505..339M}.
The distributions with the peaks of $\chi_{\rm eff}\ge0.3$ in AGNs are narrower and always greater than 0 because we assume that the misalignment angles of the BBHs are less than $\pi/2$.
The reason for this assumption is that gas accretion from the AGN disk will tend to torque the BH spin direction into alignment with the disk orbital angular momentum~\cite{Bogdanovi-2007-Reynolds-ApJ...661L.147B,McKernan-2022-Ford-MNRAS.514.3886M}.

We find that there are no differences between $\chi_{\rm eff}$ either in SCs or in AGNs if variations to the hierarchical branches are fixed because the finally spins of any merger generations lie in a stable range from $\sim$$0.5$ to $\sim$0.8~\cite{Gerosa-2021-Giacobbo-ApJ...915...56G,Zevin-2022-Holz-ApJ...935L..20Z}.
That indicates that the effective spin distribution of hierarchical mergers weakly depends on escape velocities and delay times.
In SCs, the distribution of $\chi_{\rm eff}$ of NG+NG mergers is relatively wider than that of the other two hierarchical branches, though not obvious.
In AGNs, the peaks of the distributions of $\chi_{\rm eff}$ of the mergers of NG+1G, NG+$\le$NG, NG+NG increase in turn to $\sim$$0.32$, $\sim$$0.4$, and $\sim$$0.5$, respectively, which means equal-mass BBH mergers have large effective spins.
The peak values of the distributions in AGNs broadly agree with the distributions of the 2G and 3G mergers in Ref.~\cite{Yang-2019-Bartos-PhRvL.123r1101Y}.

Figure~\ref{MyFigB} also shows that the gravitational-wave (GW) events with large $\chi_{\rm eff}$ reported by LVK~\cite{GWTC-32021arXiv211103606T} most likely originate from AGNs because $\chi_{\rm eff}$ of the merger form isolated binary evolution tend to be positive close to zero, while that from SCs also centers zero (see also Fig.~\ref{MyFigC}).
The distribution of the model of AGN\_8 is higher than others because we adopt the maximum initial BH spin is $1$.

\subsection{Comparison with the promising candidates}\label{SecC3}

We would expect that NG+1G and NG+NG mergers dominate the hierarchical BH merger rates in AGNs and SCs, respectively, because of migration traps and mass segregation.
We show 2D probability densities of the chirp mass ($\mathcal{M} = (m_1m_2)^{3/5}/(m_1+m_2)^{1/5}$) and effective spin ($\chi_{\rm eff}$) of the hierarchical BH mergers detected by LIGO-Virgo~\cite{LIGO-2015CQGra..32g4001L,Virgo-2015CQGra..32b4001A} in SCs and AGNs in Fig.~\ref{MyFigC}~\cite{Finn-1993-Chernoff-PhRvD..47.2198F}.
In the left panel, we plot the detectable mergers in SCs with the model of SC\_1 and the hierarchical branch of NG+NG, and in the right is the detectable mergers in AGNs with the model of AGN\_1 and the hierarchical branch of NG+1G.
We assume that redshifts of the mergers are drawn uniformly in comoving volume between $z\in[0,2]$, and that the generated gravitational waves conform to PhemonA~\cite{Ajith-2007-Babak-CQGra..24S.689A}.
We calculate the signal-to-noise ratio (SNR) according to $\rho^2 = \frac{16}{5} \int \frac{(2fT)S_{\rm h}(f)}{S_{\rm n}(f)}d(\ln f)$, where $f$ is frequency of the gravitational wave, $T$ is the observation time, $S_{\rm h}(f)$ is the one-sided, averaged, power spectral density of the signal, and $S_{\rm n}(f)$ is the noise sensitivity curve of LIGO~\cite{LVK-2020-LRR....23....3A}.
When SNR $>$$8$, we consider the signal to be detectable~\cite{Abadie-2010-Abbott-CQGra..27q3001A}.
We see that the distribution with the densest region located at $\mathcal{M}\sim 20\,M_{\odot}$ and $\chi_{\rm eff}\sim 0$ in SCs has a wider range than that with the densest region located at $\mathcal{M}\sim 40\,M_{\odot}$ and $\chi_{\rm eff}\sim 0.4$ in AGNs.

GW170729~\cite{GWTC-1-2019PhRvX...9c1040A},
GW170817A~\cite{GW170817A-2021PhRvD.104f3030Z},
GW190412~\cite{GW190412-2020PhRvD.102d3015A},
and GW190521~\cite{GW190521-2020PhRvL.125j1102A} are promising candidates for hierarchical mergers (e.g.,~\cite{Yang-2019-Bartos-PhRvL.123r1101Y,Gayathri-2020-Bartos-ApJ...890L..20G,Gerosa-2020-Vitale-PhRvL.125j1103G,
Abbott-2020-Abbott-ApJ...900L..13A}, see also a review of Ref.~\cite{Gerosa-2021-Fishbach-NatAs...5..749G}), which we plot them in Fig.~\ref{MyFigC}.
Moreover, GW190519, GW190602, GW190620, and GW190706 in the GWTC-2~\cite{GWTC-2LVK-2021PhRvX..11b1053A} are also promising candidate events found by Ref.~\cite{Kimball-2021-Talbot-ApJ...915L..35K}, although they used globular models that imply that these events may not be hierarchical merger candidates if they originate from other channels such as an AGN disk.
We find that most of the hierarchical merger candidates are consistent with the AGN channel because of the large chirp masses and high effective spins.
Thus, most of the hierarchical merger candidate events detected by LIGO-Virgo~\cite{LIGO-2015CQGra..32g4001L,Virgo-2015CQGra..32b4001A} may originate from the AGN channel if AGNs in all probability dominant the hierarchical BH merger rate~\cite{Yang-2019-Bartos-PhRvL.123r1101Y,Ford-2022-McKernan-MNRAS.tmp.2697S}.

We show that GW170729~\cite{Yang-2019-Bartos-PhRvL.123r1101Y,Chatziioannou-2019-Cotesta-PhRvD.100j4015C,Fishbach-2020-Farr-ApJ...891L..31F,
Sedda-2020-Mapelli-ApJ...894..133A} could be well explained in the AGN channel. Reference~\cite{Yang-2019-Bartos-PhRvL.123r1101Y} also showed that it could have originated from this channel, although not definitively (with odds ratio of $\sim$1).
It is possible that GW190412~\cite{Gerosa-2020-Vitale-PhRvL.125j1103G,Hamers-2020-Safarzadeh-ApJ...898...99H,Rodriguez-2020-Kremer-ApJ...896L..10R,
Zevin-2020-Berry-ApJ...899L..17Z} originated from SCs or AGNs.
However, GW190412 has a component BH with the mass of $\sim$$8\,M_{\odot}$ that should be a 1G BH, which implies it is more likely to come from an AGN because NG+1G mergers prefer to occur in AGNs.
GW190521~\cite{Abbott-2020-Abbott-ApJ...900L..13A,Fragione-2020-Loeb-ApJ...902L..26F,Kimball-2021-Talbot-ApJ...915L..35K,Anagnostou-2022-Trenti-ApJ...941....4A} is in disfavor of originating from the AGN channel because of $\chi_{\rm eff}$ nears zero; it has relatively symmetric masses with a total mass of $\sim$$150\,M_{\odot}$, which suggests it would be an NG+NG merger.
Therefore, GW190521 may originate from a SC, but even within SCs, it is still an extremely rare case.

\section{Discussion}\label{SecD}

The assumption that the mergers of NG+1G and NG+NG dominate the hierarchical merger rates of AGNs and SCs, respectively, relies on the efficiency of migration traps and mass segregation.
Reference~\cite{Li-2022PhRvD.105f3006L} has shown that the NG+1G binaries dominate hierarchical BH mergers in AGNs with the percentage in hierarchical mergers being at least $\sim$$90\%$ by neglecting migration times and considering that the BHs reach the migration trap region once they align with their orbits with the AGN disk.
In Ref.~\cite{Li-2022A&A...666A.194L}, we predicted that the branching ratio of the mergers of 2G+1G and 2G+2G in SCs is $\gtrsim$20 by neglecting the pairing probability.
However, this could go into reverse if the pairing probability is strongly in favor of equal-mass binaries because of mass segregation.
We expect to identify whether NG+NG or NG+1G dominates hierarchical mergers in SCs by the observation of future ground-based GW detectors, which is also a test for the efficiency of migration traps and mass segregation.

Generally, the initial BH mass function in dense stellar environment depends on metallicity~\cite{Sedda-2020-Mapelli-ApJ...894..133A,Fragione-2020-Loeb-ApJ...902L..26F,Mapelli-2021-Dall'Amico-MNRAS.505..339M}
that we have not considered in our models.
Most GCs are low-metallicity environments~\cite{Harris-1996AJ....112.1487H}, which therefore can form much more massive BHs~\cite{Vink-2001-Koter-A&A...369..574V,Spera-2017-Mapelli-MNRAS.470.4739S}.
Both low- and high-metallicity stars are in NSCs because of their complex history and various episodes of accretion and star formation (e.g.,~Ref.~\cite{Antonini-2013ApJ...763...62A}).
We also have ignored the increase in mass of BHs in AGN disks under accretion~\cite{Yi-2018-Cheng-ApJ...859L..25Y,Yang-2020-Bartos-ApJ...901L..34Y}.
These may change our results of masses slightly.

The kick velocities of merger remnants are sensitive to BH spins;
low spins are in favor of the relatively small kick velocities imparted to merger remnants~\cite{Rodriguez-2019-Zevin-PhRvD.100d3027R,Fragione-2021-Loeb-MNRAS.502.3879F}.
Possibly, the occurrence of hierarchical mergers in young star clusters if the kick velocities are small enough~\cite{Gerosa-2021-Fishbach-NatAs...5..749G,Mapelli-2021-Santoliquido-Symm...13.1678M,Mapelli-2021-Dall'Amico-MNRAS.505..339M}.
The rate of hierarchical mergers in SCs depends on the escape velocities of host clusters.
Reference~\cite{Gerosa-2019-Berti-PhRvD.100d1301G} showed that the SC with an escape velocity of $\ge$$50\,{\rm km\,s^{-1}}$ could populate the PI mass gap.
Moreover, the results of Ref.~\cite{Zevin-2022-Holz-ApJ...935L..20Z} indicated that there is a `cluster catastrophe' of an abundance of high-mass mergers if the SCs with escape velocities of $\sim$$300\,{\rm km\,s^{-1}}$ dominate the BBH merger rate.
Therefore, the kick velocities between $\sim$$50\,{\rm km\,s^{-1}}$ and $\sim$$300\,{\rm km\,s^{-1}}$ are appropriate to hierarchical mergers in SCs, although Ref.~\cite{Mahapatra-2022-Gupta-arXiv220905766M} found that two of the subdominant peaks of the predictive BH mass spectrum are consistent with the 2G and 3G mergers with escape velocities of $\sim$$500\,{\rm km\,s^{-1}}$.
In our models, the hierarchical merger efficiency with $\sim$$50\%$ of the mergers being hierarchies would be too high if the SCs with escape velocities of $\sim$$500\,{\rm km\,s^{-1}}$ dominate the BBH merger rate (see Table~\ref{MyTabB} in Appendix~\ref{appA}).

The hierarchical merger rate in AGNs is determined by delay times (i.e., migration times) in our models.
Because the kick velocities of merger remnants are always less than the escape velocity in AGN disks due to the large orbital velocities and the appropriate misalignment angle~\cite{Bogdanovi-2007-Reynolds-ApJ...661L.147B,McKernan-2012-Ford-MNRAS.425..460M,Yi-2019-Cheng-ApJ...884L..12Y,
McKernan-2020-Ford-MNRAS.494.1203M}.
If migration times are short, then the fraction of hierarchical mergers can reach up to $\sim$$50\%$ in all three hierarchical branches (see Table~\ref{MyTabB} in Appendix~\ref{appA}).
Reference~\cite{Ford-2022-McKernan-MNRAS.tmp.2697S} predicted that the BBH merger rate in AGNs is larger than that of NSCs and contributes $\sim$$25\%{-}80\%$ of the LIGO-Virgo measured rate of $\sim$$24\,{\rm Gpc^{-3}\,yr^{-1}}$~\cite{LVK-GWTC-3-2021arXiv211103634T}.
Moreover, Ref.~\cite{McKernan-2020-Ford-MNRAS.494.1203M} found that $\sim$$80\%{-}90\%$ of mergers occur away from migration traps, and $\sim$$10\%{-}20\%$ of mergers occur at traps, which means most mergers occur within migration times.
These show that multibody interactions~\cite{Secunda-2021-Hernandez-ApJ...908L..27S,Wang-2021-McKernan-ApJ...923L..23W,Samsing-2022-Bartos-Natur.603..237S,
Li-2022-Dempsey-arXiv221110357L} and/or the efficiency of migration traps~\cite{McKernan-2012-Ford-MNRAS.425..460M,Bellovary-2016-Mac-ApJ...819L..17B,Secunda-2019-Bellovary-ApJ...878...85S,Pan-2021-Yang-PhRvD.103j3018P,
Peng-2021-Chen-MNRAS.505.1324P} in AGN disks may play an important role if the efficiency of hierarchical mergers is overestimated by us, although we can constrain it by rising migration times.

We note that one of the key conclusions is that the values of $\chi_{\rm eff}$ for AGN disks are mostly positive.
This is due to the assumption about alignment made in Sec.~\ref{SecB1}.
However, Ref.~\cite{Secunda-2020-Bellovary-ApJ...903..133S} found that $68\%$ of the BBHs in their simulation orbit in the retrograde direction, which implies that BBHs would have small $\chi_{\rm eff}$.
We expect that we could probe the likely torquing by disk accretion onto the embedded objects by testing the population of BBH mergers in AGN disks.
GCs are believed to be a major contributor to the rate of dynamically formed LIGO-Virgo events~\cite{Mapelli-2022-Bouffanais-MNRAS.511.5797M,Mandel-2022-Broekgaarden-LRR....25....1M}.
However, most of the cluster models considered here have an escape speed of $\ge$$100\,{\rm km\,s^{-1}}$, which implies that the models assumed only applied to NSCs.
This would weakens the results presented in this paper and should be taken into account when interpreting our results.
\section{Conclusions}\label{SecE}

In this paper, we compare hierarchical BH mergers in SCs and AGNs using simple models.
We mainly focus on the differences of hierarchical mergers between SCs and AGNs, not on the differences within SCs or AGNs under different model parameters.
In our models, the two dynamical BBH formation channels are distinguished by initial BH distributions in mass and spin, pairing probabilities, escape velocities, and delay times.
We show that hierarchical mergers in mass and spin have significantly differences in between SCs and AGNs regardless of the model parameters.
We stress that our estimates should be seen as upper limit because of neglecting multibody interactions and the efficiency of migration traps and mass segregation.
Our conclusions are as follows:

\begin{itemize}
    \item{The primary mass distribution of the hierarchical mergers in AGNs, with the peak of $\sim$$50\,M_{\odot}$ and with a wide range, is higher than that with the peak of $\sim$$13\,M_{\odot}$ in SCs (see Fig.~\ref{MyFigA}). The hierarchical mergers in both AGNs and SCs can pollute the PI mass gap, and it is mare effective for mergers in AGN disks to fill IMBHs. Compared with SCs, the hierarchical mergers in AGNs prefer asymmetric masses (see Fig.~\ref{MyFigD} in Appendix~\ref{appB}).
        }

    \item{The effective spin distribution of hierarchical mergers in SCs is symmetrical around zero as expected, in which $\sim$$50\%$ of the mergers have $|\chi_{\rm eff}|>0.2$, while that in AGNs is narrower and prefers positive values with the peak of $\chi_{\rm eff}\ge0.3$ with the assistance of AGN disks (see Fig.~\ref{MyFigB}). The distribution of $\chi_{\rm eff}$ weakly depends on escape velocities and delay times. The effective precession parameter distribution with the peak of $\chi_{\rm p}\sim0.66$ in SCs are much narrower than that in AGNs; the distribution of $\chi_{\rm p}$ in AGNs is flat, especially for NG+1G mergers, because of the assistance of AGN disks (see Fig.~\ref{MyFigE} in Appendix~\ref{appC}).
        }

    \item{The hierarchical BH merger rate in SCs strongly depends on the escape velocities of clusters, while that in AGNs depends on the delay times between subsequent mergers. Compared with SCs, the fraction of hierarchical mergers in AGNs is higher with $\sim$$30\%{-}50\%$; the percentage in SCs is $\sim$$10\%{-}50\%$ that has great uncertainty determined by the escape velocities (see Table~\ref{MyTabB} in Appendix~\ref{appA}). As a whole, BH hierarchical growth efficiency in AGNs should be much higher than the efficiency in SCs.
        }

    \item{Most of the hierarchical merger candidate events (especially GW170729) detected by LIGO-Virgo may originate from the AGN channel (see Fig.~\ref{MyFigC}). GW190412 is more likely to come from AGNs because of a small component BH mass. GW190521 should originate from SCs due to a significantly large total mass and relatively symmetric masses, but even within SCs, it is still an extremely rare case.
        }
\end{itemize}

Our results in SCs and/or AGNs broadly agree with those in~Refs.~\cite{Rodriguez-2019-Zevin-PhRvD.100d3027R,Yang-2019-Bartos-PhRvL.123r1101Y,Fragione-2020-Silk-MNRAS.498.4591F,
Tagawa-2021-Haiman-MNRAS.507.3362T,Mapelli-2021-Santoliquido-Symm...13.1678M,Mapelli-2021-Dall'Amico-MNRAS.505..339M}.
We expect that with third-generation GW detectors in operation~\cite{Punturo-2010-Abernathy-CQGra..27s4002P,Punturo-2010-Abernathy-CQGra..27h4007P,LIGO-2017CQGra..34d4001A}, the increasing data on GW events will help us to constrain hierarchical mergers precisely in the two dynamical formation channels.

\acknowledgments{This work is supported by
the National Natural Science Foundation of China (Grant No. 12273005),
the Guangxi Science Foundation (Grant No. 2018GXNSFFA281010),
and China Manned Spaced Project (CMS-CSST-2021-B11).}

\appendix

\begin{center}
\begin{longtable*}{lcccccccccccc}
\caption{The fraction of each merger generation of the three hierarchical branches for the eighteen models and their detected fraction.}\label{MyTabB}\\
\hline
Model &Branch &1G&2G&3G&4G&5G&6G&7G&8G&9G&$\ge$10G&Detected\\
\hline
\endfirsthead
\multicolumn{12}{c}%
{{\bfseries \tablename\ \thetable{} -- continued from previous page}} \\
\hline Model &Branch &1G&2G&3G&4G&5G&6G&7G&8G&9G&$\ge$10G&Detected\\ \hline
\endhead
\hline \multicolumn{12}{l}{{Continued on next page}} \\
\endfoot
\hline
\endlastfoot
\multirow{3}{*}{SC\_1}&NG+1G&0.751&0.24&0.007&$9\times10^{-4}$&$3\times10^{-4}$&$1\times10^{-4}$&
$7\times10^{-5}$&$3\times10^{-5}$&$2\times10^{-5}$&$8\times10^{-6}$&0.006\\
&NG+NG&0.753&0.241&0.006&$1\times10^{-4}$&$6\times10^{-6}$&      0&      0&      0&      0&      0&0.008\\
&NG+$\le$NG&0.752& 0.24&0.007& $5\times10^{-4}$&$1\times10^{-4}$&$4\times10^{-5}$& $2\times10^{-5}$&$8\times10^{-6}$&0&0&0.007	\\
\multirow{3}{*}{SC\_2}&NG+1G&0.907&0.093&$3\times10^{-4}$&$1\times10^{-5}$&$7\times10^{-6}$&      0&     0&      0&     0&      0&0.006\\
&NG+NG&0.907&0.093&$2\times10^{-4}$&$7\times10^{-6}$&      0&      0&      0&      0&      0&      0&0.006\\
&NG+$\le$NG&0.907& 0.093&$3\times10^{-4}$&$7\times10^{-6}$&      0&      0&      0&     0&      0&      0&	0.006\\
\multirow{3}{*}{SC\_3}&NG+1G&0.614&0.307& 0.051& 0.015& 0.007& 0.004& 0.002&$9\times10^{-4}$&$5\times10^{-4}$&$2\times10^{-4}$& 0.007\\
&NG+NG&0.636&0.318& 0.039& 0.006& $7\times10^{-4}$&$9\times10^{-5}$& $8\times10^{-6}$&$4\times10^{-6}$&       0&       0&0.013\\
&NG+$\le$NG&0.622&0.311&0.047& 0.011&0.004& 0.002& 0.001&$5\times10^{-4}$&$2\times10^{-4}$& $1\times10^{-4}$&0.009\\
\multirow{3}{*}{SC\_4}&NG+1G&0.559&0.279&0.087& 0.038& 0.019&0.009& 0.005& 0.002& 0.001& $6\times10^{-4}$&0.008\\
&NG+NG&0.602&0.301& 0.072&  0.019&0.006& 0.001& $2\times10^{-4}$& $4\times10^{-5}$&$1\times10^{-5}$& $4\times10^{-6}$&0.012\\
&NG+$\le$NG&0.571&0.286& 0.082& 0.031& 0.015&0.008& 0.004& 0.002& 0.001&$5\times10^{-4}$&0.012\\
\multirow{3}{*}{SC\_5}&NG+1G&0.501& 0.25&  0.125& 0.063& 0.031& 0.016& 0.009& 0.004& 0.002& 0.001&0.009\\
&NG+NG&0.532&0.266&  0.115& 0.052& 0.021& 0.009& 0.003& 0.001& $4\times10^{-4}$& $2\times10^{-4}$&0.031\\
&NG+$\le$NG&0.503&0.252& 0.123& 0.062&0.031& 0.015&  0.008& 0.004& 0.002& 0.001&0.015\\
\multirow{3}{*}{SC\_6}&NG+1G&0.751& 0.24&0.008& $9\times10^{-4}$&$3\times10^{-4}$&$1\times10^{-4}$& $6\times10^{-5}$& $3\times10^{-5}$&$1\times10^{-5}$&$8\times10^{-6}$&0.006\\
&NG+NG&0.753&0.241&0.006&$1\times10^{-4}$&$6\times10^{-6}$&     0&      0&      0&     0&      0&0.008\\
&NG+$\le$NG&0.752&  0.24& 0.007& $5\times10^{-4}$&$1\times10^{-4}$& $4\times10^{-5}$& $2\times10^{-5}$& $6\times10^{-6}$& 0&0&0.007\\
\multirow{3}{*}{SC\_7}&NG+1G&0.751&  0.24& 0.007& $9\times10^{-4}$&$3\times10^{-4}$&$1\times10^{-4}$& $6\times10^{-5}$&$3\times10^{-5}$& $1\times10^{-5}$&$4\times10^{-6}$&0.006\\
&NG+NG&0.753& 0.241& 0.006&$2\times10^{-4}$& $6\times10^{-6}$& $6\times10^{-6}$&       0&       0&       0&       0&0.008\\
&NG+$\le$NG&0.752&  0.24& 0.007& $5\times10^{-4}$& $1\times10^{-4}$& $3\times10^{-5}$& $1\times10^{-5}$& $6\times10^{-6}$& 0& 0&0.007\\
\multirow{3}{*}{SC\_8}&NG+1G&0.691&0.296&0.01&0.002&$6\times10^{-4}$&$3\times10^{-4}$&$2\times10^{-4}$&$8\times10^{-5}$&$4\times10^{-5}$& $2\times10^{-5}$&0.006\\
&NG+NG &0.695 &0.298 &0.007 &$2\times10^{-4}$ &$6\times10^{-6}$ &0 &0&0&0&	0&0.008\\
&NG+$\le$NG&0.693&0.297&0.009&0.001&$2\times10^{-4}$&$8\times10^{-5}$&$3\times10^{-5}$&$2\times10^{-5}$&$7\times10^{-6}$&0&0.007\\
\multirow{3}{*}{SC\_9}&NG+1G&0.659&0.329&0.01&0.001&$4\times10^{-4}$&$2\times10^{-4}$&$1\times10^{-4}$&$5\times10^{-5}$&$2\times10^{-5}$ &$1\times10^{-5}$&0.006\\
&NG+NG&0.661&0.331&	0.008&$2\times10^{-4}$&$6\times10^{-6}$&0	&0	&0&	0&	0&0.009\\
&NG+$\le$NG&0.660&0.330&0.009&$8\times10^{-4}$&$2\times10^{-4}$&$6\times10^{-5}$&$2\times10^{-5}$&$9\times10^{-6}$&$6\times10^{-6}$&0&0.007\\
\multirow{3}{*}{SC\_10}&NG+1G&0.853&0.142&0.004&$5\times10^{-4}$&$2\times10^{-4}$&$8\times10^{-5}$&$4\times10^{-5}$&$2\times10^{-5}$ &$9\times10^{-6}$&	0&0.006\\
&NG+NG&0.855&0.142&0.003&$8\times10^{-5}$&0&	0&	0&0&	0&0&	0.007\\
&NG+$\le$NG&0.854&0.142	&0.004&$3\times10^{-4}$&$9\times10^{-5}$&$2\times10^{-5}$&$8\times10^{-6}$&0&0&	0&0.006\\
\multirow{3}{*}{AGN\_1}&NG+1G&0.501&  0.25&   0.125&  0.062&  0.031&  0.016& 0.008& 0.004& 0.002&0.001&0.027\\
&NG+NG&0.501&  0.25&   0.125&  0.062&  0.031&  0.016& 0.008& 0.004& 0.002& 0.001&0.054\\
&NG+$\le$NG&0.501&  0.25&   0.125&  0.062&  0.031&  0.016& 0.008& 0.004& 0.002&0.001&0.036\\
\multirow{3}{*}{AGN\_2}&NG+1G&0.501&  0.25&   0.125&  0.062&  0.031&  0.016& 0.008& 0.004& 0.002& 0.001&0.027\\
&NG+NG&0.506& 0.253&   0.126&   0.063&   0.029&  0.013& 0.006&  0.003& 0.001& $5\times10^{-4}$&0.055\\
&NG+$\le$NG&0.501&  0.25&   0.125&  0.062&  0.031&  0.016& 0.008& 0.004& 0.002& 0.001&0.037\\
\multirow{3}{*}{AGN\_3}&NG+1G&0.501&  0.25&   0.125&  0.063&  0.031&  0.016& 0.008& 0.004& 0.002& 0.001&0.027\\
&NG+NG&0.501&  0.25&   0.125&  0.063&  0.031&  0.016& 0.008& 0.004& 0.002& 0.001&0.054\\
&NG+$\le$NG&0.501&  0.25&   0.125&  0.063&  0.031&  0.016& 0.008& 0.004& 0.002& 0.001&0.036\\
\multirow{3}{*}{AGN\_4}&NG+1G&0.552& 0.257&   0.116&  0.049&  0.012&  0.006& 0.001& $6\times10^{-5}$&       0&       0&0.025\\
&NG+NG&0.552& 0.257&   0.116&  0.049&  0.019& 0.006& 0.001& $5\times10^{-5}$&       0&       0&0.050\\
&NG+$\le$NG&0.552& 0.257&   0.116&  0.049&  0.019& 0.006& 0.001& $6\times10^{-5}$& $4\times10^{-6}$&       0&0.032\\
\multirow{3}{*}{AGN\_5}&NG+1G&0.668&0.255&0.071&0.006&0&     0&      0&      0&     0&      0&0.020	\\
&NG+NG&0.667& 0.255& 0.071& 0.006&     0&     0&     0&     0&    0&      0&0.036\\
&NG+$\le$NG&0.668&0.255& 0.071& 0.006&      0&       0&      0&     0&      0&      0&0.024\\
\multirow{3}{*}{AGN\_6}&NG+1G&0.501&0.25&0.125	&0.062	&0.031	&0.016&	0.008&0.004&0.002&0.001&0.027\\
&NG+NG&0.501&0.25&0.125&0.062&0.031&0.016&0.008& 0.004&0.002&0.001&0.054\\
&NG+$\le$NG&0.501&0.25&0.125&0.062&0.031&0.016&0.008&0.004&0.002&0.001&0.036\\
\multirow{3}{*}{AGN\_7}&NG+1G&0.501&0.25&0.125	&0.062	&0.031	&0.016&	0.008&0.004&0.002&0.001&0.027\\
&NG+NG&0.501&0.25&0.125&0.062&0.031&0.016&0.008& 0.004&0.002&0.001&0.054\\
&NG+$\le$NG&0.501&0.25&0.125&0.062&0.031&0.016&0.008&0.004&0.002&0.001&0.036\\
\multirow{3}{*}{AGN\_8}&NG+1G&0.501&0.25&0.125	&0.062	&0.031	&0.016&	0.008&0.004&0.002&0.001&0.027\\
&NG+NG&0.501&0.25&0.125&0.062&0.031&0.016&0.008& 0.004&0.002&0.001&0.054\\
&NG+$\le$NG&0.501&0.25&0.125&0.062&0.031&0.016&0.008&0.004&0.002&0.001&0.036\\
\hline
\end{longtable*}
\end{center}

\begin{figure*}[ht]
\centering
\includegraphics[width=0.8\textwidth]{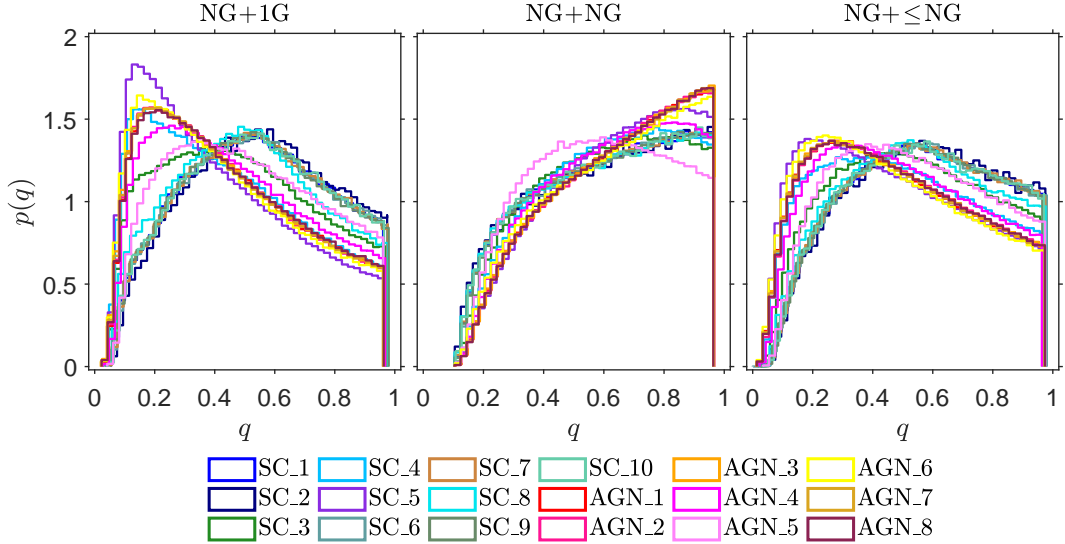}
\caption{Same as Fig.~\ref{MyFigA}, but the probability density distribution of the mass ratios ($q$) of hierarchical BH mergers.}
\label{MyFigD}
\end{figure*}

\begin{figure*}
\centering
\includegraphics[width=0.8\textwidth]{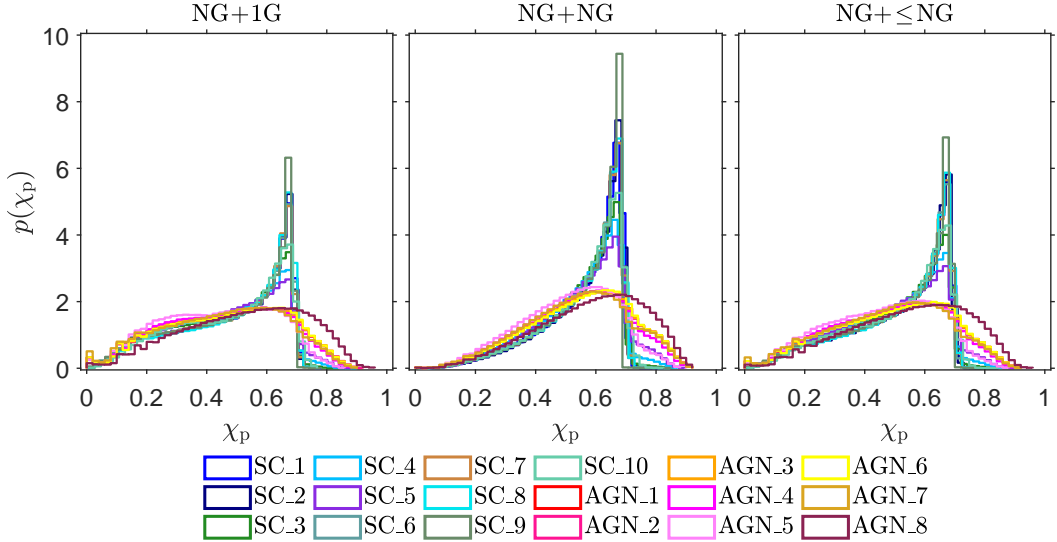}
\caption{Same as Fig.~\ref{MyFigA}, but the probability density distribution of the effective precession parameters ($\chi_{\rm p}$) of hierarchical BH mergers.}
\label{MyFigE}
\end{figure*}

\section{Fraction of each merger generation}\label{appA}

Table~\ref{MyTabB} lists the fraction of each merger generation of the three hierarchical branches for the eighteen models.
The hierarchical mergers in AGNs are more efficient  than that in SCs because almost all of the merger remnants could be retained in migration traps in AGN disks.
The kick velocities of NG+NG merges are larger than the others and therefore their fractions of hierarchical mergers are relatively low in SCs.
We also show their fraction detected by LIGO-Virgo in Table~\ref{MyTabB}~\cite{Finn-1993-Chernoff-PhRvD..47.2198F}, with a network detection threshold of SNR $>$$8$~\cite{Abadie-2010-Abbott-CQGra..27q3001A} (see more details in Sec.~\ref{SecC3}).
The detectable fractions of BBH mergers in AGNs are, on average, about three times that of BBH mergers in SCs.

\section{Mass ratio distribution}\label{appB}

In Sec.~\ref{SecC1}, we show the primary mass distribution of hierarchical mergers (see Fig.~\ref{MyFigA}).
Here, we plot their probability density distribution of the mass ratios ($q$) in Fig.~\ref{MyFigD}, which is broadly consistent with the results of Ref.~\cite{Mahapatra-2022-Gupta-arXiv220905766M} for SCs.
We find that (on average) hierarchical mergers could lead to the formation of more asymmetric binaries in dynamical formation channels.
Compared with NG+NG mergers, NG+1G mergers in both SCs and AGNs prefers unequal-mass binaries depending on hierarchical merger efficiency.
Because the higher-generation mergers, the more extreme mass ratios for the branch of NG+1G.
The mass ratio distribution of NG+$\le$NG mergers is between NG+1G and NG+NG mergers.
For NG+NG mergers, the distributions in SCs and AGNs are not very different.
In SCs, the distribution of $q$ of NG+1G mergers has large uncertainty, in which the distribution of the model of SC\_5 is the highest at the low-$q$ end.
The hierarchical mergers in AGNs would be more asymmetric that that in SCs, if NG+1G and NG+NG mergers dominate the hierarchical BH merger rates in AGNs and SCs, respectively.

\section{Effective precession parameter distribution}\label{appC}

In Sec.~\ref{SecC2}, we show the effective spin distribution of hierarchical mergers (see Fig.~\ref{MyFigB}).
Here, we show the probability density distribution of the effective precession parameters ($\chi_{\rm p}$) of hierarchical BH mergers in Fig.~\ref{MyFigE},
where $\chi_{\rm p}={\rm max}[\chi_1 {\rm sin}\theta_1,\chi_2 {\rm sin}\theta_2q(4q+3)/(4+3q)]$.
We see that the effective precession parameter distributions with the peak of $\chi_{\rm p}\sim0.66$ in SCs are much narrower than that in AGNs.
The distribution of $\chi_{\rm p}$ in AGNs is flat, especially for NG+1G mergers, because gas accretion tends to torque the BH spin into alignment with the AGN disk.
The results of the distributions of $\chi_{\rm p}$ in SCs and/or AGNs are agree with those in Refs.~\cite{Baibhav-2021-Berti-PhRvD.104h4002B,Tagawa-2021-Haiman-MNRAS.507.3362T}.


\begin{thebibliography}{100}%
\makeatletter
\providecommand \@ifxundefined [1]{%
 \@ifx{#1\undefined}
}%
\providecommand \@ifnum [1]{%
 \ifnum #1\expandafter \@firstoftwo
 \else \expandafter \@secondoftwo
 \fi
}%
\providecommand \@ifx [1]{%
 \ifx #1\expandafter \@firstoftwo
 \else \expandafter \@secondoftwo
 \fi
}%
\providecommand \natexlab [1]{#1}%
\providecommand \enquote  [1]{``#1''}%
\providecommand \bibnamefont  [1]{#1}%
\providecommand \bibfnamefont [1]{#1}%
\providecommand \citenamefont [1]{#1}%
\providecommand \href@noop [0]{\@secondoftwo}%
\providecommand \href [0]{\begingroup \@sanitize@url \@href}%
\providecommand \@href[1]{\@@startlink{#1}\@@href}%
\providecommand \@@href[1]{\endgroup#1\@@endlink}%
\providecommand \@sanitize@url [0]{\catcode `\\12\catcode `\$12\catcode
  `\&12\catcode `\#12\catcode `\^12\catcode `\_12\catcode `\%12\relax}%
\providecommand \@@startlink[1]{}%
\providecommand \@@endlink[0]{}%
\providecommand \url  [0]{\begingroup\@sanitize@url \@url }%
\providecommand \@url [1]{\endgroup\@href {#1}{\urlprefix }}%
\providecommand \urlprefix  [0]{URL }%
\providecommand \Eprint [0]{\href }%
\providecommand \doibase [0]{http://dx.doi.org/}%
\providecommand \selectlanguage [0]{\@gobble}%
\providecommand \bibinfo  [0]{\@secondoftwo}%
\providecommand \bibfield  [0]{\@secondoftwo}%
\providecommand \translation [1]{[#1]}%
\providecommand \BibitemOpen [0]{}%
\providecommand \bibitemStop [0]{}%
\providecommand \bibitemNoStop [0]{.\EOS\space}%
\providecommand \EOS [0]{\spacefactor3000\relax}%
\providecommand \BibitemShut  [1]{\csname bibitem#1\endcsname}%
\let\auto@bib@innerbib\@empty
\bibitem [{\citenamefont {{Abbott}}\ \emph {{\it et~al.}}(2019)\citenamefont
  {{Abbott}}, \citenamefont {{Abbott}}, \citenamefont {{Abbott}}, \citenamefont
  {{Abraham}}, \citenamefont {{Acernese}}, \citenamefont {{Ackley}},
  \citenamefont {{Adams}}, \citenamefont {{Adhikari}}, \citenamefont {{Adya}},
  \citenamefont {{Affeldt}},\ and\ \citenamefont
  {et~al.}}]{GWTC-1-2019PhRvX...9c1040A}%
  \BibitemOpen
  \bibfield  {author} {\bibinfo {author} {\bibfnamefont {B.~P.}\ \bibnamefont
  {{Abbott}}} {\it et~al.},\ }\bibfield  {title} {\enquote {\bibinfo {title}
  {{GWTC-1: A Gravitational-Wave Transient Catalog of Compact Binary Mergers
  Observed by LIGO and Virgo during the First and Second Observing Runs}},}\
  }\href {\doibase 10.1103/PhysRevX.9.031040} {\bibfield  {journal} {\bibinfo
  {journal} {Physical Review X}\ }\textbf {\bibinfo {volume} {9}},\ \bibinfo
  {eid} {031040} (\bibinfo {year} {2019})},\ \Eprint
  {http://arxiv.org/abs/1811.12907}{arXiv:1811.12907}\BibitemShut {NoStop}%
\bibitem [{\citenamefont {{Abbott}}\ \emph {{\it et~al.}}(2021)\citenamefont
  {{Abbott}}, \citenamefont {{Abbott}}, \citenamefont {{Abraham}},
  \citenamefont {{Acernese}}, \citenamefont {{Ackley}}, \citenamefont
  {{Adams}}, \citenamefont {{Adams}}, \citenamefont {{Adhikari}}, \citenamefont
  {{Adya}}, \citenamefont {{Affeldt}},\ and\ \citenamefont
  {et~al.}}]{GWTC-2LVK-2021PhRvX..11b1053A}%
  \BibitemOpen
  \bibfield  {author} {\bibinfo {author} {\bibfnamefont {R.}~\bibnamefont
  {{Abbott}}} {\it et~al.},\ }\bibfield  {title} {\enquote {\bibinfo {title}
  {{GWTC-2: Compact Binary Coalescences Observed by LIGO and Virgo during the
  First Half of the Third Observing Run}},}\ }\href {\doibase
  10.1103/PhysRevX.11.021053} {\bibfield  {journal} {\bibinfo  {journal}
  {Physical Review X}\ }\textbf {\bibinfo {volume} {11}},\ \bibinfo {eid}
  {021053} (\bibinfo {year} {2021})},\ \Eprint
  {http://arxiv.org/abs/2010.14527}{arXiv:2010.14527}\BibitemShut {NoStop}%
\bibitem [{\citenamefont {{The LIGO Scientific Collaboration}}\ \emph {{\it
  et~al.}}(2021{\natexlab{a}})\citenamefont {{The LIGO Scientific
  Collaboration}}, \citenamefont {{the Virgo Collaboration}}, \citenamefont
  {{Abbott}}, \citenamefont {{Abbott}}, \citenamefont {{Acernese}},
  \citenamefont {{Ackley}}, \citenamefont {{Adams}}, \citenamefont
  {{Adhikari}}, \citenamefont {{Adhikari}}, \citenamefont {{Adya}},\ and\
  \citenamefont {et~al.}}]{GWTC-2.1-2021arXiv210801045T}%
  \BibitemOpen
  \bibfield  {author} {\bibinfo {author} {\bibnamefont {{The LIGO Scientific
  Collaboration}}} {\it et~al.},\ }\bibfield  {title} {\enquote {\bibinfo
  {title} {{GWTC-2.1: Deep Extended Catalog of Compact Binary Coalescences
  Observed by LIGO and Virgo During the First Half of the Third Observing
  Run}},}\ }\href@noop {} {\bibfield  {journal} {\bibinfo  {journal} {arXiv
  e-prints}\ ,\ \bibinfo {eid} {arXiv:2108.01045}} (\bibinfo {year}
  {2021}{\natexlab{a}})},\ \Eprint
  {http://arxiv.org/abs/2108.01045}{arXiv:2108.01045}\BibitemShut {NoStop}%
\bibitem [{\citenamefont {{The LIGO Scientific Collaboration}}\ \emph {{\it
  et~al.}}(2021{\natexlab{b}})\citenamefont {{The LIGO Scientific
  Collaboration}}, \citenamefont {{the Virgo Collaboration}}, \citenamefont
  {{the KAGRA Collaboration}}, \citenamefont {{Abbott}}, \citenamefont
  {{Abbott}}, \citenamefont {{Acernese}}, \citenamefont {{Ackley}},
  \citenamefont {{Adams}}, \citenamefont {{Adhikari}}, \citenamefont
  {{Adhikari}},\ and\ \citenamefont {et~al.}}]{GWTC-32021arXiv211103606T}%
  \BibitemOpen
  \bibfield  {author} {\bibinfo {author} {\bibnamefont {{The LIGO Scientific
  Collaboration}}} {\it et~al.},\ }\bibfield  {title} {\enquote {\bibinfo
  {title} {{GWTC-3: Compact Binary Coalescences Observed by LIGO and Virgo
  During the Second Part of the Third Observing Run}},}\ }\href@noop {}
  {\bibfield  {journal} {\bibinfo  {journal} {arXiv e-prints}\ ,\ \bibinfo
  {eid} {arXiv:2111.03606}} (\bibinfo {year} {2021}{\natexlab{b}})},\ \Eprint
  {http://arxiv.org/abs/2111.03606}{arXiv:2111.03606}\BibitemShut {NoStop}%
\bibitem [{\citenamefont {{O'Leary}}\ \emph {{\it et~al.}}(2016)\citenamefont
  {{O'Leary}}, \citenamefont {{Meiron}},\ and\ \citenamefont
  {{Kocsis}}}]{O'Leary-2016-Meiron-ApJ...824L..12O}%
  \BibitemOpen
  \bibfield  {author} {\bibinfo {author} {\bibfnamefont {R.~M.}\ \bibnamefont
  {{O'Leary}}}, \bibinfo {author} {\bibfnamefont {Y.}~\bibnamefont {{Meiron}}},
  and\ \bibinfo {author} {\bibfnamefont {B.}~\bibnamefont {{Kocsis}}},\
  }\bibfield  {title} {\enquote {\bibinfo {title} {{Dynamical Formation
  Signatures of Black Hole Binaries in the First Detected Mergers by LIGO}},}\
  }\href {\doibase 10.3847/2041-8205/824/1/L12} {\bibfield  {journal} {\bibinfo
   {journal} {\apjl}\ }\textbf {\bibinfo {volume} {824}},\ \bibinfo {eid} {L12}
  (\bibinfo {year} {2016})},\ \Eprint
  {http://arxiv.org/abs/1602.02809}{arXiv:1602.02809}\BibitemShut {NoStop}%
\bibitem [{\citenamefont {{Fishbach}}\ \emph {{\it et~al.}}(2017)\citenamefont
  {{Fishbach}}, \citenamefont {{Holz}},\ and\ \citenamefont
  {{Farr}}}]{Fishbach-2017-Holz-ApJ...840L..24F}%
  \BibitemOpen
  \bibfield  {author} {\bibinfo {author} {\bibfnamefont {M.}~\bibnamefont
  {{Fishbach}}}, \bibinfo {author} {\bibfnamefont {D.~E.}\ \bibnamefont
  {{Holz}}}, and\ \bibinfo {author} {\bibfnamefont {B.}~\bibnamefont
  {{Farr}}},\ }\bibfield  {title} {\enquote {\bibinfo {title} {{Are LIGO's
  Black Holes Made from Smaller Black Holes?}}}\ }\href {\doibase
  10.3847/2041-8213/aa7045} {\bibfield  {journal} {\bibinfo  {journal} {\apjl}\
  }\textbf {\bibinfo {volume} {840}},\ \bibinfo {eid} {L24} (\bibinfo {year}
  {2017})},\ \Eprint
  {http://arxiv.org/abs/1703.06869}{arXiv:1703.06869}\BibitemShut {NoStop}%
\bibitem [{\citenamefont {{Gerosa}}\ and\ \citenamefont
  {{Berti}}(2017)}]{Gerosa-2017-Berti-PhRvD..95l4046G}%
  \BibitemOpen
  \bibfield  {author} {\bibinfo {author} {\bibfnamefont {D.}~\bibnamefont
  {{Gerosa}}} and\ \bibinfo {author} {\bibfnamefont {E.}~\bibnamefont
  {{Berti}}},\ }\bibfield  {title} {\enquote {\bibinfo {title} {{Are merging
  black holes born from stellar collapse or previous mergers?}}}\ }\href
  {\doibase 10.1103/PhysRevD.95.124046} {\bibfield  {journal} {\bibinfo
  {journal} {\prd}\ }\textbf {\bibinfo {volume} {95}},\ \bibinfo {eid} {124046}
  (\bibinfo {year} {2017})},\ \Eprint
  {http://arxiv.org/abs/1703.06223}{arXiv:1703.06223}\BibitemShut {NoStop}%
\bibitem [{\citenamefont {{Kimball}}\ \emph {{\it et~al.}}(2021)\citenamefont
  {{Kimball}}, \citenamefont {{Talbot}}, \citenamefont {{Berry}}, \citenamefont
  {{Zevin}}, \citenamefont {{Thrane}}, \citenamefont {{Kalogera}},
  \citenamefont {{Buscicchio}}, \citenamefont {{Carney}}, \citenamefont
  {{Dent}}, \citenamefont {{Middleton}}, \citenamefont {{Payne}}, \citenamefont
  {{Veitch}},\ and\ \citenamefont
  {{Williams}}}]{Kimball-2021-Talbot-ApJ...915L..35K}%
  \BibitemOpen
  \bibfield  {author} {\bibinfo {author} {\bibfnamefont {C.}~\bibnamefont
  {{Kimball}}} {\it et~al.},\ }\bibfield  {title} {\enquote {\bibinfo {title}
  {{Evidence for Hierarchical Black Hole Mergers in the Second LIGO-Virgo
  Gravitational Wave Catalog}},}\ }\href {\doibase 10.3847/2041-8213/ac0aef}
  {\bibfield  {journal} {\bibinfo  {journal} {\apjl}\ }\textbf {\bibinfo
  {volume} {915}},\ \bibinfo {eid} {L35} (\bibinfo {year} {2021})},\ \Eprint
  {http://arxiv.org/abs/2011.05332}{arXiv:2011.05332}\BibitemShut {NoStop}%
\bibitem [{\citenamefont {Mould}\ \emph {{\it et~al.}}(2022)\citenamefont
  {Mould}, \citenamefont {Gerosa},\ and\ \citenamefont
  {Taylor}}]{Mould-2022-Gerosa-PhysRevD.106.103013}%
  \BibitemOpen
  \bibfield  {author} {\bibinfo {author} {\bibfnamefont {M.}~\bibnamefont
  {Mould}}, \bibinfo {author} {\bibfnamefont {D.}~\bibnamefont {Gerosa}}, and\
  \bibinfo {author} {\bibfnamefont {S.~R.}\ \bibnamefont {Taylor}},\ }\bibfield
   {title} {\enquote {\bibinfo {title} {Deep learning and bayesian inference of
  gravitational-wave populations: Hierarchical black-hole mergers},}\ }\href
  {\doibase 10.1103/PhysRevD.106.103013} {\bibfield  {journal} {\bibinfo
  {journal} {Phys. Rev. D}\ }\textbf {\bibinfo {volume} {106}},\ \bibinfo
  {pages} {103013} (\bibinfo {year} {2022})}\BibitemShut {NoStop}%
\bibitem [{\citenamefont {{Gerosa}}\ and\ \citenamefont
  {{Fishbach}}(2021)}]{Gerosa-2021-Fishbach-NatAs...5..749G}%
  \BibitemOpen
  \bibfield  {author} {\bibinfo {author} {\bibfnamefont {D.}~\bibnamefont
  {{Gerosa}}} and\ \bibinfo {author} {\bibfnamefont {M.}~\bibnamefont
  {{Fishbach}}},\ }\bibfield  {title} {\enquote {\bibinfo {title}
  {{Hierarchical mergers of stellar-mass black holes and their
  gravitational-wave signatures}},}\ }\href {\doibase
  10.1038/s41550-021-01398-w} {\bibfield  {journal} {\bibinfo  {journal}
  {Nature Astronomy}\ }\textbf {\bibinfo {volume} {5}},\ \bibinfo {pages} {749}
  (\bibinfo {year} {2021})},\ \Eprint
  {http://arxiv.org/abs/2105.03439}{arXiv:2105.03439}\BibitemShut {NoStop}%
\bibitem [{\citenamefont {{Mahapatra}}\ \emph {{\it et~al.}}(2022)\citenamefont
  {{Mahapatra}}, \citenamefont {{Gupta}}, \citenamefont {{Favata}},
  \citenamefont {{Arun}},\ and\ \citenamefont
  {{Sathyaprakash}}}]{Mahapatra-2022-Gupta-arXiv220905766M}%
  \BibitemOpen
  \bibfield  {author} {\bibinfo {author} {\bibfnamefont {P.}~\bibnamefont
  {{Mahapatra}}}, \bibinfo {author} {\bibfnamefont {A.}~\bibnamefont
  {{Gupta}}}, \bibinfo {author} {\bibfnamefont {M.}~\bibnamefont {{Favata}}},
  \bibinfo {author} {\bibfnamefont {K.~G.}\ \bibnamefont {{Arun}}}, and\
  \bibinfo {author} {\bibfnamefont {B.~S.}\ \bibnamefont {{Sathyaprakash}}},\
  }\bibfield  {title} {\enquote {\bibinfo {title} {{Black hole hierarchical
  growth efficiency and mass spectrum predictions}},}\ }\href@noop {}
  {\bibfield  {journal} {\bibinfo  {journal} {arXiv e-prints}\ ,\ \bibinfo
  {eid} {arXiv:2209.05766}} (\bibinfo {year} {2022})},\ \Eprint
  {http://arxiv.org/abs/2209.05766}{arXiv:2209.05766}\BibitemShut {NoStop}%
\bibitem [{\citenamefont {{Rodriguez}}\ \emph {{\it et~al.}}(2019)\citenamefont
  {{Rodriguez}}, \citenamefont {{Zevin}}, \citenamefont {{Amaro-Seoane}},
  \citenamefont {{Chatterjee}}, \citenamefont {{Kremer}}, \citenamefont
  {{Rasio}},\ and\ \citenamefont
  {{Ye}}}]{Rodriguez-2019-Zevin-PhRvD.100d3027R}%
  \BibitemOpen
  \bibfield  {author} {\bibinfo {author} {\bibfnamefont {C.~L.}\ \bibnamefont
  {{Rodriguez}}}, \bibinfo {author} {\bibfnamefont {M.}~\bibnamefont
  {{Zevin}}}, \bibinfo {author} {\bibfnamefont {P.}~\bibnamefont
  {{Amaro-Seoane}}}, \bibinfo {author} {\bibfnamefont {S.}~\bibnamefont
  {{Chatterjee}}}, \bibinfo {author} {\bibfnamefont {K.}~\bibnamefont
  {{Kremer}}}, \bibinfo {author} {\bibfnamefont {F.~A.}\ \bibnamefont
  {{Rasio}}}, and\ \bibinfo {author} {\bibfnamefont {C.~S.}\ \bibnamefont
  {{Ye}}},\ }\bibfield  {title} {\enquote {\bibinfo {title} {{Black holes: The
  next generation{\textemdash}repeated mergers in dense star clusters and their
  gravitational-wave properties}},}\ }\href {\doibase
  10.1103/PhysRevD.100.043027} {\bibfield  {journal} {\bibinfo  {journal}
  {\prd}\ }\textbf {\bibinfo {volume} {100}},\ \bibinfo {eid} {043027}
  (\bibinfo {year} {2019})},\ \Eprint
  {http://arxiv.org/abs/1906.10260}{arXiv:1906.10260}\BibitemShut {NoStop}%
\bibitem [{\citenamefont {{Kimball}}\ \emph {{\it et~al.}}(2020)\citenamefont
  {{Kimball}}, \citenamefont {{Talbot}}, \citenamefont {{Berry}}, \citenamefont
  {{Carney}}, \citenamefont {{Zevin}}, \citenamefont {{Thrane}},\ and\
  \citenamefont {{Kalogera}}}]{Kimball-2020-Talbot-ApJ...900..177K}%
  \BibitemOpen
  \bibfield  {author} {\bibinfo {author} {\bibfnamefont {C.}~\bibnamefont
  {{Kimball}}}, \bibinfo {author} {\bibfnamefont {C.}~\bibnamefont {{Talbot}}},
  \bibinfo {author} {\bibfnamefont {C.~P.~L.}\ \bibnamefont {{Berry}}},
  \bibinfo {author} {\bibfnamefont {M.}~\bibnamefont {{Carney}}}, \bibinfo
  {author} {\bibfnamefont {M.}~\bibnamefont {{Zevin}}}, \bibinfo {author}
  {\bibfnamefont {E.}~\bibnamefont {{Thrane}}}, and\ \bibinfo {author}
  {\bibfnamefont {V.}~\bibnamefont {{Kalogera}}},\ }\bibfield  {title}
  {\enquote {\bibinfo {title} {{Black Hole Genealogy: Identifying Hierarchical
  Mergers with Gravitational Waves}},}\ }\href {\doibase
  10.3847/1538-4357/aba518} {\bibfield  {journal} {\bibinfo  {journal}
  {Astrophys. J.}\ }\textbf {\bibinfo {volume} {900}},\ \bibinfo {eid} {177}
  (\bibinfo {year} {2020})},\ \Eprint
  {http://arxiv.org/abs/2005.00023}{arXiv:2005.00023}\BibitemShut {NoStop}%
\bibitem [{\citenamefont {{Baibhav}}\ \emph {{\it et~al.}}(2021)\citenamefont
  {{Baibhav}}, \citenamefont {{Berti}}, \citenamefont {{Gerosa}}, \citenamefont
  {{Mould}},\ and\ \citenamefont
  {{Wong}}}]{Baibhav-2021-Berti-PhRvD.104h4002B}%
  \BibitemOpen
  \bibfield  {author} {\bibinfo {author} {\bibfnamefont {V.}~\bibnamefont
  {{Baibhav}}}, \bibinfo {author} {\bibfnamefont {E.}~\bibnamefont {{Berti}}},
  \bibinfo {author} {\bibfnamefont {D.}~\bibnamefont {{Gerosa}}}, \bibinfo
  {author} {\bibfnamefont {M.}~\bibnamefont {{Mould}}}, and\ \bibinfo {author}
  {\bibfnamefont {K.~W.~K.}\ \bibnamefont {{Wong}}},\ }\bibfield  {title}
  {\enquote {\bibinfo {title} {{Looking for the parents of LIGO's black
  holes}},}\ }\href {\doibase 10.1103/PhysRevD.104.084002} {\bibfield
  {journal} {\bibinfo  {journal} {\prd}\ }\textbf {\bibinfo {volume} {104}},\
  \bibinfo {eid} {084002} (\bibinfo {year} {2021})},\ \Eprint
  {http://arxiv.org/abs/2105.12140}{arXiv:2105.12140}\BibitemShut {NoStop}%
\bibitem [{\citenamefont {{Mapelli}}\ \emph {{\it
  et~al.}}(2021{\natexlab{a}})\citenamefont {{Mapelli}}, \citenamefont
  {{Santoliquido}}, \citenamefont {{Bouffanais}}, \citenamefont {{Arca Sedda}},
  \citenamefont {{Artale}},\ and\ \citenamefont
  {{Ballone}}}]{Mapelli-2021-Santoliquido-Symm...13.1678M}%
  \BibitemOpen
  \bibfield  {author} {\bibinfo {author} {\bibfnamefont {M.}~\bibnamefont
  {{Mapelli}}}, \bibinfo {author} {\bibfnamefont {F.}~\bibnamefont
  {{Santoliquido}}}, \bibinfo {author} {\bibfnamefont {Y.}~\bibnamefont
  {{Bouffanais}}}, \bibinfo {author} {\bibfnamefont {M.~A.}\ \bibnamefont
  {{Arca Sedda}}}, \bibinfo {author} {\bibfnamefont {M.~C.}\ \bibnamefont
  {{Artale}}}, and\ \bibinfo {author} {\bibfnamefont {A.}~\bibnamefont
  {{Ballone}}},\ }\bibfield  {title} {\enquote {\bibinfo {title} {{Mass and
  Rate of Hierarchical Black Hole Mergers in Young, Globular and Nuclear Star
  Clusters}},}\ }\href {\doibase 10.3390/sym13091678} {\bibfield  {journal}
  {\bibinfo  {journal} {Symmetry}\ }\textbf {\bibinfo {volume} {13}},\ \bibinfo
  {pages} {1678} (\bibinfo {year} {2021}{\natexlab{a}})},\ \Eprint
  {http://arxiv.org/abs/2007.15022}{arXiv:2007.15022}\BibitemShut {NoStop}%
\bibitem [{\citenamefont {{Mapelli}}\ \emph {{\it
  et~al.}}(2021{\natexlab{b}})\citenamefont {{Mapelli}}, \citenamefont
  {{Dall'Amico}}, \citenamefont {{Bouffanais}}, \citenamefont {{Giacobbo}},
  \citenamefont {{Arca Sedda}}, \citenamefont {{Artale}}, \citenamefont
  {{Ballone}}, \citenamefont {{Di Carlo}}, \citenamefont {{Iorio}},
  \citenamefont {{Santoliquido}},\ and\ \citenamefont
  {{Torniamenti}}}]{Mapelli-2021-Dall'Amico-MNRAS.505..339M}%
  \BibitemOpen
  \bibfield  {author} {\bibinfo {author} {\bibfnamefont {M.}~\bibnamefont
  {{Mapelli}}} {\it et~al.},\ }\bibfield  {title} {\enquote {\bibinfo {title}
  {{Hierarchical black hole mergers in young, globular and nuclear star
  clusters: the effect of metallicity, spin and cluster properties}},}\ }\href
  {\doibase 10.1093/mnras/stab1334} {\bibfield  {journal} {\bibinfo  {journal}
  {\mnras}\ }\textbf {\bibinfo {volume} {505}},\ \bibinfo {pages} {339}
  (\bibinfo {year} {2021}{\natexlab{b}})},\ \Eprint
  {http://arxiv.org/abs/2103.05016}{arXiv:2103.05016}\BibitemShut {NoStop}%
\bibitem [{\citenamefont {{Li}}(2022{\natexlab{a}})}]{Li-2022A&A...666A.194L}%
  \BibitemOpen
  \bibfield  {author} {\bibinfo {author} {\bibfnamefont {G.-P.}\ \bibnamefont
  {{Li}}},\ }\bibfield  {title} {\enquote {\bibinfo {title} {{Constraining
  hierarchical mergers of binary black holes detectable with LIGO-Virgo}},}\
  }\href {\doibase 10.1051/0004-6361/202244257} {\bibfield  {journal} {\bibinfo
   {journal} {\aap}\ }\textbf {\bibinfo {volume} {666}},\ \bibinfo {eid} {A194}
  (\bibinfo {year} {2022}{\natexlab{a}})},\ \Eprint
  {http://arxiv.org/abs/2208.11894}{arXiv:2208.11894}\BibitemShut {NoStop}%
\bibitem [{\citenamefont {{Yang}}\ \emph {{\it
  et~al.}}(2019{\natexlab{a}})\citenamefont {{Yang}}, \citenamefont {{Bartos}},
  \citenamefont {{Gayathri}}, \citenamefont {{Ford}}, \citenamefont {{Haiman}},
  \citenamefont {{Klimenko}}, \citenamefont {{Kocsis}}, \citenamefont
  {{M{\'a}rka}}, \citenamefont {{M{\'a}rka}}, \citenamefont {{McKernan}},\ and\
  \citenamefont {{O'Shaughnessy}}}]{Yang-2019-Bartos-PhRvL.123r1101Y}%
  \BibitemOpen
  \bibfield  {author} {\bibinfo {author} {\bibfnamefont {Y.}~\bibnamefont
  {{Yang}}} {\it et~al.},\ }\bibfield  {title} {\enquote {\bibinfo {title}
  {{Hierarchical Black Hole Mergers in Active Galactic Nuclei}},}\ }\href
  {\doibase 10.1103/PhysRevLett.123.181101} {\bibfield  {journal} {\bibinfo
  {journal} {\prl}\ }\textbf {\bibinfo {volume} {123}},\ \bibinfo {eid}
  {181101} (\bibinfo {year} {2019}{\natexlab{a}})},\ \Eprint
  {http://arxiv.org/abs/1906.09281}{arXiv:1906.09281}\BibitemShut {NoStop}%
\bibitem [{\citenamefont {{Gayathri}}\ \emph {{\it et~al.}}(2020)\citenamefont
  {{Gayathri}}, \citenamefont {{Bartos}}, \citenamefont {{Haiman}},
  \citenamefont {{Klimenko}}, \citenamefont {{Kocsis}}, \citenamefont
  {{M{\'a}rka}},\ and\ \citenamefont
  {{Yang}}}]{Gayathri-2020-Bartos-ApJ...890L..20G}%
  \BibitemOpen
  \bibfield  {author} {\bibinfo {author} {\bibfnamefont {V.}~\bibnamefont
  {{Gayathri}}}, \bibinfo {author} {\bibfnamefont {I.}~\bibnamefont
  {{Bartos}}}, \bibinfo {author} {\bibfnamefont {Z.}~\bibnamefont {{Haiman}}},
  \bibinfo {author} {\bibfnamefont {S.}~\bibnamefont {{Klimenko}}}, \bibinfo
  {author} {\bibfnamefont {B.}~\bibnamefont {{Kocsis}}}, \bibinfo {author}
  {\bibfnamefont {S.}~\bibnamefont {{M{\'a}rka}}}, and\ \bibinfo {author}
  {\bibfnamefont {Y.}~\bibnamefont {{Yang}}},\ }\bibfield  {title} {\enquote
  {\bibinfo {title} {{GW170817A as a Hierarchical Black Hole Merger}},}\ }\href
  {\doibase 10.3847/2041-8213/ab745d} {\bibfield  {journal} {\bibinfo
  {journal} {\apjl}\ }\textbf {\bibinfo {volume} {890}},\ \bibinfo {eid} {L20}
  (\bibinfo {year} {2020})},\ \Eprint
  {http://arxiv.org/abs/1911.11142}{arXiv:1911.11142}\BibitemShut {NoStop}%
\bibitem [{\citenamefont {{Tagawa}}\ \emph {{\it
  et~al.}}(2021{\natexlab{a}})\citenamefont {{Tagawa}}, \citenamefont
  {{Kocsis}}, \citenamefont {{Haiman}}, \citenamefont {{Bartos}}, \citenamefont
  {{Omukai}},\ and\ \citenamefont
  {{Samsing}}}]{Tagawa-2021-Kocsis-ApJ...908..194T}%
  \BibitemOpen
  \bibfield  {author} {\bibinfo {author} {\bibfnamefont {H.}~\bibnamefont
  {{Tagawa}}}, \bibinfo {author} {\bibfnamefont {B.}~\bibnamefont {{Kocsis}}},
  \bibinfo {author} {\bibfnamefont {Z.}~\bibnamefont {{Haiman}}}, \bibinfo
  {author} {\bibfnamefont {I.}~\bibnamefont {{Bartos}}}, \bibinfo {author}
  {\bibfnamefont {K.}~\bibnamefont {{Omukai}}}, and\ \bibinfo {author}
  {\bibfnamefont {J.}~\bibnamefont {{Samsing}}},\ }\bibfield  {title} {\enquote
  {\bibinfo {title} {{Mass-gap Mergers in Active Galactic Nuclei}},}\ }\href
  {\doibase 10.3847/1538-4357/abd555} {\bibfield  {journal} {\bibinfo
  {journal} {Astrophys. J.}\ }\textbf {\bibinfo {volume} {908}},\ \bibinfo
  {eid} {194} (\bibinfo {year} {2021}{\natexlab{a}})},\ \Eprint
  {http://arxiv.org/abs/2012.00011}{arXiv:2012.00011}\BibitemShut {NoStop}%
\bibitem [{\citenamefont {{Li}}(2022{\natexlab{b}})}]{Li-2022PhRvD.105f3006L}%
  \BibitemOpen
  \bibfield  {author} {\bibinfo {author} {\bibfnamefont {G.-P.}\ \bibnamefont
  {{Li}}},\ }\bibfield  {title} {\enquote {\bibinfo {title} {{Time-dependent
  stellar-mass binary black hole mergers in AGN disks: Mass distribution of
  hierarchical mergers}},}\ }\href {\doibase 10.1103/PhysRevD.105.063006}
  {\bibfield  {journal} {\bibinfo  {journal} {\prd}\ }\textbf {\bibinfo
  {volume} {105}},\ \bibinfo {eid} {063006} (\bibinfo {year}
  {2022}{\natexlab{b}})},\ \Eprint
  {http://arxiv.org/abs/2202.09961}{arXiv:2202.09961}\BibitemShut {NoStop}%
\bibitem [{\citenamefont {{Heger}}\ \emph {{\it et~al.}}(2003)\citenamefont
  {{Heger}}, \citenamefont {{Fryer}}, \citenamefont {{Woosley}}, \citenamefont
  {{Langer}},\ and\ \citenamefont
  {{Hartmann}}}]{Heger-2003-Fryer-ApJ...591..288H}%
  \BibitemOpen
  \bibfield  {author} {\bibinfo {author} {\bibfnamefont {A.}~\bibnamefont
  {{Heger}}}, \bibinfo {author} {\bibfnamefont {C.~L.}\ \bibnamefont
  {{Fryer}}}, \bibinfo {author} {\bibfnamefont {S.~E.}\ \bibnamefont
  {{Woosley}}}, \bibinfo {author} {\bibfnamefont {N.}~\bibnamefont {{Langer}}},
  and\ \bibinfo {author} {\bibfnamefont {D.~H.}\ \bibnamefont {{Hartmann}}},\
  }\bibfield  {title} {\enquote {\bibinfo {title} {{How Massive Single Stars
  End Their Life}},}\ }\href {\doibase 10.1086/375341} {\bibfield  {journal}
  {\bibinfo  {journal} {Astrophys. J.}\ }\textbf {\bibinfo {volume} {591}},\
  \bibinfo {pages} {288} (\bibinfo {year} {2003})},\ \Eprint
  {http://arxiv.org/abs/astro-ph/0212469}{arXiv:astro-ph/0212469}\BibitemShut
  {NoStop}%
\bibitem [{\citenamefont {{Woosley}}\ \emph {{\it et~al.}}(2007)\citenamefont
  {{Woosley}}, \citenamefont {{Blinnikov}},\ and\ \citenamefont
  {{Heger}}}]{Woosley-2007-Blinnikov-Natur.450..390W}%
  \BibitemOpen
  \bibfield  {author} {\bibinfo {author} {\bibfnamefont {S.~E.}\ \bibnamefont
  {{Woosley}}}, \bibinfo {author} {\bibfnamefont {S.}~\bibnamefont
  {{Blinnikov}}}, and\ \bibinfo {author} {\bibfnamefont {A.}~\bibnamefont
  {{Heger}}},\ }\bibfield  {title} {\enquote {\bibinfo {title} {{Pulsational
  pair instability as an explanation for the most luminous supernovae}},}\
  }\href {\doibase 10.1038/nature06333} {\bibfield  {journal} {\bibinfo
  {journal} {Nature}\ }\textbf {\bibinfo {volume} {450}},\ \bibinfo {pages}
  {390} (\bibinfo {year} {2007})},\ \Eprint
  {http://arxiv.org/abs/0710.3314}{arXiv:0710.3314}\BibitemShut {NoStop}%
\bibitem [{\citenamefont {{Quinlan}}\ and\ \citenamefont
  {{Shapiro}}(1987)}]{Quinlan-1987-Shapiro-ApJ...321..199Q}%
  \BibitemOpen
  \bibfield  {author} {\bibinfo {author} {\bibfnamefont {G.~D.}\ \bibnamefont
  {{Quinlan}}} and\ \bibinfo {author} {\bibfnamefont {S.~L.}\ \bibnamefont
  {{Shapiro}}},\ }\bibfield  {title} {\enquote {\bibinfo {title} {{The Collapse
  of Dense Star Clusters to Supermassive Black Holes: Binaries and
  Gravitational Radiation}},}\ }\href {\doibase 10.1086/165624} {\bibfield
  {journal} {\bibinfo  {journal} {Astrophys. J.}\ }\textbf {\bibinfo {volume}
  {321}},\ \bibinfo {pages} {199} (\bibinfo {year} {1987})}\BibitemShut
  {NoStop}%
\bibitem [{\citenamefont {{Fragione}}\ \emph {{\it et~al.}}(2020)\citenamefont
  {{Fragione}}, \citenamefont {{Loeb}},\ and\ \citenamefont
  {{Rasio}}}]{Fragione-2020-Loeb-ApJ...902L..26F}%
  \BibitemOpen
  \bibfield  {author} {\bibinfo {author} {\bibfnamefont {G.}~\bibnamefont
  {{Fragione}}}, \bibinfo {author} {\bibfnamefont {A.}~\bibnamefont {{Loeb}}},
  and\ \bibinfo {author} {\bibfnamefont {F.~A.}\ \bibnamefont {{Rasio}}},\
  }\bibfield  {title} {\enquote {\bibinfo {title} {{On the Origin of
  GW190521-like Events from Repeated Black Hole Mergers in Star Clusters}},}\
  }\href {\doibase 10.3847/2041-8213/abbc0a} {\bibfield  {journal} {\bibinfo
  {journal} {\apjl}\ }\textbf {\bibinfo {volume} {902}},\ \bibinfo {eid} {L26}
  (\bibinfo {year} {2020})},\ \Eprint
  {http://arxiv.org/abs/2009.05065}{arXiv:2009.05065}\BibitemShut {NoStop}%
\bibitem [{\citenamefont {{Fragione}}\ \emph {{\it et~al.}}(2022)\citenamefont
  {{Fragione}}, \citenamefont {{Kocsis}}, \citenamefont {{Rasio}},\ and\
  \citenamefont {{Silk}}}]{Fragione-2022-Kocsis-ApJ...927..231F}%
  \BibitemOpen
  \bibfield  {author} {\bibinfo {author} {\bibfnamefont {G.}~\bibnamefont
  {{Fragione}}}, \bibinfo {author} {\bibfnamefont {B.}~\bibnamefont
  {{Kocsis}}}, \bibinfo {author} {\bibfnamefont {F.~A.}\ \bibnamefont
  {{Rasio}}}, and\ \bibinfo {author} {\bibfnamefont {J.}~\bibnamefont
  {{Silk}}},\ }\bibfield  {title} {\enquote {\bibinfo {title} {{Repeated
  Mergers, Mass-gap Black Holes, and Formation of Intermediate-mass Black Holes
  in Dense Massive Star Clusters}},}\ }\href {\doibase
  10.3847/1538-4357/ac5026} {\bibfield  {journal} {\bibinfo  {journal}
  {Astrophys. J.}\ }\textbf {\bibinfo {volume} {927}},\ \bibinfo {eid} {231}
  (\bibinfo {year} {2022})},\ \Eprint
  {http://arxiv.org/abs/2107.04639}{arXiv:2107.04639}\BibitemShut {NoStop}%
\bibitem [{\citenamefont {{Gonz{\'a}lez Prieto}}\ \emph {{\it
  et~al.}}(2022)\citenamefont {{Gonz{\'a}lez Prieto}}, \citenamefont
  {{Kremer}}, \citenamefont {{Fragione}}, \citenamefont {{Martinez}},
  \citenamefont {{Weatherford}}, \citenamefont {{Zevin}},\ and\ \citenamefont
  {{Rasio}}}]{Prieto-2022-Kremer-arXiv220807881G}%
  \BibitemOpen
  \bibfield  {author} {\bibinfo {author} {\bibfnamefont {E.}~\bibnamefont
  {{Gonz{\'a}lez Prieto}}}, \bibinfo {author} {\bibfnamefont {K.}~\bibnamefont
  {{Kremer}}}, \bibinfo {author} {\bibfnamefont {G.}~\bibnamefont
  {{Fragione}}}, \bibinfo {author} {\bibfnamefont {M.~A.~S.}\ \bibnamefont
  {{Martinez}}}, \bibinfo {author} {\bibfnamefont {N.~C.}\ \bibnamefont
  {{Weatherford}}}, \bibinfo {author} {\bibfnamefont {M.}~\bibnamefont
  {{Zevin}}}, and\ \bibinfo {author} {\bibfnamefont {F.~A.}\ \bibnamefont
  {{Rasio}}},\ }\bibfield  {title} {\enquote {\bibinfo {title}
  {{Intermediate-mass Black Holes on the Run from Young Star Clusters}},}\
  }\href@noop {} {\bibfield  {journal} {\bibinfo  {journal} {arXiv e-prints}\
  ,\ \bibinfo {eid} {arXiv:2208.07881}} (\bibinfo {year} {2022})},\ \Eprint
  {http://arxiv.org/abs/2208.07881}{arXiv:2208.07881}\BibitemShut {NoStop}%
\bibitem [{\citenamefont {{Zevin}}\ and\ \citenamefont
  {{Holz}}(2022)}]{Zevin-2022-Holz-ApJ...935L..20Z}%
  \BibitemOpen
  \bibfield  {author} {\bibinfo {author} {\bibfnamefont {M.}~\bibnamefont
  {{Zevin}}} and\ \bibinfo {author} {\bibfnamefont {D.~E.}\ \bibnamefont
  {{Holz}}},\ }\bibfield  {title} {\enquote {\bibinfo {title} {{Avoiding a
  Cluster Catastrophe: Retention Efficiency and the Binary Black Hole Mass
  Spectrum}},}\ }\href {\doibase 10.3847/2041-8213/ac853d} {\bibfield
  {journal} {\bibinfo  {journal} {\apjl}\ }\textbf {\bibinfo {volume} {935}},\
  \bibinfo {eid} {L20} (\bibinfo {year} {2022})},\ \Eprint
  {http://arxiv.org/abs/2205.08549}{arXiv:2205.08549}\BibitemShut {NoStop}%
\bibitem [{\citenamefont {{The LIGO Scientific Collaboration}}\ \emph {{\it
  et~al.}}(2021{\natexlab{c}})\citenamefont {{The LIGO Scientific
  Collaboration}}, \citenamefont {{the Virgo Collaboration}}, \citenamefont
  {{the KAGRA Collaboration}}, \citenamefont {{Abbott}}, \citenamefont
  {{Abbott}}, \citenamefont {{Acernese}}, \citenamefont {{Ackley}},
  \citenamefont {{Adams}}, \citenamefont {{Adhikari}}, \citenamefont
  {{Adhikari}},\ and\ \citenamefont {et~al.}}]{LVK-GWTC-3-2021arXiv211103634T}%
  \BibitemOpen
  \bibfield  {author} {\bibinfo {author} {\bibnamefont {{The LIGO Scientific
  Collaboration}}} {\it et~al.},\ }\bibfield  {title} {\enquote {\bibinfo
  {title} {{The population of merging compact binaries inferred using
  gravitational waves through GWTC-3}},}\ }\href@noop {} {\bibfield  {journal}
  {\bibinfo  {journal} {arXiv e-prints}\ ,\ \bibinfo {eid} {arXiv:2111.03634}}
  (\bibinfo {year} {2021}{\natexlab{c}})},\ \Eprint
  {http://arxiv.org/abs/2111.03634}{arXiv:2111.03634}\BibitemShut {NoStop}%
\bibitem [{\citenamefont {{McKernan}}\ \emph {{\it et~al.}}(2012)\citenamefont
  {{McKernan}}, \citenamefont {{Ford}}, \citenamefont {{Lyra}},\ and\
  \citenamefont {{Perets}}}]{McKernan-2012-Ford-MNRAS.425..460M}%
  \BibitemOpen
  \bibfield  {author} {\bibinfo {author} {\bibfnamefont {B.}~\bibnamefont
  {{McKernan}}}, \bibinfo {author} {\bibfnamefont {K.~E.~S.}\ \bibnamefont
  {{Ford}}}, \bibinfo {author} {\bibfnamefont {W.}~\bibnamefont {{Lyra}}}, and\
  \bibinfo {author} {\bibfnamefont {H.~B.}\ \bibnamefont {{Perets}}},\
  }\bibfield  {title} {\enquote {\bibinfo {title} {{Intermediate mass black
  holes in AGN discs - I. Production and growth}},}\ }\href {\doibase
  10.1111/j.1365-2966.2012.21486.x} {\bibfield  {journal} {\bibinfo  {journal}
  {\mnras}\ }\textbf {\bibinfo {volume} {425}},\ \bibinfo {pages} {460}
  (\bibinfo {year} {2012})},\ \Eprint
  {http://arxiv.org/abs/1206.2309}{arXiv:1206.2309}\BibitemShut {NoStop}%
\bibitem [{\citenamefont {{Bellovary}}\ \emph {{\it et~al.}}(2016)\citenamefont
  {{Bellovary}}, \citenamefont {{Mac Low}}, \citenamefont {{McKernan}},\ and\
  \citenamefont {{Ford}}}]{Bellovary-2016-Mac-ApJ...819L..17B}%
  \BibitemOpen
  \bibfield  {author} {\bibinfo {author} {\bibfnamefont {J.~M.}\ \bibnamefont
  {{Bellovary}}}, \bibinfo {author} {\bibfnamefont {M.-M.}\ \bibnamefont {{Mac
  Low}}}, \bibinfo {author} {\bibfnamefont {B.}~\bibnamefont {{McKernan}}},
  and\ \bibinfo {author} {\bibfnamefont {K.~E.~S.}\ \bibnamefont {{Ford}}},\
  }\bibfield  {title} {\enquote {\bibinfo {title} {{Migration Traps in Disks
  around Supermassive Black Holes}},}\ }\href {\doibase
  10.3847/2041-8205/819/2/L17} {\bibfield  {journal} {\bibinfo  {journal}
  {\apjl}\ }\textbf {\bibinfo {volume} {819}},\ \bibinfo {eid} {L17} (\bibinfo
  {year} {2016})},\ \Eprint
  {http://arxiv.org/abs/1511.00005}{arXiv:1511.00005}\BibitemShut {NoStop}%
\bibitem [{\citenamefont {{Secunda}}\ \emph {{\it et~al.}}(2019)\citenamefont
  {{Secunda}}, \citenamefont {{Bellovary}}, \citenamefont {{Mac Low}},
  \citenamefont {{Ford}}, \citenamefont {{McKernan}}, \citenamefont {{Leigh}},
  \citenamefont {{Lyra}},\ and\ \citenamefont
  {{S{\'a}ndor}}}]{Secunda-2019-Bellovary-ApJ...878...85S}%
  \BibitemOpen
  \bibfield  {author} {\bibinfo {author} {\bibfnamefont {A.}~\bibnamefont
  {{Secunda}}}, \bibinfo {author} {\bibfnamefont {J.}~\bibnamefont
  {{Bellovary}}}, \bibinfo {author} {\bibfnamefont {M.-M.}\ \bibnamefont {{Mac
  Low}}}, \bibinfo {author} {\bibfnamefont {K.~E.~S.}\ \bibnamefont {{Ford}}},
  \bibinfo {author} {\bibfnamefont {B.}~\bibnamefont {{McKernan}}}, \bibinfo
  {author} {\bibfnamefont {N.~W.~C.}\ \bibnamefont {{Leigh}}}, \bibinfo
  {author} {\bibfnamefont {W.}~\bibnamefont {{Lyra}}}, and\ \bibinfo {author}
  {\bibfnamefont {Z.}~\bibnamefont {{S{\'a}ndor}}},\ }\bibfield  {title}
  {\enquote {\bibinfo {title} {{Orbital Migration of Interacting Stellar Mass
  Black Holes in Disks around Supermassive Black Holes}},}\ }\href {\doibase
  10.3847/1538-4357/ab20ca} {\bibfield  {journal} {\bibinfo  {journal}
  {Astrophys. J.}\ }\textbf {\bibinfo {volume} {878}},\ \bibinfo {eid} {85}
  (\bibinfo {year} {2019})},\ \Eprint
  {http://arxiv.org/abs/1807.02859}{arXiv:1807.02859}\BibitemShut {NoStop}%
\bibitem [{\citenamefont {{McKernan}}\ \emph {{\it et~al.}}(2018)\citenamefont
  {{McKernan}}, \citenamefont {{Ford}}, \citenamefont {{Bellovary}},
  \citenamefont {{Leigh}}, \citenamefont {{Haiman}}, \citenamefont {{Kocsis}},
  \citenamefont {{Lyra}}, \citenamefont {{Mac Low}}, \citenamefont {{Metzger}},
  \citenamefont {{O'Dowd}}, \citenamefont {{Endlich}},\ and\ \citenamefont
  {{Rosen}}}]{McKernan-2018-Ford-ApJ...866...66M}%
  \BibitemOpen
  \bibfield  {author} {\bibinfo {author} {\bibfnamefont {B.}~\bibnamefont
  {{McKernan}}} {\it et~al.},\ }\bibfield  {title} {\enquote {\bibinfo {title}
  {{Constraining Stellar-mass Black Hole Mergers in AGN Disks Detectable with
  LIGO}},}\ }\href {\doibase 10.3847/1538-4357/aadae5} {\bibfield  {journal}
  {\bibinfo  {journal} {Astrophys. J.}\ }\textbf {\bibinfo {volume} {866}},\
  \bibinfo {eid} {66} (\bibinfo {year} {2018})},\ \Eprint
  {http://arxiv.org/abs/1702.07818}{arXiv:1702.07818}\BibitemShut {NoStop}%
\bibitem [{\citenamefont {{Scaria}}\ and\ \citenamefont
  {{Bappu}}(1981)}]{Scaria-1981-Bappu-JApA....2..215S}%
  \BibitemOpen
  \bibfield  {author} {\bibinfo {author} {\bibfnamefont {K.~K.}\ \bibnamefont
  {{Scaria}}} and\ \bibinfo {author} {\bibfnamefont {M.~K.~V.}\ \bibnamefont
  {{Bappu}}},\ }\bibfield  {title} {\enquote {\bibinfo {title} {{Mass
  segregation in globular clusters}},}\ }\href {\doibase 10.1007/BF02714549}
  {\bibfield  {journal} {\bibinfo  {journal} {Journal of Astrophysics and
  Astronomy}\ }\textbf {\bibinfo {volume} {2}},\ \bibinfo {pages} {215}
  (\bibinfo {year} {1981})}\BibitemShut {NoStop}%
\bibitem [{\citenamefont {{Nony}}\ \emph {{\it et~al.}}(2021)\citenamefont
  {{Nony}}, \citenamefont {{Robitaille}}, \citenamefont {{Motte}},
  \citenamefont {{Gonzalez}}, \citenamefont {{Joncour}}, \citenamefont
  {{Moraux}}, \citenamefont {{Men'shchikov}}, \citenamefont {{Didelon}},
  \citenamefont {{Louvet}}, \citenamefont {{Buckner}}, \citenamefont
  {{Schneider}}, \citenamefont {{Lumsden}}, \citenamefont {{Bontemps}},
  \citenamefont {{Pouteau}}, \citenamefont {{Cunningham}}, \citenamefont
  {{Fiorellino}}, \citenamefont {{Oudmaijer}}, \citenamefont {{Andr{\'e}}},\
  and\ \citenamefont {{Thomasson}}}]{Nony-2021-Robitaille-A&A...645A..94N}%
  \BibitemOpen
  \bibfield  {author} {\bibinfo {author} {\bibfnamefont {T.}~\bibnamefont
  {{Nony}}} {\it et~al.},\ }\bibfield  {title} {\enquote {\bibinfo {title}
  {{Mass segregation and sequential star formation in NGC 2264 revealed by
  Herschel}},}\ }\href {\doibase 10.1051/0004-6361/202039353} {\bibfield
  {journal} {\bibinfo  {journal} {\aap}\ }\textbf {\bibinfo {volume} {645}},\
  \bibinfo {eid} {A94} (\bibinfo {year} {2021})},\ \Eprint
  {http://arxiv.org/abs/2011.05939}{arXiv:2011.05939}\BibitemShut {NoStop}%
\bibitem [{\citenamefont {{Pavl{\'\i}k}}\ and\ \citenamefont
  {{Vesperini}}(2022)}]{Pavl-2022-Vesperini-MNRAS.515.1830P}%
  \BibitemOpen
  \bibfield  {author} {\bibinfo {author} {\bibfnamefont {V.}~\bibnamefont
  {{Pavl{\'\i}k}}} and\ \bibinfo {author} {\bibfnamefont {E.}~\bibnamefont
  {{Vesperini}}},\ }\bibfield  {title} {\enquote {\bibinfo {title} {{Mass
  segregation and dynamics of primordial binaries in star clusters with a
  radially anisotropic velocity distribution}},}\ }\href {\doibase
  10.1093/mnras/stac1776} {\bibfield  {journal} {\bibinfo  {journal} {\mnras}\
  }\textbf {\bibinfo {volume} {515}},\ \bibinfo {pages} {1830} (\bibinfo {year}
  {2022})},\ \Eprint
  {http://arxiv.org/abs/2206.11905}{arXiv:2206.11905}\BibitemShut {NoStop}%
\bibitem [{\citenamefont {{Vitral}}\ \emph {{\it et~al.}}(2022)\citenamefont
  {{Vitral}}, \citenamefont {{Kremer}}, \citenamefont {{Libralato}},
  \citenamefont {{Mamon}},\ and\ \citenamefont
  {{Bellini}}}]{Vitral-2022-Kremer-MNRAS.514..806V}%
  \BibitemOpen
  \bibfield  {author} {\bibinfo {author} {\bibfnamefont {E.}~\bibnamefont
  {{Vitral}}}, \bibinfo {author} {\bibfnamefont {K.}~\bibnamefont {{Kremer}}},
  \bibinfo {author} {\bibfnamefont {M.}~\bibnamefont {{Libralato}}}, \bibinfo
  {author} {\bibfnamefont {G.~A.}\ \bibnamefont {{Mamon}}}, and\ \bibinfo
  {author} {\bibfnamefont {A.}~\bibnamefont {{Bellini}}},\ }\bibfield  {title}
  {\enquote {\bibinfo {title} {{Stellar graveyards: clustering of compact
  objects in globular clusters NGC 3201 and NGC 6397}},}\ }\href {\doibase
  10.1093/mnras/stac1337} {\bibfield  {journal} {\bibinfo  {journal} {\mnras}\
  }\textbf {\bibinfo {volume} {514}},\ \bibinfo {pages} {806} (\bibinfo {year}
  {2022})},\ \Eprint
  {http://arxiv.org/abs/2202.01599}{arXiv:2202.01599}\BibitemShut {NoStop}%
\bibitem [{\citenamefont {{Gerosa}}\ and\ \citenamefont
  {{Berti}}(2019)}]{Gerosa-2019-Berti-PhRvD.100d1301G}%
  \BibitemOpen
  \bibfield  {author} {\bibinfo {author} {\bibfnamefont {D.}~\bibnamefont
  {{Gerosa}}} and\ \bibinfo {author} {\bibfnamefont {E.}~\bibnamefont
  {{Berti}}},\ }\bibfield  {title} {\enquote {\bibinfo {title} {{Escape speed
  of stellar clusters from multiple-generation black-hole mergers in the upper
  mass gap}},}\ }\href {\doibase 10.1103/PhysRevD.100.041301} {\bibfield
  {journal} {\bibinfo  {journal} {\prd}\ }\textbf {\bibinfo {volume} {100}},\
  \bibinfo {eid} {041301} (\bibinfo {year} {2019})},\ \Eprint
  {http://arxiv.org/abs/1906.05295}{arXiv:1906.05295}\BibitemShut {NoStop}%
\bibitem [{\citenamefont {{Fragione}}\ and\ \citenamefont
  {{Silk}}(2020)}]{Fragione-2020-Silk-MNRAS.498.4591F}%
  \BibitemOpen
  \bibfield  {author} {\bibinfo {author} {\bibfnamefont {G.}~\bibnamefont
  {{Fragione}}} and\ \bibinfo {author} {\bibfnamefont {J.}~\bibnamefont
  {{Silk}}},\ }\bibfield  {title} {\enquote {\bibinfo {title} {{Repeated
  mergers and ejection of black holes within nuclear star clusters}},}\ }\href
  {\doibase 10.1093/mnras/staa2629} {\bibfield  {journal} {\bibinfo  {journal}
  {\mnras}\ }\textbf {\bibinfo {volume} {498}},\ \bibinfo {pages} {4591}
  (\bibinfo {year} {2020})},\ \Eprint
  {http://arxiv.org/abs/2006.01867}{arXiv:2006.01867}\BibitemShut {NoStop}%
\bibitem [{\citenamefont {{Liu}}\ and\ \citenamefont
  {{Lai}}(2021)}]{Liu-2021-Lai-MNRAS.502.2049L}%
  \BibitemOpen
  \bibfield  {author} {\bibinfo {author} {\bibfnamefont {B.}~\bibnamefont
  {{Liu}}} and\ \bibinfo {author} {\bibfnamefont {D.}~\bibnamefont {{Lai}}},\
  }\bibfield  {title} {\enquote {\bibinfo {title} {{Hierarchical black hole
  mergers in multiple systems: constrain the formation of GW190412-, GW190814-,
  and GW190521-like events}},}\ }\href {\doibase 10.1093/mnras/stab178}
  {\bibfield  {journal} {\bibinfo  {journal} {\mnras}\ }\textbf {\bibinfo
  {volume} {502}},\ \bibinfo {pages} {2049} (\bibinfo {year} {2021})},\ \Eprint
  {http://arxiv.org/abs/2009.10068}{arXiv:2009.10068}\BibitemShut {NoStop}%
\bibitem [{\citenamefont {{Mahapatra}}\ \emph {{\it et~al.}}(2021)\citenamefont
  {{Mahapatra}}, \citenamefont {{Gupta}}, \citenamefont {{Favata}},
  \citenamefont {{Arun}},\ and\ \citenamefont
  {{Sathyaprakash}}}]{Mahapatra-2021-Gupta-ApJ...918L..31M}%
  \BibitemOpen
  \bibfield  {author} {\bibinfo {author} {\bibfnamefont {P.}~\bibnamefont
  {{Mahapatra}}}, \bibinfo {author} {\bibfnamefont {A.}~\bibnamefont
  {{Gupta}}}, \bibinfo {author} {\bibfnamefont {M.}~\bibnamefont {{Favata}}},
  \bibinfo {author} {\bibfnamefont {K.~G.}\ \bibnamefont {{Arun}}}, and\
  \bibinfo {author} {\bibfnamefont {B.~S.}\ \bibnamefont {{Sathyaprakash}}},\
  }\bibfield  {title} {\enquote {\bibinfo {title} {{Remnant Black Hole Kicks
  and Implications for Hierarchical Mergers}},}\ }\href {\doibase
  10.3847/2041-8213/ac20db} {\bibfield  {journal} {\bibinfo  {journal} {\apjl}\
  }\textbf {\bibinfo {volume} {918}},\ \bibinfo {eid} {L31} (\bibinfo {year}
  {2021})},\ \Eprint
  {http://arxiv.org/abs/2106.07179}{arXiv:2106.07179}\BibitemShut {NoStop}%
\bibitem [{\citenamefont {{Doctor}}\ \emph {{\it et~al.}}(2020)\citenamefont
  {{Doctor}}, \citenamefont {{Wysocki}}, \citenamefont {{O'Shaughnessy}},
  \citenamefont {{Holz}},\ and\ \citenamefont
  {{Farr}}}]{Doctor-2020-Wysocki-ApJ...893...35D}%
  \BibitemOpen
  \bibfield  {author} {\bibinfo {author} {\bibfnamefont {Z.}~\bibnamefont
  {{Doctor}}}, \bibinfo {author} {\bibfnamefont {D.}~\bibnamefont {{Wysocki}}},
  \bibinfo {author} {\bibfnamefont {R.}~\bibnamefont {{O'Shaughnessy}}},
  \bibinfo {author} {\bibfnamefont {D.~E.}\ \bibnamefont {{Holz}}}, and\
  \bibinfo {author} {\bibfnamefont {B.}~\bibnamefont {{Farr}}},\ }\bibfield
  {title} {\enquote {\bibinfo {title} {{Black Hole Coagulation: Modeling
  Hierarchical Mergers in Black Hole Populations}},}\ }\href {\doibase
  10.3847/1538-4357/ab7fac} {\bibfield  {journal} {\bibinfo  {journal}
  {Astrophys. J.}\ }\textbf {\bibinfo {volume} {893}},\ \bibinfo {eid} {35}
  (\bibinfo {year} {2020})},\ \Eprint
  {http://arxiv.org/abs/1911.04424}{arXiv:1911.04424}\BibitemShut {NoStop}%
\bibitem [{\citenamefont {{Tagawa}}\ \emph {{\it
  et~al.}}(2021{\natexlab{b}})\citenamefont {{Tagawa}}, \citenamefont
  {{Haiman}}, \citenamefont {{Bartos}}, \citenamefont {{Kocsis}},\ and\
  \citenamefont {{Omukai}}}]{Tagawa-2021-Haiman-MNRAS.507.3362T}%
  \BibitemOpen
  \bibfield  {author} {\bibinfo {author} {\bibfnamefont {H.}~\bibnamefont
  {{Tagawa}}}, \bibinfo {author} {\bibfnamefont {Z.}~\bibnamefont {{Haiman}}},
  \bibinfo {author} {\bibfnamefont {I.}~\bibnamefont {{Bartos}}}, \bibinfo
  {author} {\bibfnamefont {B.}~\bibnamefont {{Kocsis}}}, and\ \bibinfo {author}
  {\bibfnamefont {K.}~\bibnamefont {{Omukai}}},\ }\bibfield  {title} {\enquote
  {\bibinfo {title} {{Signatures of hierarchical mergers in black hole spin and
  mass distribution}},}\ }\href {\doibase 10.1093/mnras/stab2315} {\bibfield
  {journal} {\bibinfo  {journal} {\mnras}\ }\textbf {\bibinfo {volume} {507}},\
  \bibinfo {pages} {3362} (\bibinfo {year} {2021}{\natexlab{b}})},\ \Eprint
  {http://arxiv.org/abs/2104.09510}{arXiv:2104.09510}\BibitemShut {NoStop}%
\bibitem [{\citenamefont {{Gerosa}}\ \emph {{\it et~al.}}(2021)\citenamefont
  {{Gerosa}}, \citenamefont {{Giacobbo}},\ and\ \citenamefont
  {{Vecchio}}}]{Gerosa-2021-Giacobbo-ApJ...915...56G}%
  \BibitemOpen
  \bibfield  {author} {\bibinfo {author} {\bibfnamefont {D.}~\bibnamefont
  {{Gerosa}}}, \bibinfo {author} {\bibfnamefont {N.}~\bibnamefont
  {{Giacobbo}}}, and\ \bibinfo {author} {\bibfnamefont {A.}~\bibnamefont
  {{Vecchio}}},\ }\bibfield  {title} {\enquote {\bibinfo {title} {{High Mass
  but Low Spin: An Exclusion Region to Rule Out Hierarchical Black Hole Mergers
  as a Mechanism to Populate the Pair-instability Mass Gap}},}\ }\href
  {\doibase 10.3847/1538-4357/ac00bb} {\bibfield  {journal} {\bibinfo
  {journal} {Astrophys. J.}\ }\textbf {\bibinfo {volume} {915}},\ \bibinfo
  {eid} {56} (\bibinfo {year} {2021})},\ \Eprint
  {http://arxiv.org/abs/2104.11247}{arXiv:2104.11247}\BibitemShut {NoStop}%
\bibitem [{\citenamefont {{Barausse}}\ \emph {{\it et~al.}}(2012)\citenamefont
  {{Barausse}}, \citenamefont {{Morozova}},\ and\ \citenamefont
  {{Rezzolla}}}]{Barausse-2012-Morozova-ApJ...758...63B}%
  \BibitemOpen
  \bibfield  {author} {\bibinfo {author} {\bibfnamefont {E.}~\bibnamefont
  {{Barausse}}}, \bibinfo {author} {\bibfnamefont {V.}~\bibnamefont
  {{Morozova}}}, and\ \bibinfo {author} {\bibfnamefont {L.}~\bibnamefont
  {{Rezzolla}}},\ }\bibfield  {title} {\enquote {\bibinfo {title} {{On the Mass
  Radiated by Coalescing Black Hole Binaries}},}\ }\href {\doibase
  10.1088/0004-637X/758/1/63} {\bibfield  {journal} {\bibinfo  {journal}
  {Astrophys. J.}\ }\textbf {\bibinfo {volume} {758}},\ \bibinfo {eid} {63}
  (\bibinfo {year} {2012})},\ \Eprint
  {http://arxiv.org/abs/1206.3803}{arXiv:1206.3803}\BibitemShut {NoStop}%
\bibitem [{\citenamefont {{Hofmann}}\ \emph {{\it et~al.}}(2016)\citenamefont
  {{Hofmann}}, \citenamefont {{Barausse}},\ and\ \citenamefont
  {{Rezzolla}}}]{Hofmann-2016-Barausse-ApJ...825L..19H}%
  \BibitemOpen
  \bibfield  {author} {\bibinfo {author} {\bibfnamefont {F.}~\bibnamefont
  {{Hofmann}}}, \bibinfo {author} {\bibfnamefont {E.}~\bibnamefont
  {{Barausse}}}, and\ \bibinfo {author} {\bibfnamefont {L.}~\bibnamefont
  {{Rezzolla}}},\ }\bibfield  {title} {\enquote {\bibinfo {title} {{The Final
  Spin from Binary Black Holes in Quasi-circular Orbits}},}\ }\href {\doibase
  10.3847/2041-8205/825/2/L19} {\bibfield  {journal} {\bibinfo  {journal}
  {\apjl}\ }\textbf {\bibinfo {volume} {825}},\ \bibinfo {eid} {L19} (\bibinfo
  {year} {2016})},\ \Eprint
  {http://arxiv.org/abs/1605.01938}{arXiv:1605.01938}\BibitemShut {NoStop}%
\bibitem [{\citenamefont {{Campanelli}}\ \emph {{\it
  et~al.}}(2007)\citenamefont {{Campanelli}}, \citenamefont {{Lousto}},
  \citenamefont {{Zlochower}},\ and\ \citenamefont
  {{Merritt}}}]{Campanelli-2007-Lousto-ApJ...659L...5C}%
  \BibitemOpen
  \bibfield  {author} {\bibinfo {author} {\bibfnamefont {M.}~\bibnamefont
  {{Campanelli}}}, \bibinfo {author} {\bibfnamefont {C.}~\bibnamefont
  {{Lousto}}}, \bibinfo {author} {\bibfnamefont {Y.}~\bibnamefont
  {{Zlochower}}}, and\ \bibinfo {author} {\bibfnamefont {D.}~\bibnamefont
  {{Merritt}}},\ }\bibfield  {title} {\enquote {\bibinfo {title} {{Large Merger
  Recoils and Spin Flips from Generic Black Hole Binaries}},}\ }\href {\doibase
  10.1086/516712} {\bibfield  {journal} {\bibinfo  {journal} {\apjl}\ }\textbf
  {\bibinfo {volume} {659}},\ \bibinfo {pages} {L5} (\bibinfo {year} {2007})},\
  \Eprint {http://arxiv.org/abs/gr-qc/0701164}{arXiv:gr-qc/0701164}\BibitemShut
  {NoStop}%
\bibitem [{\citenamefont {{Gerosa}}\ and\ \citenamefont
  {{Kesden}}(2016)}]{Gerosa-2016-Kesden-PhRvD..93l4066G}%
  \BibitemOpen
  \bibfield  {author} {\bibinfo {author} {\bibfnamefont {D.}~\bibnamefont
  {{Gerosa}}} and\ \bibinfo {author} {\bibfnamefont {M.}~\bibnamefont
  {{Kesden}}},\ }\bibfield  {title} {\enquote {\bibinfo {title} {{precession:
  Dynamics of spinning black-hole binaries with python}},}\ }\href {\doibase
  10.1103/PhysRevD.93.124066} {\bibfield  {journal} {\bibinfo  {journal}
  {\prd}\ }\textbf {\bibinfo {volume} {93}},\ \bibinfo {eid} {124066} (\bibinfo
  {year} {2016})},\ \Eprint
  {http://arxiv.org/abs/1605.01067}{arXiv:1605.01067}\BibitemShut {NoStop}%
\bibitem [{\citenamefont {{Kroupa}}(2001)}]{Kroupa-2001MNRAS.322..231K}%
  \BibitemOpen
  \bibfield  {author} {\bibinfo {author} {\bibfnamefont {P.}~\bibnamefont
  {{Kroupa}}},\ }\bibfield  {title} {\enquote {\bibinfo {title} {{On the
  variation of the initial mass function}},}\ }\href {\doibase
  10.1046/j.1365-8711.2001.04022.x} {\bibfield  {journal} {\bibinfo  {journal}
  {\mnras}\ }\textbf {\bibinfo {volume} {322}},\ \bibinfo {pages} {231}
  (\bibinfo {year} {2001})},\ \Eprint
  {http://arxiv.org/abs/astro-ph/0009005}{arXiv:astro-ph/0009005}\BibitemShut
  {NoStop}%
\bibitem [{\citenamefont {{Yang}}\ \emph {{\it
  et~al.}}(2019{\natexlab{b}})\citenamefont {{Yang}}, \citenamefont {{Bartos}},
  \citenamefont {{Haiman}}, \citenamefont {{Kocsis}}, \citenamefont
  {{M{\'a}rka}}, \citenamefont {{Stone}},\ and\ \citenamefont
  {{M{\'a}rka}}}]{Yang-2019-Bartos-ApJ...876..122Y}%
  \BibitemOpen
  \bibfield  {author} {\bibinfo {author} {\bibfnamefont {Y.}~\bibnamefont
  {{Yang}}}, \bibinfo {author} {\bibfnamefont {I.}~\bibnamefont {{Bartos}}},
  \bibinfo {author} {\bibfnamefont {Z.}~\bibnamefont {{Haiman}}}, \bibinfo
  {author} {\bibfnamefont {B.}~\bibnamefont {{Kocsis}}}, \bibinfo {author}
  {\bibfnamefont {Z.}~\bibnamefont {{M{\'a}rka}}}, \bibinfo {author}
  {\bibfnamefont {N.~C.}\ \bibnamefont {{Stone}}}, and\ \bibinfo {author}
  {\bibfnamefont {S.}~\bibnamefont {{M{\'a}rka}}},\ }\bibfield  {title}
  {\enquote {\bibinfo {title} {{AGN Disks Harden the Mass Distribution of
  Stellar-mass Binary Black Hole Mergers}},}\ }\href {\doibase
  10.3847/1538-4357/ab16e3} {\bibfield  {journal} {\bibinfo  {journal}
  {Astrophys. J.}\ }\textbf {\bibinfo {volume} {876}},\ \bibinfo {eid} {122}
  (\bibinfo {year} {2019}{\natexlab{b}})},\ \Eprint
  {http://arxiv.org/abs/1903.01405}{arXiv:1903.01405}\BibitemShut {NoStop}%
\bibitem [{\citenamefont {{Yi}}\ and\ \citenamefont
  {{Cheng}}(2019)}]{Yi-2019-Cheng-ApJ...884L..12Y}%
  \BibitemOpen
  \bibfield  {author} {\bibinfo {author} {\bibfnamefont {S.-X.}\ \bibnamefont
  {{Yi}}} and\ \bibinfo {author} {\bibfnamefont {K.~S.}\ \bibnamefont
  {{Cheng}}},\ }\bibfield  {title} {\enquote {\bibinfo {title} {{Where Are the
  Electromagnetic-wave Counterparts of Stellar-mass Binary Black Hole
  Mergers?}}}\ }\href {\doibase 10.3847/2041-8213/ab459a} {\bibfield  {journal}
  {\bibinfo  {journal} {\apjl}\ }\textbf {\bibinfo {volume} {884}},\ \bibinfo
  {eid} {L12} (\bibinfo {year} {2019})},\ \Eprint
  {http://arxiv.org/abs/1909.08384}{arXiv:1909.08384}\BibitemShut {NoStop}%
\bibitem [{\citenamefont {{Bogdanovi{\'c}}}\ \emph {{\it
  et~al.}}(2007)\citenamefont {{Bogdanovi{\'c}}}, \citenamefont {{Reynolds}},\
  and\ \citenamefont {{Miller}}}]{Bogdanovi-2007-Reynolds-ApJ...661L.147B}%
  \BibitemOpen
  \bibfield  {author} {\bibinfo {author} {\bibfnamefont {T.}~\bibnamefont
  {{Bogdanovi{\'c}}}}, \bibinfo {author} {\bibfnamefont {C.~S.}\ \bibnamefont
  {{Reynolds}}}, and\ \bibinfo {author} {\bibfnamefont {M.~C.}\ \bibnamefont
  {{Miller}}},\ }\bibfield  {title} {\enquote {\bibinfo {title} {{Alignment of
  the Spins of Supermassive Black Holes Prior to Coalescence}},}\ }\href
  {\doibase 10.1086/518769} {\bibfield  {journal} {\bibinfo  {journal} {\apjl}\
  }\textbf {\bibinfo {volume} {661}},\ \bibinfo {pages} {L147} (\bibinfo {year}
  {2007})},\ \Eprint
  {http://arxiv.org/abs/astro-ph/0703054}{arXiv:astro-ph/0703054}\BibitemShut
  {NoStop}%
\bibitem [{\citenamefont {{McKernan}}\ \emph {{\it et~al.}}(2022)\citenamefont
  {{McKernan}}, \citenamefont {{Ford}}, \citenamefont {{Callister}},
  \citenamefont {{Farr}}, \citenamefont {{O'Shaughnessy}}, \citenamefont
  {{Smith}}, \citenamefont {{Thrane}},\ and\ \citenamefont
  {{Vajpeyi}}}]{McKernan-2022-Ford-MNRAS.514.3886M}%
  \BibitemOpen
  \bibfield  {author} {\bibinfo {author} {\bibfnamefont {B.}~\bibnamefont
  {{McKernan}}}, \bibinfo {author} {\bibfnamefont {K.~E.~S.}\ \bibnamefont
  {{Ford}}}, \bibinfo {author} {\bibfnamefont {T.}~\bibnamefont {{Callister}}},
  \bibinfo {author} {\bibfnamefont {W.~M.}\ \bibnamefont {{Farr}}}, \bibinfo
  {author} {\bibfnamefont {R.}~\bibnamefont {{O'Shaughnessy}}}, \bibinfo
  {author} {\bibfnamefont {R.}~\bibnamefont {{Smith}}}, \bibinfo {author}
  {\bibfnamefont {E.}~\bibnamefont {{Thrane}}}, and\ \bibinfo {author}
  {\bibfnamefont {A.}~\bibnamefont {{Vajpeyi}}},\ }\bibfield  {title} {\enquote
  {\bibinfo {title} {{LIGO-Virgo correlations between mass ratio and effective
  inspiral spin: testing the active galactic nuclei channel}},}\ }\href
  {\doibase 10.1093/mnras/stac1570} {\bibfield  {journal} {\bibinfo  {journal}
  {\mnras}\ }\textbf {\bibinfo {volume} {514}},\ \bibinfo {pages} {3886}
  (\bibinfo {year} {2022})},\ \Eprint
  {http://arxiv.org/abs/2107.07551}{arXiv:2107.07551}\BibitemShut {NoStop}%
\bibitem [{\citenamefont {{McKernan}}\ \emph {{\it et~al.}}(2020)\citenamefont
  {{McKernan}}, \citenamefont {{Ford}}, \citenamefont {{O'Shaugnessy}},\ and\
  \citenamefont {{Wysocki}}}]{McKernan-2020-Ford-MNRAS.494.1203M}%
  \BibitemOpen
  \bibfield  {author} {\bibinfo {author} {\bibfnamefont {B.}~\bibnamefont
  {{McKernan}}}, \bibinfo {author} {\bibfnamefont {K.~E.~S.}\ \bibnamefont
  {{Ford}}}, \bibinfo {author} {\bibfnamefont {R.}~\bibnamefont
  {{O'Shaugnessy}}}, and\ \bibinfo {author} {\bibfnamefont {D.}~\bibnamefont
  {{Wysocki}}},\ }\bibfield  {title} {\enquote {\bibinfo {title} {{Monte Carlo
  simulations of black hole mergers in AGN discs: Low
  {\ensuremath{\chi}}$_{eff}$ mergers and predictions for LIGO}},}\ }\href
  {\doibase 10.1093/mnras/staa740} {\bibfield  {journal} {\bibinfo  {journal}
  {\mnras}\ }\textbf {\bibinfo {volume} {494}},\ \bibinfo {pages} {1203}
  (\bibinfo {year} {2020})},\ \Eprint
  {http://arxiv.org/abs/1907.04356}{arXiv:1907.04356}\BibitemShut {NoStop}%
\bibitem [{\citenamefont {{Tiwari}}(2021)}]{Tiwari-2021CQGra..38o5007T}%
  \BibitemOpen
  \bibfield  {author} {\bibinfo {author} {\bibfnamefont {V.}~\bibnamefont
  {{Tiwari}}},\ }\bibfield  {title} {\enquote {\bibinfo {title} {{VAMANA:
  modeling binary black hole population with minimal assumptions}},}\ }\href
  {\doibase 10.1088/1361-6382/ac0b54} {\bibfield  {journal} {\bibinfo
  {journal} {Classical and Quantum Gravity}\ }\textbf {\bibinfo {volume}
  {38}},\ \bibinfo {eid} {155007} (\bibinfo {year} {2021})},\ \Eprint
  {http://arxiv.org/abs/2006.15047}{arXiv:2006.15047}\BibitemShut {NoStop}%
\bibitem [{\citenamefont {{Tiwari}}(2022)}]{Tiwari-2022ApJ...928..155T}%
  \BibitemOpen
  \bibfield  {author} {\bibinfo {author} {\bibfnamefont {V.}~\bibnamefont
  {{Tiwari}}},\ }\bibfield  {title} {\enquote {\bibinfo {title} {{Exploring
  Features in the Binary Black Hole Population}},}\ }\href {\doibase
  10.3847/1538-4357/ac589a} {\bibfield  {journal} {\bibinfo  {journal}
  {Astrophys. J.}\ }\textbf {\bibinfo {volume} {928}},\ \bibinfo {eid} {155}
  (\bibinfo {year} {2022})},\ \Eprint
  {http://arxiv.org/abs/2111.13991}{arXiv:2111.13991}\BibitemShut {NoStop}%
\bibitem [{\citenamefont {{Varma}}\ \emph {{\it et~al.}}(2022)\citenamefont
  {{Varma}}, \citenamefont {{Biscoveanu}}, \citenamefont {{Islam}},
  \citenamefont {{Shaik}}, \citenamefont {{Haster}}, \citenamefont {{Isi}},
  \citenamefont {{Farr}}, \citenamefont {{Field}},\ and\ \citenamefont
  {{Vitale}}}]{Varma-2022-Biscoveanu-PhRvL.128s1102V}%
  \BibitemOpen
  \bibfield  {author} {\bibinfo {author} {\bibfnamefont {V.}~\bibnamefont
  {{Varma}}}, \bibinfo {author} {\bibfnamefont {S.}~\bibnamefont
  {{Biscoveanu}}}, \bibinfo {author} {\bibfnamefont {T.}~\bibnamefont
  {{Islam}}}, \bibinfo {author} {\bibfnamefont {F.~H.}\ \bibnamefont
  {{Shaik}}}, \bibinfo {author} {\bibfnamefont {C.-J.}\ \bibnamefont
  {{Haster}}}, \bibinfo {author} {\bibfnamefont {M.}~\bibnamefont {{Isi}}},
  \bibinfo {author} {\bibfnamefont {W.~M.}\ \bibnamefont {{Farr}}}, \bibinfo
  {author} {\bibfnamefont {S.~E.}\ \bibnamefont {{Field}}}, and\ \bibinfo
  {author} {\bibfnamefont {S.}~\bibnamefont {{Vitale}}},\ }\bibfield  {title}
  {\enquote {\bibinfo {title} {{Evidence of Large Recoil Velocity from a Black
  Hole Merger Signal}},}\ }\href {\doibase 10.1103/PhysRevLett.128.191102}
  {\bibfield  {journal} {\bibinfo  {journal} {\prl}\ }\textbf {\bibinfo
  {volume} {128}},\ \bibinfo {eid} {191102} (\bibinfo {year} {2022})},\ \Eprint
  {http://arxiv.org/abs/2201.01302}{arXiv:2201.01302}\BibitemShut {NoStop}%
\bibitem [{\citenamefont {{Antonini}}\ and\ \citenamefont
  {{Rasio}}(2016)}]{Antonini-2016-Rasio-ApJ...831..187A}%
  \BibitemOpen
  \bibfield  {author} {\bibinfo {author} {\bibfnamefont {F.}~\bibnamefont
  {{Antonini}}} and\ \bibinfo {author} {\bibfnamefont {F.~A.}\ \bibnamefont
  {{Rasio}}},\ }\bibfield  {title} {\enquote {\bibinfo {title} {{Merging Black
  Hole Binaries in Galactic Nuclei: Implications for Advanced-LIGO
  Detections}},}\ }\href {\doibase 10.3847/0004-637X/831/2/187} {\bibfield
  {journal} {\bibinfo  {journal} {Astrophys. J.}\ }\textbf {\bibinfo {volume}
  {831}},\ \bibinfo {eid} {187} (\bibinfo {year} {2016})},\ \Eprint
  {http://arxiv.org/abs/1606.04889}{arXiv:1606.04889}\BibitemShut {NoStop}%
\bibitem [{\citenamefont {{Tagawa}}\ \emph {{\it et~al.}}(2020)\citenamefont
  {{Tagawa}}, \citenamefont {{Haiman}},\ and\ \citenamefont
  {{Kocsis}}}]{Tagawa-2020-Haiman-ApJ...898...25T}%
  \BibitemOpen
  \bibfield  {author} {\bibinfo {author} {\bibfnamefont {H.}~\bibnamefont
  {{Tagawa}}}, \bibinfo {author} {\bibfnamefont {Z.}~\bibnamefont {{Haiman}}},
  and\ \bibinfo {author} {\bibfnamefont {B.}~\bibnamefont {{Kocsis}}},\
  }\bibfield  {title} {\enquote {\bibinfo {title} {{Formation and Evolution of
  Compact-object Binaries in AGN Disks}},}\ }\href {\doibase
  10.3847/1538-4357/ab9b8c} {\bibfield  {journal} {\bibinfo  {journal}
  {Astrophys. J.}\ }\textbf {\bibinfo {volume} {898}},\ \bibinfo {eid} {25}
  (\bibinfo {year} {2020})},\ \Eprint
  {http://arxiv.org/abs/1912.08218}{arXiv:1912.08218}\BibitemShut {NoStop}%
\bibitem [{\citenamefont {{Yang}}\ \emph {{\it et~al.}}(2020)\citenamefont
  {{Yang}}, \citenamefont {{Gayathri}}, \citenamefont {{Bartos}}, \citenamefont
  {{Haiman}}, \citenamefont {{Safarzadeh}},\ and\ \citenamefont
  {{Tagawa}}}]{Yang-2020-Bartos-ApJ...901L..34Y}%
  \BibitemOpen
  \bibfield  {author} {\bibinfo {author} {\bibfnamefont {Y.}~\bibnamefont
  {{Yang}}}, \bibinfo {author} {\bibfnamefont {V.}~\bibnamefont {{Gayathri}}},
  \bibinfo {author} {\bibfnamefont {I.}~\bibnamefont {{Bartos}}}, \bibinfo
  {author} {\bibfnamefont {Z.}~\bibnamefont {{Haiman}}}, \bibinfo {author}
  {\bibfnamefont {M.}~\bibnamefont {{Safarzadeh}}}, and\ \bibinfo {author}
  {\bibfnamefont {H.}~\bibnamefont {{Tagawa}}},\ }\bibfield  {title} {\enquote
  {\bibinfo {title} {{Black Hole Formation in the Lower Mass Gap through
  Mergers and Accretion in AGN Disks}},}\ }\href {\doibase
  10.3847/2041-8213/abb940} {\bibfield  {journal} {\bibinfo  {journal} {\apjl}\
  }\textbf {\bibinfo {volume} {901}},\ \bibinfo {eid} {L34} (\bibinfo {year}
  {2020})},\ \Eprint
  {http://arxiv.org/abs/2007.04781}{arXiv:2007.04781}\BibitemShut {NoStop}%
\bibitem [{\citenamefont {{Dominik}}\ \emph {{\it et~al.}}(2012)\citenamefont
  {{Dominik}}, \citenamefont {{Belczynski}}, \citenamefont {{Fryer}},
  \citenamefont {{Holz}}, \citenamefont {{Berti}}, \citenamefont {{Bulik}},
  \citenamefont {{Mandel}},\ and\ \citenamefont
  {{O'Shaughnessy}}}]{Dominik-2012-Belczynski-ApJ...759...52D}%
  \BibitemOpen
  \bibfield  {author} {\bibinfo {author} {\bibfnamefont {M.}~\bibnamefont
  {{Dominik}}}, \bibinfo {author} {\bibfnamefont {K.}~\bibnamefont
  {{Belczynski}}}, \bibinfo {author} {\bibfnamefont {C.}~\bibnamefont
  {{Fryer}}}, \bibinfo {author} {\bibfnamefont {D.~E.}\ \bibnamefont {{Holz}}},
  \bibinfo {author} {\bibfnamefont {E.}~\bibnamefont {{Berti}}}, \bibinfo
  {author} {\bibfnamefont {T.}~\bibnamefont {{Bulik}}}, \bibinfo {author}
  {\bibfnamefont {I.}~\bibnamefont {{Mandel}}}, and\ \bibinfo {author}
  {\bibfnamefont {R.}~\bibnamefont {{O'Shaughnessy}}},\ }\bibfield  {title}
  {\enquote {\bibinfo {title} {{Double Compact Objects. I. The Significance of
  the Common Envelope on Merger Rates}},}\ }\href {\doibase
  10.1088/0004-637X/759/1/52} {\bibfield  {journal} {\bibinfo  {journal}
  {Astrophys. J.}\ }\textbf {\bibinfo {volume} {759}},\ \bibinfo {eid} {52}
  (\bibinfo {year} {2012})},\ \Eprint
  {http://arxiv.org/abs/1202.4901}{arXiv:1202.4901}\BibitemShut {NoStop}%
\bibitem [{\citenamefont {{Di Carlo}}\ \emph {{\it et~al.}}(2020)\citenamefont
  {{Di Carlo}}, \citenamefont {{Mapelli}}, \citenamefont {{Bouffanais}},
  \citenamefont {{Giacobbo}}, \citenamefont {{Santoliquido}}, \citenamefont
  {{Bressan}}, \citenamefont {{Spera}},\ and\ \citenamefont
  {{Haardt}}}]{Carlo-2020-Mapelli-MNRAS.497.1043D}%
  \BibitemOpen
  \bibfield  {author} {\bibinfo {author} {\bibfnamefont {U.~N.}\ \bibnamefont
  {{Di Carlo}}}, \bibinfo {author} {\bibfnamefont {M.}~\bibnamefont
  {{Mapelli}}}, \bibinfo {author} {\bibfnamefont {Y.}~\bibnamefont
  {{Bouffanais}}}, \bibinfo {author} {\bibfnamefont {N.}~\bibnamefont
  {{Giacobbo}}}, \bibinfo {author} {\bibfnamefont {F.}~\bibnamefont
  {{Santoliquido}}}, \bibinfo {author} {\bibfnamefont {A.}~\bibnamefont
  {{Bressan}}}, \bibinfo {author} {\bibfnamefont {M.}~\bibnamefont {{Spera}}},
  and\ \bibinfo {author} {\bibfnamefont {F.}~\bibnamefont {{Haardt}}},\
  }\bibfield  {title} {\enquote {\bibinfo {title} {{Binary black holes in the
  pair instability mass gap}},}\ }\href {\doibase 10.1093/mnras/staa1997}
  {\bibfield  {journal} {\bibinfo  {journal} {\mnras}\ }\textbf {\bibinfo
  {volume} {497}},\ \bibinfo {pages} {1043} (\bibinfo {year} {2020})},\ \Eprint
  {http://arxiv.org/abs/1911.01434}{arXiv:1911.01434}\BibitemShut {NoStop}%
\bibitem [{\citenamefont {{Bartos}}\ \emph {{\it et~al.}}(2017)\citenamefont
  {{Bartos}}, \citenamefont {{Kocsis}}, \citenamefont {{Haiman}},\ and\
  \citenamefont {{M{\'a}rka}}}]{Bartos-2017-Kocsis-ApJ...835..165B}%
  \BibitemOpen
  \bibfield  {author} {\bibinfo {author} {\bibfnamefont {I.}~\bibnamefont
  {{Bartos}}}, \bibinfo {author} {\bibfnamefont {B.}~\bibnamefont {{Kocsis}}},
  \bibinfo {author} {\bibfnamefont {Z.}~\bibnamefont {{Haiman}}}, and\ \bibinfo
  {author} {\bibfnamefont {S.}~\bibnamefont {{M{\'a}rka}}},\ }\bibfield
  {title} {\enquote {\bibinfo {title} {{Rapid and Bright Stellar-mass Binary
  Black Hole Mergers in Active Galactic Nuclei}},}\ }\href {\doibase
  10.3847/1538-4357/835/2/165} {\bibfield  {journal} {\bibinfo  {journal}
  {Astrophys. J.}\ }\textbf {\bibinfo {volume} {835}},\ \bibinfo {eid} {165}
  (\bibinfo {year} {2017})},\ \Eprint
  {http://arxiv.org/abs/1602.03831}{arXiv:1602.03831}\BibitemShut {NoStop}%
\bibitem [{\citenamefont {{Abbott}}\ \emph {{\it
  et~al.}}(2020{\natexlab{a}})\citenamefont {{Abbott}}, \citenamefont
  {{Abbott}}, \citenamefont {{Abraham}}, \citenamefont {{Acernese}},
  \citenamefont {{Ackley}}, \citenamefont {{Adams}}, \citenamefont
  {{Adhikari}}, \citenamefont {{Adya}}, \citenamefont {{Affeldt}},
  \citenamefont {{Agathos}},\ and\ \citenamefont
  {et~al.}}]{GW190412-2020PhRvD.102d3015A}%
  \BibitemOpen
  \bibfield  {author} {\bibinfo {author} {\bibfnamefont {R.}~\bibnamefont
  {{Abbott}}} {\it et~al.},\ }\bibfield  {title} {\enquote {\bibinfo {title}
  {{GW190412: Observation of a binary-black-hole coalescence with asymmetric
  masses}},}\ }\href {\doibase 10.1103/PhysRevD.102.043015} {\bibfield
  {journal} {\bibinfo  {journal} {\prd}\ }\textbf {\bibinfo {volume} {102}},\
  \bibinfo {eid} {043015} (\bibinfo {year} {2020}{\natexlab{a}})},\ \Eprint
  {http://arxiv.org/abs/2004.08342}{arXiv:2004.08342}\BibitemShut {NoStop}%
\bibitem [{\citenamefont {{Abbott}}\ \emph {{\it
  et~al.}}(2020{\natexlab{b}})\citenamefont {{Abbott}}, \citenamefont
  {{Abbott}}, \citenamefont {{Abraham}}, \citenamefont {{Acernese}},
  \citenamefont {{Ackley}}, \citenamefont {{Adams}}, \citenamefont
  {{Adhikari}}, \citenamefont {{Adya}}, \citenamefont {{Affeldt}},
  \citenamefont {{Agathos}},\ and\ \citenamefont
  {et~al.}}]{GW190521-2020PhRvL.125j1102A}%
  \BibitemOpen
  \bibfield  {author} {\bibinfo {author} {\bibfnamefont {R.}~\bibnamefont
  {{Abbott}}} {\it et~al.},\ }\bibfield  {title} {\enquote {\bibinfo {title}
  {{GW190521: A Binary Black Hole Merger with a Total Mass of 150
  M$_{{\ensuremath{\odot}}}$}},}\ }\href {\doibase
  10.1103/PhysRevLett.125.101102} {\bibfield  {journal} {\bibinfo  {journal}
  {\prl}\ }\textbf {\bibinfo {volume} {125}},\ \bibinfo {eid} {101102}
  (\bibinfo {year} {2020}{\natexlab{b}})},\ \Eprint
  {http://arxiv.org/abs/2009.01075}{arXiv:2009.01075}\BibitemShut {NoStop}%
\bibitem [{\citenamefont {{LIGO Scientific Collaboration}}\ \emph {{\it
  et~al.}}(2015)\citenamefont {{LIGO Scientific Collaboration}}, \citenamefont
  {{Aasi}}, \citenamefont {{Abbott}}, \citenamefont {{Abbott}}, \citenamefont
  {{Abbott}}, \citenamefont {{Abernathy}}, \citenamefont {{Ackley}},
  \citenamefont {{Adams}}, \citenamefont {{Adams}}, \citenamefont {{Addesso}},\
  and\ \citenamefont {et~al.}}]{LIGO-2015CQGra..32g4001L}%
  \BibitemOpen
  \bibfield  {author} {\bibinfo {author} {\bibnamefont {{LIGO Scientific
  Collaboration}}} {\it et~al.},\ }\bibfield  {title} {\enquote {\bibinfo
  {title} {{Advanced LIGO}},}\ }\href {\doibase 10.1088/0264-9381/32/7/074001}
  {\bibfield  {journal} {\bibinfo  {journal} {Classical and Quantum Gravity}\
  }\textbf {\bibinfo {volume} {32}},\ \bibinfo {eid} {074001} (\bibinfo {year}
  {2015})},\ \Eprint
  {http://arxiv.org/abs/1411.4547}{arXiv:1411.4547}\BibitemShut {NoStop}%
\bibitem [{\citenamefont {{Acernese}}\ \emph {{\it et~al.}}(2015)\citenamefont
  {{Acernese}}, \citenamefont {{Agathos}}, \citenamefont {{Agatsuma}},
  \citenamefont {{Aisa}}, \citenamefont {{Allemandou}}, \citenamefont
  {{Allocca}}, \citenamefont {{Amarni}}, \citenamefont {{Astone}},
  \citenamefont {{Balestri}}, \citenamefont {{Ballardin}},\ and\ \citenamefont
  {et~al.}}]{Virgo-2015CQGra..32b4001A}%
  \BibitemOpen
  \bibfield  {author} {\bibinfo {author} {\bibfnamefont {F.}~\bibnamefont
  {{Acernese}}} {\it et~al.},\ }\bibfield  {title} {\enquote {\bibinfo {title}
  {{Advanced Virgo: a second-generation interferometric gravitational wave
  detector}},}\ }\href {\doibase 10.1088/0264-9381/32/2/024001} {\bibfield
  {journal} {\bibinfo  {journal} {Classical and Quantum Gravity}\ }\textbf
  {\bibinfo {volume} {32}},\ \bibinfo {eid} {024001} (\bibinfo {year}
  {2015})},\ \Eprint
  {http://arxiv.org/abs/1408.3978}{arXiv:1408.3978}\BibitemShut {NoStop}%
\bibitem [{\citenamefont {{Finn}}\ and\ \citenamefont
  {{Chernoff}}(1993)}]{Finn-1993-Chernoff-PhRvD..47.2198F}%
  \BibitemOpen
  \bibfield  {author} {\bibinfo {author} {\bibfnamefont {L.~S.}\ \bibnamefont
  {{Finn}}} and\ \bibinfo {author} {\bibfnamefont {D.~F.}\ \bibnamefont
  {{Chernoff}}},\ }\bibfield  {title} {\enquote {\bibinfo {title} {{Observing
  binary inspiral in gravitational radiation: One interferometer}},}\ }\href
  {\doibase 10.1103/PhysRevD.47.2198} {\bibfield  {journal} {\bibinfo
  {journal} {\prd}\ }\textbf {\bibinfo {volume} {47}},\ \bibinfo {pages} {2198}
  (\bibinfo {year} {1993})},\ \Eprint
  {http://arxiv.org/abs/gr-qc/9301003}{arXiv:gr-qc/9301003}\BibitemShut
  {NoStop}%
\bibitem [{\citenamefont {{Ajith}}\ \emph {{\it et~al.}}(2007)\citenamefont
  {{Ajith}}, \citenamefont {{Babak}}, \citenamefont {{Chen}}, \citenamefont
  {{Hewitson}}, \citenamefont {{Krishnan}}, \citenamefont {{Whelan}},
  \citenamefont {{Br{\"u}gmann}}, \citenamefont {{Diener}}, \citenamefont
  {{Gonzalez}}, \citenamefont {{Hannam}}, \citenamefont {{Husa}}, \citenamefont
  {{Koppitz}}, \citenamefont {{Pollney}}, \citenamefont {{Rezzolla}},
  \citenamefont {{Santamar{\'\i}a}}, \citenamefont {{Sintes}}, \citenamefont
  {{Sperhake}},\ and\ \citenamefont
  {{Thornburg}}}]{Ajith-2007-Babak-CQGra..24S.689A}%
  \BibitemOpen
  \bibfield  {author} {\bibinfo {author} {\bibfnamefont {P.}~\bibnamefont
  {{Ajith}}} {\it et~al.},\ }\bibfield  {title} {\enquote {\bibinfo {title} {{A
  phenomenological template family for black-hole coalescence waveforms}},}\
  }\href {\doibase 10.1088/0264-9381/24/19/S31} {\bibfield  {journal} {\bibinfo
   {journal} {Classical and Quantum Gravity}\ }\textbf {\bibinfo {volume}
  {24}},\ \bibinfo {pages} {S689} (\bibinfo {year} {2007})},\ \Eprint
  {http://arxiv.org/abs/0704.3764}{arXiv:0704.3764}\BibitemShut {NoStop}%
\bibitem [{\citenamefont {{Abbott}}\ \emph {{\it
  et~al.}}(2020{\natexlab{c}})\citenamefont {{Abbott}}, \citenamefont
  {{Abbott}}, \citenamefont {{Abbott}}, \citenamefont {{Abraham}},
  \citenamefont {{Acernese}}, \citenamefont {{Ackley}}, \citenamefont
  {{Adams}}, \citenamefont {{Adya}}, \citenamefont {{Affeldt}}, \citenamefont
  {{Agathos}},\ and\ \citenamefont {et~al.}}]{LVK-2020-LRR....23....3A}%
  \BibitemOpen
  \bibfield  {author} {\bibinfo {author} {\bibfnamefont {B.~P.}\ \bibnamefont
  {{Abbott}}} {\it et~al.},\ }\bibfield  {title} {\enquote {\bibinfo {title}
  {{Prospects for observing and localizing gravitational-wave transients with
  Advanced LIGO, Advanced Virgo and KAGRA}},}\ }\href {\doibase
  10.1007/s41114-020-00026-9} {\bibfield  {journal} {\bibinfo  {journal}
  {Living Reviews in Relativity}\ }\textbf {\bibinfo {volume} {23}},\ \bibinfo
  {eid} {3} (\bibinfo {year} {2020}{\natexlab{c}})}\BibitemShut {NoStop}%
\bibitem [{\citenamefont {{Abadie}}\ \emph {{\it et~al.}}(2010)\citenamefont
  {{Abadie}}, \citenamefont {{Abbott}}, \citenamefont {{Abbott}}, \citenamefont
  {{Abernathy}}, \citenamefont {{Accadia}}, \citenamefont {{Acernese}},
  \citenamefont {{Adams}}, \citenamefont {{Adhikari}}, \citenamefont {{Ajith}},
  \citenamefont {{Allen}},\ and\ \citenamefont
  {et~al.}}]{Abadie-2010-Abbott-CQGra..27q3001A}%
  \BibitemOpen
  \bibfield  {author} {\bibinfo {author} {\bibfnamefont {J.}~\bibnamefont
  {{Abadie}}} {\it et~al.},\ }\bibfield  {title} {\enquote {\bibinfo {title}
  {{TOPICAL REVIEW: Predictions for the rates of compact binary coalescences
  observable by ground-based gravitational-wave detectors}},}\ }\href {\doibase
  10.1088/0264-9381/27/17/173001} {\bibfield  {journal} {\bibinfo  {journal}
  {Classical and Quantum Gravity}\ }\textbf {\bibinfo {volume} {27}},\ \bibinfo
  {eid} {173001} (\bibinfo {year} {2010})},\ \Eprint
  {http://arxiv.org/abs/1003.2480}{arXiv:1003.2480}\BibitemShut {NoStop}%
\bibitem [{\citenamefont {{Zackay}}\ \emph {{\it et~al.}}(2021)\citenamefont
  {{Zackay}}, \citenamefont {{Dai}}, \citenamefont {{Venumadhav}},
  \citenamefont {{Roulet}},\ and\ \citenamefont
  {{Zaldarriaga}}}]{GW170817A-2021PhRvD.104f3030Z}%
  \BibitemOpen
  \bibfield  {author} {\bibinfo {author} {\bibfnamefont {B.}~\bibnamefont
  {{Zackay}}}, \bibinfo {author} {\bibfnamefont {L.}~\bibnamefont {{Dai}}},
  \bibinfo {author} {\bibfnamefont {T.}~\bibnamefont {{Venumadhav}}}, \bibinfo
  {author} {\bibfnamefont {J.}~\bibnamefont {{Roulet}}}, and\ \bibinfo {author}
  {\bibfnamefont {M.}~\bibnamefont {{Zaldarriaga}}},\ }\bibfield  {title}
  {\enquote {\bibinfo {title} {{Detecting gravitational waves with disparate
  detector responses: Two new binary black hole mergers}},}\ }\href {\doibase
  10.1103/PhysRevD.104.063030} {\bibfield  {journal} {\bibinfo  {journal}
  {\prd}\ }\textbf {\bibinfo {volume} {104}},\ \bibinfo {eid} {063030}
  (\bibinfo {year} {2021})}\BibitemShut {NoStop}%
\bibitem [{\citenamefont {{Gerosa}}\ \emph {{\it et~al.}}(2020)\citenamefont
  {{Gerosa}}, \citenamefont {{Vitale}},\ and\ \citenamefont
  {{Berti}}}]{Gerosa-2020-Vitale-PhRvL.125j1103G}%
  \BibitemOpen
  \bibfield  {author} {\bibinfo {author} {\bibfnamefont {D.}~\bibnamefont
  {{Gerosa}}}, \bibinfo {author} {\bibfnamefont {S.}~\bibnamefont {{Vitale}}},
  and\ \bibinfo {author} {\bibfnamefont {E.}~\bibnamefont {{Berti}}},\
  }\bibfield  {title} {\enquote {\bibinfo {title} {{Astrophysical Implications
  of GW190412 as a Remnant of a Previous Black-Hole Merger}},}\ }\href
  {\doibase 10.1103/PhysRevLett.125.101103} {\bibfield  {journal} {\bibinfo
  {journal} {\prl}\ }\textbf {\bibinfo {volume} {125}},\ \bibinfo {eid}
  {101103} (\bibinfo {year} {2020})},\ \Eprint
  {http://arxiv.org/abs/2005.04243}{arXiv:2005.04243}\BibitemShut {NoStop}%
\bibitem [{\citenamefont {{Abbott}}\ \emph {{\it
  et~al.}}(2020{\natexlab{d}})\citenamefont {{Abbott}}, \citenamefont
  {{Abbott}}, \citenamefont {{Abraham}}, \citenamefont {{Acernese}},
  \citenamefont {{Ackley}}, \citenamefont {{Adams}}, \citenamefont
  {{Adhikari}}, \citenamefont {{Adya}}, \citenamefont {{Affeldt}},
  \citenamefont {{Agathos}},\ and\ \citenamefont
  {et~al.}}]{Abbott-2020-Abbott-ApJ...900L..13A}%
  \BibitemOpen
  \bibfield  {author} {\bibinfo {author} {\bibfnamefont {R.}~\bibnamefont
  {{Abbott}}} {\it et~al.},\ }\bibfield  {title} {\enquote {\bibinfo {title}
  {{Properties and Astrophysical Implications of the 150
  M$_{{\ensuremath{\odot}}}$ Binary Black Hole Merger GW190521}},}\ }\href
  {\doibase 10.3847/2041-8213/aba493} {\bibfield  {journal} {\bibinfo
  {journal} {\apjl}\ }\textbf {\bibinfo {volume} {900}},\ \bibinfo {eid} {L13}
  (\bibinfo {year} {2020}{\natexlab{d}})},\ \Eprint
  {http://arxiv.org/abs/2009.01190}{arXiv:2009.01190}\BibitemShut {NoStop}%
\bibitem [{\citenamefont {{Saavik Ford}}\ and\ \citenamefont
  {{McKernan}}(2022)}]{Ford-2022-McKernan-MNRAS.tmp.2697S}%
  \BibitemOpen
  \bibfield  {author} {\bibinfo {author} {\bibfnamefont {K.~E.}\ \bibnamefont
  {{Saavik Ford}}} and\ \bibinfo {author} {\bibfnamefont {B.}~\bibnamefont
  {{McKernan}}},\ }\bibfield  {title} {\enquote {\bibinfo {title} {{Binary
  black hole merger rates in AGN disks versus nuclear star clusters: Loud beats
  quiet}},}\ }\href {\doibase 10.1093/mnras/stac2861} {\bibfield  {journal}
  {\bibinfo  {journal} {\mnras}\ } (\bibinfo {year} {2022}),\
  10.1093/mnras/stac2861},\ \Eprint
  {http://arxiv.org/abs/2109.03212}{arXiv:2109.03212}\BibitemShut {NoStop}%
\bibitem [{\citenamefont {{Chatziioannou}}\ \emph {{\it
  et~al.}}(2019)\citenamefont {{Chatziioannou}}, \citenamefont {{Cotesta}},
  \citenamefont {{Ghonge}}, \citenamefont {{Lange}}, \citenamefont {{Ng}},
  \citenamefont {{Calder{\'o}n Bustillo}}, \citenamefont {{Clark}},
  \citenamefont {{Haster}}, \citenamefont {{Khan}}, \citenamefont
  {{P{\"u}rrer}}, \citenamefont {{Raymond}}, \citenamefont {{Vitale}},
  \citenamefont {{Afshari}}, \citenamefont {{Babak}}, \citenamefont
  {{Barkett}}, \citenamefont {{Blackman}}, \citenamefont {{Boh{\'e}}},
  \citenamefont {{Boyle}}, \citenamefont {{Buonanno}}, \citenamefont
  {{Campanelli}}, \citenamefont {{Carullo}}, \citenamefont {{Chu}},
  \citenamefont {{Flynn}}, \citenamefont {{Fong}}, \citenamefont {{Garcia}},
  \citenamefont {{Giesler}}, \citenamefont {{Haney}}, \citenamefont {{Hannam}},
  \citenamefont {{Harry}}, \citenamefont {{Healy}}, \citenamefont
  {{Hemberger}}, \citenamefont {{Hinder}}, \citenamefont {{Jani}},
  \citenamefont {{Khamersa}}, \citenamefont {{Kidder}}, \citenamefont
  {{Kumar}}, \citenamefont {{Laguna}}, \citenamefont {{Lousto}}, \citenamefont
  {{Lovelace}}, \citenamefont {{Littenberg}}, \citenamefont {{London}},
  \citenamefont {{Millhouse}}, \citenamefont {{Nuttall}}, \citenamefont
  {{Ohme}}, \citenamefont {{O'Shaughnessy}}, \citenamefont {{Ossokine}},
  \citenamefont {{Pannarale}}, \citenamefont {{Schmidt}}, \citenamefont
  {{Pfeiffer}}, \citenamefont {{Scheel}}, \citenamefont {{Shao}}, \citenamefont
  {{Shoemaker}}, \citenamefont {{Szilagyi}}, \citenamefont {{Taracchini}},
  \citenamefont {{Teukolsky}},\ and\ \citenamefont
  {{Zlochower}}}]{Chatziioannou-2019-Cotesta-PhRvD.100j4015C}%
  \BibitemOpen
  \bibfield  {author} {\bibinfo {author} {\bibfnamefont {K.}~\bibnamefont
  {{Chatziioannou}}} {\it et~al.},\ }\bibfield  {title} {\enquote {\bibinfo
  {title} {{On the properties of the massive binary black hole merger
  GW170729}},}\ }\href {\doibase 10.1103/PhysRevD.100.104015} {\bibfield
  {journal} {\bibinfo  {journal} {\prd}\ }\textbf {\bibinfo {volume} {100}},\
  \bibinfo {eid} {104015} (\bibinfo {year} {2019})},\ \Eprint
  {http://arxiv.org/abs/1903.06742}{arXiv:1903.06742}\BibitemShut {NoStop}%
\bibitem [{\citenamefont {{Fishbach}}\ \emph {{\it et~al.}}(2020)\citenamefont
  {{Fishbach}}, \citenamefont {{Farr}},\ and\ \citenamefont
  {{Holz}}}]{Fishbach-2020-Farr-ApJ...891L..31F}%
  \BibitemOpen
  \bibfield  {author} {\bibinfo {author} {\bibfnamefont {M.}~\bibnamefont
  {{Fishbach}}}, \bibinfo {author} {\bibfnamefont {W.~M.}\ \bibnamefont
  {{Farr}}}, and\ \bibinfo {author} {\bibfnamefont {D.~E.}\ \bibnamefont
  {{Holz}}},\ }\bibfield  {title} {\enquote {\bibinfo {title} {{The Most
  Massive Binary Black Hole Detections and the Identification of Population
  Outliers}},}\ }\href {\doibase 10.3847/2041-8213/ab77c9} {\bibfield
  {journal} {\bibinfo  {journal} {\apjl}\ }\textbf {\bibinfo {volume} {891}},\
  \bibinfo {eid} {L31} (\bibinfo {year} {2020})},\ \Eprint
  {http://arxiv.org/abs/1911.05882}{arXiv:1911.05882}\BibitemShut {NoStop}%
\bibitem [{\citenamefont {{Arca Sedda}}\ \emph {{\it
  et~al.}}(2020)\citenamefont {{Arca Sedda}}, \citenamefont {{Mapelli}},
  \citenamefont {{Spera}}, \citenamefont {{Benacquista}},\ and\ \citenamefont
  {{Giacobbo}}}]{Sedda-2020-Mapelli-ApJ...894..133A}%
  \BibitemOpen
  \bibfield  {author} {\bibinfo {author} {\bibfnamefont {M.}~\bibnamefont
  {{Arca Sedda}}}, \bibinfo {author} {\bibfnamefont {M.}~\bibnamefont
  {{Mapelli}}}, \bibinfo {author} {\bibfnamefont {M.}~\bibnamefont {{Spera}}},
  \bibinfo {author} {\bibfnamefont {M.}~\bibnamefont {{Benacquista}}}, and\
  \bibinfo {author} {\bibfnamefont {N.}~\bibnamefont {{Giacobbo}}},\ }\bibfield
   {title} {\enquote {\bibinfo {title} {{Fingerprints of Binary Black Hole
  Formation Channels Encoded in the Mass and Spin of Merger Remnants}},}\
  }\href {\doibase 10.3847/1538-4357/ab88b2} {\bibfield  {journal} {\bibinfo
  {journal} {Astrophys. J.}\ }\textbf {\bibinfo {volume} {894}},\ \bibinfo
  {eid} {133} (\bibinfo {year} {2020})},\ \Eprint
  {http://arxiv.org/abs/2003.07409}{arXiv:2003.07409}\BibitemShut {NoStop}%
\bibitem [{\citenamefont {{Hamers}}\ and\ \citenamefont
  {{Safarzadeh}}(2020)}]{Hamers-2020-Safarzadeh-ApJ...898...99H}%
  \BibitemOpen
  \bibfield  {author} {\bibinfo {author} {\bibfnamefont {A.~S.}\ \bibnamefont
  {{Hamers}}} and\ \bibinfo {author} {\bibfnamefont {M.}~\bibnamefont
  {{Safarzadeh}}},\ }\bibfield  {title} {\enquote {\bibinfo {title} {{Was
  GW190412 Born from a Hierarchical 3 + 1 Quadruple Configuration?}}}\ }\href
  {\doibase 10.3847/1538-4357/ab9b27} {\bibfield  {journal} {\bibinfo
  {journal} {Astrophys. J.}\ }\textbf {\bibinfo {volume} {898}},\ \bibinfo
  {eid} {99} (\bibinfo {year} {2020})},\ \Eprint
  {http://arxiv.org/abs/2005.03045}{arXiv:2005.03045}\BibitemShut {NoStop}%
\bibitem [{\citenamefont {{Rodriguez}}\ \emph {{\it et~al.}}(2020)\citenamefont
  {{Rodriguez}}, \citenamefont {{Kremer}}, \citenamefont {{Grudi{\'c}}},
  \citenamefont {{Hafen}}, \citenamefont {{Chatterjee}}, \citenamefont
  {{Fragione}}, \citenamefont {{Lamberts}}, \citenamefont {{Martinez}},
  \citenamefont {{Rasio}}, \citenamefont {{Weatherford}},\ and\ \citenamefont
  {{Ye}}}]{Rodriguez-2020-Kremer-ApJ...896L..10R}%
  \BibitemOpen
  \bibfield  {author} {\bibinfo {author} {\bibfnamefont {C.~L.}\ \bibnamefont
  {{Rodriguez}}} {\it et~al.},\ }\bibfield  {title} {\enquote {\bibinfo {title}
  {{GW190412 as a Third-generation Black Hole Merger from a Super Star
  Cluster}},}\ }\href {\doibase 10.3847/2041-8213/ab961d} {\bibfield  {journal}
  {\bibinfo  {journal} {\apjl}\ }\textbf {\bibinfo {volume} {896}},\ \bibinfo
  {eid} {L10} (\bibinfo {year} {2020})},\ \Eprint
  {http://arxiv.org/abs/2005.04239}{arXiv:2005.04239}\BibitemShut {NoStop}%
\bibitem [{\citenamefont {{Zevin}}\ \emph {{\it et~al.}}(2020)\citenamefont
  {{Zevin}}, \citenamefont {{Berry}}, \citenamefont {{Coughlin}}, \citenamefont
  {{Chatziioannou}},\ and\ \citenamefont
  {{Vitale}}}]{Zevin-2020-Berry-ApJ...899L..17Z}%
  \BibitemOpen
  \bibfield  {author} {\bibinfo {author} {\bibfnamefont {M.}~\bibnamefont
  {{Zevin}}}, \bibinfo {author} {\bibfnamefont {C.~P.~L.}\ \bibnamefont
  {{Berry}}}, \bibinfo {author} {\bibfnamefont {S.}~\bibnamefont {{Coughlin}}},
  \bibinfo {author} {\bibfnamefont {K.}~\bibnamefont {{Chatziioannou}}}, and\
  \bibinfo {author} {\bibfnamefont {S.}~\bibnamefont {{Vitale}}},\ }\bibfield
  {title} {\enquote {\bibinfo {title} {{You Can't Always Get What You Want: The
  Impact of Prior Assumptions on Interpreting GW190412}},}\ }\href {\doibase
  10.3847/2041-8213/aba8ef} {\bibfield  {journal} {\bibinfo  {journal} {\apjl}\
  }\textbf {\bibinfo {volume} {899}},\ \bibinfo {eid} {L17} (\bibinfo {year}
  {2020})},\ \Eprint
  {http://arxiv.org/abs/2006.11293}{arXiv:2006.11293}\BibitemShut {NoStop}%
\bibitem [{\citenamefont {{Anagnostou}}\ \emph {{\it
  et~al.}}(2022)\citenamefont {{Anagnostou}}, \citenamefont {{Trenti}},\ and\
  \citenamefont {{Melatos}}}]{Anagnostou-2022-Trenti-ApJ...941....4A}%
  \BibitemOpen
  \bibfield  {author} {\bibinfo {author} {\bibfnamefont {O.}~\bibnamefont
  {{Anagnostou}}}, \bibinfo {author} {\bibfnamefont {M.}~\bibnamefont
  {{Trenti}}}, and\ \bibinfo {author} {\bibfnamefont {A.}~\bibnamefont
  {{Melatos}}},\ }\bibfield  {title} {\enquote {\bibinfo {title} {{Repeated
  Mergers of Black Hole Binaries: Implications for GW190521}},}\ }\href
  {\doibase 10.3847/1538-4357/ac9d95} {\bibfield  {journal} {\bibinfo
  {journal} {Astrophys. J.}\ }\textbf {\bibinfo {volume} {941}},\ \bibinfo
  {eid} {4} (\bibinfo {year} {2022})}\BibitemShut {NoStop}%
\bibitem [{\citenamefont {{Harris}}(1996)}]{Harris-1996AJ....112.1487H}%
  \BibitemOpen
  \bibfield  {author} {\bibinfo {author} {\bibfnamefont {W.~E.}\ \bibnamefont
  {{Harris}}},\ }\bibfield  {title} {\enquote {\bibinfo {title} {{A Catalog of
  Parameters for Globular Clusters in the Milky Way}},}\ }\href {\doibase
  10.1086/118116} {\bibfield  {journal} {\bibinfo  {journal} {\aj}\ }\textbf
  {\bibinfo {volume} {112}},\ \bibinfo {pages} {1487} (\bibinfo {year}
  {1996})}\BibitemShut {NoStop}%
\bibitem [{\citenamefont {{Vink}}\ \emph {{\it et~al.}}(2001)\citenamefont
  {{Vink}}, \citenamefont {{de Koter}},\ and\ \citenamefont
  {{Lamers}}}]{Vink-2001-Koter-A&A...369..574V}%
  \BibitemOpen
  \bibfield  {author} {\bibinfo {author} {\bibfnamefont {J.~S.}\ \bibnamefont
  {{Vink}}}, \bibinfo {author} {\bibfnamefont {A.}~\bibnamefont {{de Koter}}},
  and\ \bibinfo {author} {\bibfnamefont {H.~J.~G.~L.~M.}\ \bibnamefont
  {{Lamers}}},\ }\bibfield  {title} {\enquote {\bibinfo {title} {{Mass-loss
  predictions for O and B stars as a function of metallicity}},}\ }\href
  {\doibase 10.1051/0004-6361:20010127} {\bibfield  {journal} {\bibinfo
  {journal} {\aap}\ }\textbf {\bibinfo {volume} {369}},\ \bibinfo {pages} {574}
  (\bibinfo {year} {2001})},\ \Eprint
  {http://arxiv.org/abs/astro-ph/0101509}{arXiv:astro-ph/0101509}\BibitemShut
  {NoStop}%
\bibitem [{\citenamefont {{Spera}}\ and\ \citenamefont
  {{Mapelli}}(2017)}]{Spera-2017-Mapelli-MNRAS.470.4739S}%
  \BibitemOpen
  \bibfield  {author} {\bibinfo {author} {\bibfnamefont {M.}~\bibnamefont
  {{Spera}}} and\ \bibinfo {author} {\bibfnamefont {M.}~\bibnamefont
  {{Mapelli}}},\ }\bibfield  {title} {\enquote {\bibinfo {title} {{Very massive
  stars, pair-instability supernovae and intermediate-mass black holes with the
  sevn code}},}\ }\href {\doibase 10.1093/mnras/stx1576} {\bibfield  {journal}
  {\bibinfo  {journal} {\mnras}\ }\textbf {\bibinfo {volume} {470}},\ \bibinfo
  {pages} {4739} (\bibinfo {year} {2017})},\ \Eprint
  {http://arxiv.org/abs/1706.06109}{arXiv:1706.06109}\BibitemShut {NoStop}%
\bibitem [{\citenamefont {{Antonini}}(2013)}]{Antonini-2013ApJ...763...62A}%
  \BibitemOpen
  \bibfield  {author} {\bibinfo {author} {\bibfnamefont {F.}~\bibnamefont
  {{Antonini}}},\ }\bibfield  {title} {\enquote {\bibinfo {title} {{Origin and
  Growth of Nuclear Star Clusters around Massive Black Holes}},}\ }\href
  {\doibase 10.1088/0004-637X/763/1/62} {\bibfield  {journal} {\bibinfo
  {journal} {Astrophys. J.}\ }\textbf {\bibinfo {volume} {763}},\ \bibinfo
  {eid} {62} (\bibinfo {year} {2013})},\ \Eprint
  {http://arxiv.org/abs/1207.6589}{arXiv:1207.6589}\BibitemShut {NoStop}%
\bibitem [{\citenamefont {{Yi}}\ \emph {{\it et~al.}}(2018)\citenamefont
  {{Yi}}, \citenamefont {{Cheng}},\ and\ \citenamefont
  {{Taam}}}]{Yi-2018-Cheng-ApJ...859L..25Y}%
  \BibitemOpen
  \bibfield  {author} {\bibinfo {author} {\bibfnamefont {S.-X.}\ \bibnamefont
  {{Yi}}}, \bibinfo {author} {\bibfnamefont {K.~S.}\ \bibnamefont {{Cheng}}},
  and\ \bibinfo {author} {\bibfnamefont {R.~E.}\ \bibnamefont {{Taam}}},\
  }\bibfield  {title} {\enquote {\bibinfo {title} {{The Growth of Stellar Mass
  Black Hole Binaries Trapped in the Accretion Disks of Active Galactic
  Nuclei}},}\ }\href {\doibase 10.3847/2041-8213/aac649} {\bibfield  {journal}
  {\bibinfo  {journal} {\apjl}\ }\textbf {\bibinfo {volume} {859}},\ \bibinfo
  {eid} {L25} (\bibinfo {year} {2018})},\ \Eprint
  {http://arxiv.org/abs/1805.07026}{arXiv:1805.07026}\BibitemShut {NoStop}%
\bibitem [{\citenamefont {{Fragione}}\ and\ \citenamefont
  {{Loeb}}(2021)}]{Fragione-2021-Loeb-MNRAS.502.3879F}%
  \BibitemOpen
  \bibfield  {author} {\bibinfo {author} {\bibfnamefont {G.}~\bibnamefont
  {{Fragione}}} and\ \bibinfo {author} {\bibfnamefont {A.}~\bibnamefont
  {{Loeb}}},\ }\bibfield  {title} {\enquote {\bibinfo {title} {{Implications of
  recoil kicks for black hole mergers from LIGO/Virgo catalogs}},}\ }\href
  {\doibase 10.1093/mnras/stab247} {\bibfield  {journal} {\bibinfo  {journal}
  {\mnras}\ }\textbf {\bibinfo {volume} {502}},\ \bibinfo {pages} {3879}
  (\bibinfo {year} {2021})},\ \Eprint
  {http://arxiv.org/abs/2011.08935}{arXiv:2011.08935}\BibitemShut {NoStop}%
\bibitem [{\citenamefont {{Secunda}}\ \emph {{\it et~al.}}(2021)\citenamefont
  {{Secunda}}, \citenamefont {{Hernandez}}, \citenamefont {{Goodman}},
  \citenamefont {{Leigh}}, \citenamefont {{McKernan}}, \citenamefont {{Ford}},\
  and\ \citenamefont {{Adorno}}}]{Secunda-2021-Hernandez-ApJ...908L..27S}%
  \BibitemOpen
  \bibfield  {author} {\bibinfo {author} {\bibfnamefont {A.}~\bibnamefont
  {{Secunda}}}, \bibinfo {author} {\bibfnamefont {B.}~\bibnamefont
  {{Hernandez}}}, \bibinfo {author} {\bibfnamefont {J.}~\bibnamefont
  {{Goodman}}}, \bibinfo {author} {\bibfnamefont {N.~W.~C.}\ \bibnamefont
  {{Leigh}}}, \bibinfo {author} {\bibfnamefont {B.}~\bibnamefont {{McKernan}}},
  \bibinfo {author} {\bibfnamefont {K.~E.~S.}\ \bibnamefont {{Ford}}}, and\
  \bibinfo {author} {\bibfnamefont {J.~I.}\ \bibnamefont {{Adorno}}},\
  }\bibfield  {title} {\enquote {\bibinfo {title} {{Evolution of Retrograde
  Orbiters in an Active Galactic Nucleus Disk}},}\ }\href {\doibase
  10.3847/2041-8213/abe11d} {\bibfield  {journal} {\bibinfo  {journal} {\apjl}\
  }\textbf {\bibinfo {volume} {908}},\ \bibinfo {eid} {L27} (\bibinfo {year}
  {2021})},\ \Eprint
  {http://arxiv.org/abs/2009.03910}{arXiv:2009.03910}\BibitemShut {NoStop}%
\bibitem [{\citenamefont {{Wang}}\ \emph {{\it et~al.}}(2021)\citenamefont
  {{Wang}}, \citenamefont {{McKernan}}, \citenamefont {{Ford}}, \citenamefont
  {{Perna}}, \citenamefont {{Leigh}},\ and\ \citenamefont {{Mac
  Low}}}]{Wang-2021-McKernan-ApJ...923L..23W}%
  \BibitemOpen
  \bibfield  {author} {\bibinfo {author} {\bibfnamefont {Y.-H.}\ \bibnamefont
  {{Wang}}}, \bibinfo {author} {\bibfnamefont {B.}~\bibnamefont {{McKernan}}},
  \bibinfo {author} {\bibfnamefont {S.}~\bibnamefont {{Ford}}}, \bibinfo
  {author} {\bibfnamefont {R.}~\bibnamefont {{Perna}}}, \bibinfo {author}
  {\bibfnamefont {N.~W.~C.}\ \bibnamefont {{Leigh}}}, and\ \bibinfo {author}
  {\bibfnamefont {M.-M.}\ \bibnamefont {{Mac Low}}},\ }\bibfield  {title}
  {\enquote {\bibinfo {title} {{Symmetry Breaking in Dynamical Encounters in
  the Disks of Active Galactic Nuclei}},}\ }\href {\doibase
  10.3847/2041-8213/ac400a} {\bibfield  {journal} {\bibinfo  {journal} {\apjl}\
  }\textbf {\bibinfo {volume} {923}},\ \bibinfo {eid} {L23} (\bibinfo {year}
  {2021})},\ \Eprint
  {http://arxiv.org/abs/2110.03698}{arXiv:2110.03698}\BibitemShut {NoStop}%
\bibitem [{\citenamefont {{Samsing}}\ \emph {{\it et~al.}}(2022)\citenamefont
  {{Samsing}}, \citenamefont {{Bartos}}, \citenamefont {{D'Orazio}},
  \citenamefont {{Haiman}}, \citenamefont {{Kocsis}}, \citenamefont {{Leigh}},
  \citenamefont {{Liu}}, \citenamefont {{Pessah}},\ and\ \citenamefont
  {{Tagawa}}}]{Samsing-2022-Bartos-Natur.603..237S}%
  \BibitemOpen
  \bibfield  {author} {\bibinfo {author} {\bibfnamefont {J.}~\bibnamefont
  {{Samsing}}}, \bibinfo {author} {\bibfnamefont {I.}~\bibnamefont {{Bartos}}},
  \bibinfo {author} {\bibfnamefont {D.~J.}\ \bibnamefont {{D'Orazio}}},
  \bibinfo {author} {\bibfnamefont {Z.}~\bibnamefont {{Haiman}}}, \bibinfo
  {author} {\bibfnamefont {B.}~\bibnamefont {{Kocsis}}}, \bibinfo {author}
  {\bibfnamefont {N.~W.~C.}\ \bibnamefont {{Leigh}}}, \bibinfo {author}
  {\bibfnamefont {B.}~\bibnamefont {{Liu}}}, \bibinfo {author} {\bibfnamefont
  {M.~E.}\ \bibnamefont {{Pessah}}}, and\ \bibinfo {author} {\bibfnamefont
  {H.}~\bibnamefont {{Tagawa}}},\ }\bibfield  {title} {\enquote {\bibinfo
  {title} {{AGN as potential factories for eccentric black hole mergers}},}\
  }\href {\doibase 10.1038/s41586-021-04333-1} {\bibfield  {journal} {\bibinfo
  {journal} {Nature}\ }\textbf {\bibinfo {volume} {603}},\ \bibinfo {pages}
  {237} (\bibinfo {year} {2022})}\BibitemShut {NoStop}%
\bibitem [{\citenamefont {{Li}}\ \emph {{\it et~al.}}(2022)\citenamefont
  {{Li}}, \citenamefont {{Dempsey}}, \citenamefont {{Li}}, \citenamefont
  {{Lai}},\ and\ \citenamefont {{Li}}}]{Li-2022-Dempsey-arXiv221110357L}%
  \BibitemOpen
  \bibfield  {author} {\bibinfo {author} {\bibfnamefont {J.}~\bibnamefont
  {{Li}}}, \bibinfo {author} {\bibfnamefont {A.~M.}\ \bibnamefont {{Dempsey}}},
  \bibinfo {author} {\bibfnamefont {H.}~\bibnamefont {{Li}}}, \bibinfo {author}
  {\bibfnamefont {D.}~\bibnamefont {{Lai}}}, and\ \bibinfo {author}
  {\bibfnamefont {S.}~\bibnamefont {{Li}}},\ }\bibfield  {title} {\enquote
  {\bibinfo {title} {{Hydrodynamical Simulations of Black-Hole Binary Formation
  in AGN Disks}},}\ }\href@noop {} {\bibfield  {journal} {\bibinfo  {journal}
  {arXiv e-prints}\ ,\ \bibinfo {eid} {arXiv:2211.10357}} (\bibinfo {year}
  {2022})},\ \Eprint
  {http://arxiv.org/abs/2211.10357}{arXiv:2211.10357}\BibitemShut {NoStop}%
\bibitem [{\citenamefont {{Pan}}\ and\ \citenamefont
  {{Yang}}(2021)}]{Pan-2021-Yang-PhRvD.103j3018P}%
  \BibitemOpen
  \bibfield  {author} {\bibinfo {author} {\bibfnamefont {Z.}~\bibnamefont
  {{Pan}}} and\ \bibinfo {author} {\bibfnamefont {H.}~\bibnamefont {{Yang}}},\
  }\bibfield  {title} {\enquote {\bibinfo {title} {{Formation rate of extreme
  mass ratio inspirals in active galactic nuclei}},}\ }\href {\doibase
  10.1103/PhysRevD.103.103018} {\bibfield  {journal} {\bibinfo  {journal}
  {\prd}\ }\textbf {\bibinfo {volume} {103}},\ \bibinfo {eid} {103018}
  (\bibinfo {year} {2021})},\ \Eprint
  {http://arxiv.org/abs/2101.09146}{arXiv:2101.09146}\BibitemShut {NoStop}%
\bibitem [{\citenamefont {{Peng}}\ and\ \citenamefont
  {{Chen}}(2021)}]{Peng-2021-Chen-MNRAS.505.1324P}%
  \BibitemOpen
  \bibfield  {author} {\bibinfo {author} {\bibfnamefont {P.}~\bibnamefont
  {{Peng}}} and\ \bibinfo {author} {\bibfnamefont {X.}~\bibnamefont {{Chen}}},\
  }\bibfield  {title} {\enquote {\bibinfo {title} {{The last migration trap of
  compact objects in AGN accretion disc}},}\ }\href {\doibase
  10.1093/mnras/stab1419} {\bibfield  {journal} {\bibinfo  {journal} {\mnras}\
  }\textbf {\bibinfo {volume} {505}},\ \bibinfo {pages} {1324} (\bibinfo {year}
  {2021})},\ \Eprint
  {http://arxiv.org/abs/2104.07685}{arXiv:2104.07685}\BibitemShut {NoStop}%
\bibitem [{\citenamefont {{Secunda}}\ \emph {{\it et~al.}}(2020)\citenamefont
  {{Secunda}}, \citenamefont {{Bellovary}}, \citenamefont {{Mac Low}},
  \citenamefont {{Ford}}, \citenamefont {{McKernan}}, \citenamefont {{Leigh}},
  \citenamefont {{Lyra}}, \citenamefont {{S{\'a}ndor}},\ and\ \citenamefont
  {{Adorno}}}]{Secunda-2020-Bellovary-ApJ...903..133S}%
  \BibitemOpen
  \bibfield  {author} {\bibinfo {author} {\bibfnamefont {A.}~\bibnamefont
  {{Secunda}}}, \bibinfo {author} {\bibfnamefont {J.}~\bibnamefont
  {{Bellovary}}}, \bibinfo {author} {\bibfnamefont {M.-M.}\ \bibnamefont {{Mac
  Low}}}, \bibinfo {author} {\bibfnamefont {K.~E.~S.}\ \bibnamefont {{Ford}}},
  \bibinfo {author} {\bibfnamefont {B.}~\bibnamefont {{McKernan}}}, \bibinfo
  {author} {\bibfnamefont {N.~W.~C.}\ \bibnamefont {{Leigh}}}, \bibinfo
  {author} {\bibfnamefont {W.}~\bibnamefont {{Lyra}}}, \bibinfo {author}
  {\bibfnamefont {Z.}~\bibnamefont {{S{\'a}ndor}}}, and\ \bibinfo {author}
  {\bibfnamefont {J.~I.}\ \bibnamefont {{Adorno}}},\ }\bibfield  {title}
  {\enquote {\bibinfo {title} {{Orbital Migration of Interacting Stellar Mass
  Black Holes in Disks around Supermassive Black Holes. II. Spins and Incoming
  Objects}},}\ }\href {\doibase 10.3847/1538-4357/abbc1d} {\bibfield  {journal}
  {\bibinfo  {journal} {Astrophys. J.}\ }\textbf {\bibinfo {volume} {903}},\
  \bibinfo {eid} {133} (\bibinfo {year} {2020})},\ \Eprint
  {http://arxiv.org/abs/2004.11936}{arXiv:2004.11936}\BibitemShut {NoStop}%
\bibitem [{\citenamefont {{Mapelli}}\ \emph {{\it et~al.}}(2022)\citenamefont
  {{Mapelli}}, \citenamefont {{Bouffanais}}, \citenamefont {{Santoliquido}},
  \citenamefont {{Arca Sedda}},\ and\ \citenamefont
  {{Artale}}}]{Mapelli-2022-Bouffanais-MNRAS.511.5797M}%
  \BibitemOpen
  \bibfield  {author} {\bibinfo {author} {\bibfnamefont {M.}~\bibnamefont
  {{Mapelli}}}, \bibinfo {author} {\bibfnamefont {Y.}~\bibnamefont
  {{Bouffanais}}}, \bibinfo {author} {\bibfnamefont {F.}~\bibnamefont
  {{Santoliquido}}}, \bibinfo {author} {\bibfnamefont {M.}~\bibnamefont {{Arca
  Sedda}}}, and\ \bibinfo {author} {\bibfnamefont {M.~C.}\ \bibnamefont
  {{Artale}}},\ }\bibfield  {title} {\enquote {\bibinfo {title} {{The cosmic
  evolution of binary black holes in young, globular, and nuclear star
  clusters: rates, masses, spins, and mixing fractions}},}\ }\href {\doibase
  10.1093/mnras/stac422} {\bibfield  {journal} {\bibinfo  {journal} {\mnras}\
  }\textbf {\bibinfo {volume} {511}},\ \bibinfo {pages} {5797} (\bibinfo {year}
  {2022})},\ \Eprint
  {http://arxiv.org/abs/2109.06222}{arXiv:2109.06222}\BibitemShut {NoStop}%
\bibitem [{\citenamefont {{Mandel}}\ and\ \citenamefont
  {{Broekgaarden}}(2022)}]{Mandel-2022-Broekgaarden-LRR....25....1M}%
  \BibitemOpen
  \bibfield  {author} {\bibinfo {author} {\bibfnamefont {I.}~\bibnamefont
  {{Mandel}}} and\ \bibinfo {author} {\bibfnamefont {F.~S.}\ \bibnamefont
  {{Broekgaarden}}},\ }\bibfield  {title} {\enquote {\bibinfo {title} {{Rates
  of compact object coalescences}},}\ }\href {\doibase
  10.1007/s41114-021-00034-3} {\bibfield  {journal} {\bibinfo  {journal}
  {Living Reviews in Relativity}\ }\textbf {\bibinfo {volume} {25}},\ \bibinfo
  {eid} {1} (\bibinfo {year} {2022})},\ \Eprint
  {http://arxiv.org/abs/2107.14239}{arXiv:2107.14239}\BibitemShut {NoStop}%
\bibitem [{\citenamefont {{Punturo}}\ \emph {{\it
  et~al.}}(2010{\natexlab{a}})\citenamefont {{Punturo}}, \citenamefont
  {{Abernathy}}, \citenamefont {{Acernese}}, \citenamefont {{Allen}},
  \citenamefont {{Andersson}}, \citenamefont {{Arun}}, \citenamefont
  {{Barone}}, \citenamefont {{Barr}}, \citenamefont {{Barsuglia}},
  \citenamefont {{Beker}}, \citenamefont {{Beveridge}}, \citenamefont
  {{Birindelli}}, \citenamefont {{Bose}}, \citenamefont {{Bosi}}, \citenamefont
  {{Braccini}}, \citenamefont {{Bradaschia}}, \citenamefont {{Bulik}},
  \citenamefont {{Calloni}}, \citenamefont {{Cella}}, \citenamefont {{Chassande
  Mottin}}, \citenamefont {{Chelkowski}}, \citenamefont {{Chincarini}},
  \citenamefont {{Clark}}, \citenamefont {{Coccia}}, \citenamefont
  {{Colacino}}, \citenamefont {{Colas}}, \citenamefont {{Cumming}},
  \citenamefont {{Cunningham}}, \citenamefont {{Cuoco}}, \citenamefont
  {{Danilishin}}, \citenamefont {{Danzmann}}, \citenamefont {{De Luca}},
  \citenamefont {{De Salvo}}, \citenamefont {{Dent}}, \citenamefont {{De
  Rosa}}, \citenamefont {{Di Fiore}}, \citenamefont {{Di Virgilio}},
  \citenamefont {{Doets}}, \citenamefont {{Fafone}}, \citenamefont {{Falferi}},
  \citenamefont {{Flaminio}}, \citenamefont {{Franc}}, \citenamefont
  {{Frasconi}}, \citenamefont {{Freise}}, \citenamefont {{Fulda}},
  \citenamefont {{Gair}}, \citenamefont {{Gemme}}, \citenamefont {{Gennai}},
  \citenamefont {{Giazotto}}, \citenamefont {{Glampedakis}}, \citenamefont
  {{Granata}}, \citenamefont {{Grote}}, \citenamefont {{Guidi}}, \citenamefont
  {{Hammond}}, \citenamefont {{Hannam}}, \citenamefont {{Harms}}, \citenamefont
  {{Heinert}}, \citenamefont {{Hendry}}, \citenamefont {{Heng}}, \citenamefont
  {{Hennes}}, \citenamefont {{Hild}}, \citenamefont {{Hough}}, \citenamefont
  {{Husa}}, \citenamefont {{Huttner}}, \citenamefont {{Jones}}, \citenamefont
  {{Khalili}}, \citenamefont {{Kokeyama}}, \citenamefont {{Kokkotas}},
  \citenamefont {{Krishnan}}, \citenamefont {{Lorenzini}}, \citenamefont
  {{L{\"u}ck}}, \citenamefont {{Majorana}}, \citenamefont {{Mandel}},
  \citenamefont {{Mandic}}, \citenamefont {{Martin}}, \citenamefont {{Michel}},
  \citenamefont {{Minenkov}}, \citenamefont {{Morgado}}, \citenamefont
  {{Mosca}}, \citenamefont {{Mours}}, \citenamefont
  {{M{\"u}ller{\textendash}Ebhardt}}, \citenamefont {{Murray}}, \citenamefont
  {{Nawrodt}}, \citenamefont {{Nelson}}, \citenamefont {{Oshaughnessy}},
  \citenamefont {{Ott}}, \citenamefont {{Palomba}}, \citenamefont {{Paoli}},
  \citenamefont {{Parguez}}, \citenamefont {{Pasqualetti}}, \citenamefont
  {{Passaquieti}}, \citenamefont {{Passuello}}, \citenamefont {{Pinard}},
  \citenamefont {{Poggiani}}, \citenamefont {{Popolizio}}, \citenamefont
  {{Prato}}, \citenamefont {{Puppo}}, \citenamefont {{Rabeling}}, \citenamefont
  {{Rapagnani}}, \citenamefont {{Read}}, \citenamefont {{Regimbau}},
  \citenamefont {{Rehbein}}, \citenamefont {{Reid}}, \citenamefont
  {{Rezzolla}}, \citenamefont {{Ricci}}, \citenamefont {{Richard}},
  \citenamefont {{Rocchi}}, \citenamefont {{Rowan}}, \citenamefont
  {{R{\"u}diger}}, \citenamefont {{Sassolas}}, \citenamefont {{Sathyaprakash}},
  \citenamefont {{Schnabel}}, \citenamefont {{Schwarz}}, \citenamefont
  {{Seidel}}, \citenamefont {{Sintes}}, \citenamefont {{Somiya}}, \citenamefont
  {{Speirits}}, \citenamefont {{Strain}}, \citenamefont {{Strigin}},
  \citenamefont {{Sutton}}, \citenamefont {{Tarabrin}}, \citenamefont
  {{Th{\"u}ring}}, \citenamefont {{van den Brand}}, \citenamefont {{van
  Leewen}}, \citenamefont {{van Veggel}}, \citenamefont {{van den Broeck}},
  \citenamefont {{Vecchio}}, \citenamefont {{Veitch}}, \citenamefont
  {{Vetrano}}, \citenamefont {{Vicere}}, \citenamefont {{Vyatchanin}},
  \citenamefont {{Willke}}, \citenamefont {{Woan}}, \citenamefont
  {{Wolfango}},\ and\ \citenamefont
  {{Yamamoto}}}]{Punturo-2010-Abernathy-CQGra..27s4002P}%
  \BibitemOpen
  \bibfield  {author} {\bibinfo {author} {\bibfnamefont {M.}~\bibnamefont
  {{Punturo}}} {\it et~al.},\ }\bibfield  {title} {\enquote {\bibinfo {title}
  {{The Einstein Telescope: a third-generation gravitational wave
  observatory}},}\ }\href {\doibase 10.1088/0264-9381/27/19/194002} {\bibfield
  {journal} {\bibinfo  {journal} {Classical and Quantum Gravity}\ }\textbf
  {\bibinfo {volume} {27}},\ \bibinfo {eid} {194002} (\bibinfo {year}
  {2010}{\natexlab{a}})}\BibitemShut {NoStop}%
\bibitem [{\citenamefont {{Punturo}}\ \emph {{\it
  et~al.}}(2010{\natexlab{b}})\citenamefont {{Punturo}}, \citenamefont
  {{Abernathy}}, \citenamefont {{Acernese}}, \citenamefont {{Allen}},
  \citenamefont {{Andersson}}, \citenamefont {{Arun}}, \citenamefont
  {{Barone}}, \citenamefont {{Barr}}, \citenamefont {{Barsuglia}},
  \citenamefont {{Beker}}, \citenamefont {{Beveridge}}, \citenamefont
  {{Birindelli}}, \citenamefont {{Bose}}, \citenamefont {{Bosi}}, \citenamefont
  {{Braccini}}, \citenamefont {{Bradaschia}}, \citenamefont {{Bulik}},
  \citenamefont {{Calloni}}, \citenamefont {{Cella}}, \citenamefont {{Chassande
  Mottin}}, \citenamefont {{Chelkowski}}, \citenamefont {{Chincarini}},
  \citenamefont {{Clark}}, \citenamefont {{Coccia}}, \citenamefont
  {{Colacino}}, \citenamefont {{Colas}}, \citenamefont {{Cumming}},
  \citenamefont {{Cunningham}}, \citenamefont {{Cuoco}}, \citenamefont
  {{Danilishin}}, \citenamefont {{Danzmann}}, \citenamefont {{De Luca}},
  \citenamefont {{De Salvo}}, \citenamefont {{Dent}}, \citenamefont {{Derosa}},
  \citenamefont {{Di Fiore}}, \citenamefont {{Di Virgilio}}, \citenamefont
  {{Doets}}, \citenamefont {{Fafone}}, \citenamefont {{Falferi}}, \citenamefont
  {{Flaminio}}, \citenamefont {{Franc}}, \citenamefont {{Frasconi}},
  \citenamefont {{Freise}}, \citenamefont {{Fulda}}, \citenamefont {{Gair}},
  \citenamefont {{Gemme}}, \citenamefont {{Gennai}}, \citenamefont
  {{Giazotto}}, \citenamefont {{Glampedakis}}, \citenamefont {{Granata}},
  \citenamefont {{Grote}}, \citenamefont {{Guidi}}, \citenamefont {{Hammond}},
  \citenamefont {{Hannam}}, \citenamefont {{Harms}}, \citenamefont {{Heinert}},
  \citenamefont {{Hendry}}, \citenamefont {{Heng}}, \citenamefont {{Hennes}},
  \citenamefont {{Hild}}, \citenamefont {{Hough}}, \citenamefont {{Husa}},
  \citenamefont {{Huttner}}, \citenamefont {{Jones}}, \citenamefont
  {{Khalili}}, \citenamefont {{Kokeyama}}, \citenamefont {{Kokkotas}},
  \citenamefont {{Krishnan}}, \citenamefont {{Lorenzini}}, \citenamefont
  {{L{\"u}ck}}, \citenamefont {{Majorana}}, \citenamefont {{Mandel}},
  \citenamefont {{Mandic}}, \citenamefont {{Martin}}, \citenamefont {{Michel}},
  \citenamefont {{Minenkov}}, \citenamefont {{Morgado}}, \citenamefont
  {{Mosca}}, \citenamefont {{Mours}}, \citenamefont {{M{\"u}ller-Ebhardt}},
  \citenamefont {{Murray}}, \citenamefont {{Nawrodt}}, \citenamefont
  {{Nelson}}, \citenamefont {{Oshaughnessy}}, \citenamefont {{Ott}},
  \citenamefont {{Palomba}}, \citenamefont {{Paoli}}, \citenamefont
  {{Parguez}}, \citenamefont {{Pasqualetti}}, \citenamefont {{Passaquieti}},
  \citenamefont {{Passuello}}, \citenamefont {{Pinard}}, \citenamefont
  {{Poggiani}}, \citenamefont {{Popolizio}}, \citenamefont {{Prato}},
  \citenamefont {{Puppo}}, \citenamefont {{Rabeling}}, \citenamefont
  {{Rapagnani}}, \citenamefont {{Read}}, \citenamefont {{Regimbau}},
  \citenamefont {{Rehbein}}, \citenamefont {{Reid}}, \citenamefont
  {{Rezzolla}}, \citenamefont {{Ricci}}, \citenamefont {{Richard}},
  \citenamefont {{Rocchi}}, \citenamefont {{Rowan}}, \citenamefont
  {{R{\"u}diger}}, \citenamefont {{Sassolas}}, \citenamefont {{Sathyaprakash}},
  \citenamefont {{Schnabel}}, \citenamefont {{Schwarz}}, \citenamefont
  {{Seidel}}, \citenamefont {{Sintes}}, \citenamefont {{Somiya}}, \citenamefont
  {{Speirits}}, \citenamefont {{Strain}}, \citenamefont {{Strigin}},
  \citenamefont {{Sutton}}, \citenamefont {{Tarabrin}}, \citenamefont {{van den
  Brand}}, \citenamefont {{van Leewen}}, \citenamefont {{van Veggel}},
  \citenamefont {{van den Broeck}}, \citenamefont {{Vecchio}}, \citenamefont
  {{Veitch}}, \citenamefont {{Vetrano}}, \citenamefont {{Vicere}},
  \citenamefont {{Vyatchanin}}, \citenamefont {{Willke}}, \citenamefont
  {{Woan}}, \citenamefont {{Wolfango}},\ and\ \citenamefont
  {{Yamamoto}}}]{Punturo-2010-Abernathy-CQGra..27h4007P}%
  \BibitemOpen
  \bibfield  {author} {\bibinfo {author} {\bibfnamefont {M.}~\bibnamefont
  {{Punturo}}} {\it et~al.},\ }\bibfield  {title} {\enquote {\bibinfo {title}
  {{The third generation of gravitational wave observatories and their science
  reach}},}\ }\href {\doibase 10.1088/0264-9381/27/8/084007} {\bibfield
  {journal} {\bibinfo  {journal} {Classical and Quantum Gravity}\ }\textbf
  {\bibinfo {volume} {27}},\ \bibinfo {eid} {084007} (\bibinfo {year}
  {2010}{\natexlab{b}})}\BibitemShut {NoStop}%
\bibitem [{\citenamefont {{Abbott}}\ \emph {{\it et~al.}}(2017)\citenamefont
  {{Abbott}}, \citenamefont {{Abbott}}, \citenamefont {{Abbott}}, \citenamefont
  {{Abernathy}}, \citenamefont {{Ackley}}, \citenamefont {{Adams}},
  \citenamefont {{Addesso}}, \citenamefont {{Adhikari}}, \citenamefont
  {{Adya}}, \citenamefont {{Affeldt}},\ and\ \citenamefont
  {et~al.}}]{LIGO-2017CQGra..34d4001A}%
  \BibitemOpen
  \bibfield  {author} {\bibinfo {author} {\bibfnamefont {B.~P.}\ \bibnamefont
  {{Abbott}}} {\it et~al.},\ }\bibfield  {title} {\enquote {\bibinfo {title}
  {{Exploring the sensitivity of next generation gravitational wave
  detectors}},}\ }\href {\doibase 10.1088/1361-6382/aa51f4} {\bibfield
  {journal} {\bibinfo  {journal} {Classical and Quantum Gravity}\ }\textbf
  {\bibinfo {volume} {34}},\ \bibinfo {eid} {044001} (\bibinfo {year}
  {2017})},\ \Eprint
  {http://arxiv.org/abs/1607.08697}{arXiv:1607.08697}\BibitemShut {NoStop}%
\end{thebibliography}
%

\end{document}